\documentclass[a4paper,fleqn,usenatbib]{mnras}
\usepackage{mathptmx}

\usepackage[T1]{fontenc}
\usepackage{ae,aecompl}

\usepackage{pdflscape}
\usepackage{longtable}
\usepackage{rotating}

\usepackage{graphicx}	
\usepackage{amsmath}	
\usepackage{amssymb}	
\usepackage{booktabs}
\usepackage{float}
\usepackage{array}
\usepackage{wrapfig}
\usepackage{multirow}
\usepackage{tabu}
\usepackage{adjustbox}
\usepackage{tabularx}
\usepackage{booktabs,caption,fixltx2e}
\usepackage[flushleft]{threeparttable}
\usepackage{subcaption}
\usepackage{mathtools}

\usepackage[dvipsnames]{xcolor}
\hypersetup{colorlinks=true,linkcolor=Magenta,urlcolor=Plum,citecolor=DarkOrchid}

\newcommand\T{\rule{0pt}{2.6ex}}       
\newcommand\B{\rule[-1.2ex]{0pt}{0pt}} 
\newcommand{\HI}{\mbox {\sc H\thinspace{i}}}
\newcommand{\HIX}{\mbox {\sc H\thinspace{ix}}}

\title{The Neutral Hydrogen Properties of Galaxies in Gas-rich Groups}

\author[D\v{z}ud\v{z}ar et al.]{
Robert D\v{z}ud\v{z}ar$^{1}$,\thanks{E-mail: rdzudzar@swin.edu.au}
  Virginia Kilborn$^1$,
  Gerhardt Meurer$^{2}$,
  Sarah M. Sweet$^{1,3}$,
  \newauthor Michael Drinkwater$^4$,
  Kenji Bekki$^{2}$,
  Fiona Audcent-Ross$^{2}$,
  Baerbel Koribalski$^{5}$,
  \newauthor Ji Hoon Kim$^{6}$,
  Mary Putman$^{7}$,
  Emma Ryan-Weber$^{1,3}$,
  Martin Zwaan$^{8}$,
 \newauthor Joss Bland-Hawthorn$^{9}$,
  Michael Dopita$^{10}$,
  Marianne T. Doyle-Pegg$^{11}$,
  Ed Elson$^{12}$,
  \newauthor Kenneth Freeman$^{10}$,
  Dan Hanish$^{13}$,
  Tim Heckman$^{14}$,
  Robert Kennicutt$^{14}$,
  Pat Knezek$^{15}$,
  \newauthor Martin Meyer$^{2,3}$,
  Chris Smith$^{16}$,
  Lister Staveley-Smith$^{2}$,
  Rachel Webster$^{17}$,
  Jessica Werk$^{18}$ \\
  $^1$ Centre for Astrophysics and Supercomputing, Swinburne University of Technology, PO Box 218, Hawthorn, VIC 3122, Australia\\
  $^2$ International Centre for Radio Astronomy Research, ICRAR M468, 35 Stirling Highway, Crawley, WA 6009, Australia\\
  $^3$ ARC Centre of Excellence for All Sky Astrophysics in 3 Dimensions (ASTRO 3D)\\
  $^4$ School of Mathematics and Physics, University of Queensland, Brisbane, QLD 4072, Australia\\
  $^{5}$ Australia Telescope National Facility, CSIRO, P.O. Box 76, Epping, NSW 1710, Australia  \\
  $^{6}$ Subaru Telescope, National Astronomical Observatory of Japan, 650 North A'ohoku Place, Hilo, HI 96720, USA\\
  $^{7}$ Department of Astronomy, Columbia University, 550 West 120th Street, New York, 10027, USA  \\
  $^{8}$ European Southern Observatory, Karl-Schwarzschild-Strasse 2, D-85748 Garching bei M\"{u}nchen, Germany \\
  $^{9}$ Sydney Institute for Astronomy, University of Sydney, Sydney, NSW 2006, Australia \\
  $^{10}$ Research School of Astronomy and Astrophysics (RSAA), Australian National University, Cotter Road, Weston Creek, ACT 2611, Australia \\
  $^{11}$ Department of Physics, University of Queensland, Brisbane, QLD 4072, Australia \\
  $^{12}$ Department of Physics \& Astronomy, University of the Western Cape, Cape Town 7535, South Africa\\
  $^{13}$ Department of Physics and Astronomy, John Hopkins University, 3400 North Charles Street, Baltimore, MD 21218, USA\\  
  $^{14}$ Institute of Astronomy, University of Cambridge, Madingley Road, Cambridge CB3 0HA, UK  \\
  $^{15}$ National Science Foundation, 4201 Wilson Boulevard, Arlington, Virginia 22230, USA     \\
  $^{16}$ Cerro Tololo Inter-American Observatory (CTIO), Casilla 603, La Serena, Chile  \\
  $^{17}$ School of Physics, University of Melbourne, VIC 3010, Australia   \\
  $^{18}$ Astronomy Department, University of Washington, 3910 15th Ave. NE, Seattle, WA 98195-0002, USA }
\date{Accepted XXX. Received YYY; in original form ZZZ}

\pubyear{2018}

\begin{document}
\label{firstpage}
\pagerange{\pageref{firstpage}--\pageref{lastpage}}
\maketitle
\begin{abstract}
We present an analysis of the integrated neutral hydrogen (\HI) properties for 27 galaxies within nine low mass, gas-rich, late-type dominated groups which we denote ``Choirs". We find that majority of the central Choir galaxies have average \HI\ content: they have a normal gas-mass fraction with respect to isolated galaxies of the same stellar mass. In contrast, we find more satellite galaxies with a lower gas-mass fraction than isolated galaxies of the same stellar mass. A likely reason for the lower gas content in these galaxies is tidal stripping. Both the specific star formation rate and the star formation efficiency of the central group galaxies are similar to galaxies in isolation. The Choir satellite galaxies have similar specific star formation rate as galaxies in isolation, therefore satellites that exhibit a higher star formation efficiency simply owe it to their lower gas-mass fractions. We find that the most \HI\ massive galaxies have the largest \HI\ discs and fall neatly onto the \HI\ size-mass relation, while outliers are galaxies that are experiencing interactions. We find that high specific angular momentum could be a reason for galaxies to retain the large fraction of \HI\ gas in their discs. This shows that for the Choir groups with no evidence of interactions, as well as those with traces of minor mergers, the internal galaxy properties dominate over the effects of residing in a group. The probed galaxy properties strengthen evidence that the Choir groups represent the early stages of group assembly.

\end{abstract}



\section{Introduction}

Today, we observe the hierarchical structure of the Universe \citep{Press1974, White1978} ranging from sparse voids to the densest of galaxy clusters. Small groups are the most common environment where galaxies reside (e.g. \citealt{Tully1987, Eke2004, Knobel2015, Saulder2016}). Since environment has an impact on galaxy evolution, it is important to understand how galaxies evolve within such structures. Environmental impact can be traced via galaxy properties such as the observed morphology-density relation (the decline of the fraction of late-type galaxies with increasing environmental density) in clusters \citep{Dressler1980, Goto2003} as well as in groups \citep{Postman1984, Serra2012}. Furthermore, studies have shown that galaxy's color and morphology are connected with the galaxy environment \citep{Whitmore1993, Kauffmann2004, Blanton2005, Brough2006, Melnyk2014}, and that the star formation rate suppression becomes evident at group densities \citep{Lewis2002, Gomez2003, Davies2015, Barsanti2018}. Nevertheless, the relative contribution of nature vs. nurture (internal properties of a galaxy versus an environment in which galaxy resides) in small gas-rich groups remains an open question with arguments on both sides (e.g. \citealt{Serra2012, Knobel2015, Odekon2016, Carollo2016, Brown2016, Janowiecki2017, Spindler2018}).

Environmental effects can be probed by mapping the distribution of the neutral hydrogen (\HI) content. The \HI\ gas in spiral galaxies usually extends from 1.5--2 times the size of their observed stellar disc measured at the \textit{R}-band 25 mag arcsec$^{-2}$ surface brightness level (e.g. \citealt{Bosma1981, Hoffman1996, Pisano2000, Bigiel2012}). Due to the extent of the \HI\ disc, it is more susceptible than the stellar disc to disturbances via external gravitational (tidal interactions, galaxy mergers) and hydrodynamical (ram-pressure stripping) interactions. These properties make the \HI\ gas an excellent tracer of the physical processes that are affecting galaxies (e.g. \citealt{GunnGott1972, Yun1994, Barnes2001, VerdesMontenegro2001, Hess2017}).

Tidal interactions and galaxy-galaxy mergers are often observed in galaxy groups (e.g. \citealt{Yun1994, Rots1990}). Such features are prominent in  Hickson Compact Groups (HCG, compact configurations with four or more galaxy members within a magnitude difference of 3.0, obtained using red prints from the Palomar Observatory Sky Survey, \citealt{Hickson1982}) and theoretically they are often connected to a transformation of late-type galaxies into early-type galaxies \citep{Coziol2007}. Galaxy interactions are also observed in small (fewer members, \citealt{Barnes2001}) and loose (sparser, \citealt{Osterloo2018}) groups. There are also indications that ram-pressure stripping can remove gas from galaxies in galaxy groups \citep{Rasmussen2006, Westmeier2011, Wolter2015, Brown2017, Stevens2017} and not just in clusters \citep{GunnGott1972, Chung2009}. Enormous effort is aimed towards understanding how such interactions relate to the observed \HI\ content in galaxies and conversely how \HI\ distribution and kinematics uncover the physics of interaction (e.g. \citealt{Bekki2008, Pisano2011, Mihos2012, Elagali2018, Bosma2017}, and references therein).

Spiral galaxies have a lower \HI\ fraction in high density environments, for instance, near the centers of galaxy clusters (\citealt{Giovanelli1985, Chung2009}) or in HCGs \citep{VerdesMontenegro2001}, than spiral galaxies that reside in the field. \citet{VerdesMontenegro2001} proposed an evolutionary path for gas in compact galaxy groups. They proposed that the \HI\ gas evolution in the compact groups occurs through a gradual removal of the gas from galaxies, increasing its content in the intragroup medium. This evolutionary picture should be expanded since gas poor galaxies are also found in intermediate density environments \citep{Denes2016, Hess2013} as well as in loose groups \citep{Kilborn2005, Kilborn2009}. With the observations of the gas content in group galaxies we are gathering evidence that gas pre-processing starts to be evident. However, it is still unclear at which density scale the environment starts affecting the gas content of galaxies. 

Referring back to hierarchical structure formation theory, a galaxy embedded in the centre of a dark matter halo (central galaxy) grows as it accretes satellite galaxies, over time becoming the most massive and the brightest galaxy \citep{White1978, Brough2006}. In this scenario, the satellite galaxies are moving with respect to the centre of group potential and experiencing environmental effects, thus in theory they are going through a different evolutionary path from that of the central galaxy \citep{White1978, Yang2007, Brown2017, White1978, Brough2006, Stevens2017}. Observationally, and in simulations, it is often difficult to distinguish between the central and satellite galaxy, especially in a small groups \citep{Berlind2006, Campbell2015}. 

Comparing central galaxies in groups with isolated galaxies, \citet{Janowiecki2017} found that in low mass groups, central galaxies have a higher gas content than similar galaxies in isolation which they attributed either to inflows of gas from the cosmic web, or, mergers with \HI-rich galaxies; whilst in higher mass groups central galaxies have similar gas content as galaxies in isolation. This picture is consistent with the \HI-rich galaxies from the Bluedisk sample \citep{Wang2015} for which they found that \HI-rich galaxies are embedded within \HI-rich environments. Whether accretion from gas-rich mergers or accretion from the cosmic web is the main reservoir from where galaxies accrete the majority of their gas is still unclear. The contribution from gas-rich mergers is thought to be rather small \citep{Sancisi2008, DiTeodoro2014}, and the current observations are not sensitive enough \citep{Popping2009} to probe the cold mode accretion from cosmic web filaments, which in theoretical models is the dominant reservoir from where galaxies obtain their gas \citep{Keres2005}.

In this paper we investigate the extent to which the galaxy group environment influences evolution of the galaxies in small gas-rich groups, through a comparison of the group galaxies to a sample of isolated galaxies. 

This paper is structured as follows. Section \ref{Sample} describes our Choirs sample and shows Choir group HIPASSJ1250-20 in detail, describes the selection of isolated galaxies and summarises the observations of group galaxies and data reduction. Section \ref{Results} describes the main results of this paper: comparison of group galaxies with isolated galaxies in terms of gas-mass fraction, specific star formation rate, star formation efficiency, atomic depletion time and deficiency. This section also presents the size-mass relation and examines global stability parameter of the Choir galaxies. Section \ref{Discussion} discusses the implications of the obtained results, and Section \ref{Summary} summarises the results and concludes. \\
Throughout this paper the assumed cosmology is H$_{0}$ = 70 km s$^{-1}$ Mpc$^{−1}$, $\Omega_{\text{m}}$ = 0.3, and $\Omega_{\Lambda}$ = 0.7.

\section{Sample and data}
\label{Sample}

In this work we make use of resolved interferometric \HI\ data for 9 Choir groups, obtained with the Australia Telescope Compact Array (ATCA) and the Karl G. Jansky Very Large Array (VLA). A brief summary of the observations is provided in Table \ref{tab:Choir_ATCA}. We examine the properties of the individual galaxies within the Choir groups and compare them to the sample of isolated galaxies. 

\subsection{Choir group galaxies}
Our groups are selected from the Survey of Ionization in Neutral Gas Galaxies (SINGG), the H$\alpha$ narrow-band imaging follow-up program \citep{Meurer2006} for the \HI\ Parkes All Sky Survey (HIPASS; \citealt{Barnes2001}). 
Galaxies for the SINGG survey were selected based on their \HI\ mass and thus the sample is biased towards more distant, massive galaxies \citep{Meurer2006, Sweet2013}. The HIPASS galaxies were separated into log(M$_{\HI}$) bins with width of 0.2 dex and then the nearest galaxies in each log(M$_{\HI}$) bin were selected. Additional cuts were made in the selection: galaxies were required to have a HIPASS flux with S/N larger than 3.8; galaxies with Galactic latitude lower than 30 deg were not selected to avoid Galactic plane; galaxies with systemic velocities smaller than 200 km s$^{-1}$ were not selected to avoid the Milky Way and high-velocity clouds. Detailed explanations can be found in \citet{Meurer2006, Audcent2018} and Meurer et al. (in prep.).

27 per cent of the HIPASS detections in the SINGG sample correlate with multiple H$\alpha$ emitting galaxies\footnote{It should be noted that the number of the single and multiple detections relates only to the field of view of 14.7$\arcmin$} (due to the large beamsize of the Parkes telescope at 21cm). We are investigating those cases where four or more H$\alpha$ emitting galaxies were detected within a SINGG field-of-view (around seven per cent of the SINGG fields). 15 groups were denoted as SINGG Choirs in \citet{Sweet2013} and they are small groups of four to 10 members. Choir groups masses range between 10$^{11}$ and 10$^{12}$ M$_{\odot}$ (measured as the sum of \HI\ mass and stellar mass). Choir group properties such as total group mass, specific star formation rates, surface brightness, H$\alpha$ equivalent width and group \HI\ deficiency were examined with respect to the entire SINGG sample in \citet{Sweet2013}. They show that the galaxies in the Choir groups are very similar to the overall SINGG sample. This paper focuses on the integrated properties of nine Choir groups for which neutral hydrogen content was mapped.

\subsection{Isolated SINGG galaxies}

In order to examine whether the galaxies within the Choir groups are affected by the group environment, we compile a sub-sample of SINGG galaxies that are isolated and thus unlikely to have been affected by neighbouring galaxies within the past few Gyr. We compile such a sample because we found that outside the SINGG field of view (14.7$\arcmin$), single detected galaxies are members of different environments ranging from being isolated to being in clusters. Galaxies can reside in the different environments due to the fact that nearby groups may break up into multiple \HI\ detections and may in part be captured in SINGG as single galaxies \citep{Sweet2013}.

Multiple catalogues of isolated galaxies exist; these have been created by imposing different isolation criteria. The Catalogue of Isolated Galaxies (CIG) \citep{Karachentseva1973} imposed the criteria that a galaxy is isolated if it is a single galaxy within a projected distance relative to neighbours diameter: $\textrm{D}_{\textrm{p}}/4 \leqslant \textrm{D}_{\textrm{i}} \leqslant 4\textrm{D}_{\textrm{p}}$, where D$_{\textrm{p}}$ diameter of the primary galaxy, D$_{\textrm{i}}$ is diameter of the neighbouring galaxies; thus, the projected distance should satisfy criteria: $R_{\textrm{i,p}} \geqslant 20D_{\textrm{i}}$.

Applying similar criteria to the one from CIG sample, albeit using \textit{Ks}-band (instead of Zwicky magnitudes) for determining galaxy angular diameters, \citet{Karachentseva2010} made the 2MIG catalogue of isolated galaxies. \citet{VM2005} created the AMIGA (Analysis of the interstellar Medium of Isolated GAlaxies) sample of isolated galaxies which have not been affected by other galaxies in the past 3 Gyr (assuming a field velocity of 190 km s$^{-1}$, as adapted from \citet{Tonry2000}). The AMIGA sample was later revised by \citet{Argundo2013} who imposed additional constraints to those from \citet{Karachentseva1973} and \citet{VM2005} with the advantage of having the photometric SDSS-DR9 catalogue, thus creating a sample of 636 isolated galaxies. The SIGRID sample (the Small Isolated Gas-Rich Irregular Dwarfs), compiled by \citet{Nicholls2011}, catalogued galaxies that have not had interactions within the last 5 Gyr. Using their own isolation criteria, \citet{Reda2004} created a sample of isolated early type galaxies with the constraints that there are no galaxies within 700 km s$^{-1}$ in recession velocity, within 0.67 Mpc in the plane of the sky, and within 2 \textit{B}-band magnitudes of the isolated galaxy.

To compare the Choir group galaxies to those in isolation, we compiled a catalogue of isolated galaxies from the SINGG sample. Our isolation criteria closely follows those found in the literature, including a projected distance from the neighbouring galaxies, recessional velocity difference with neighbouring galaxies, as well as proximity to a galaxy cluster. 
For each SINGG galaxy we firstly determine galaxies which are single within 0.67 Mpc in the plane of the sky and within a velocity of $\pm$750 km s$^{-1}$, as imposed in \citet{Reda2004}. For this analysis we used NED (NASA/IPAC Extragalactic Database) to find all known galaxies with redshifts around each galaxy from the SINGG sample, thus the data varies in sensitivity. Applying restrictions on projected distance and velocity offset we obtained a sample of 25 galaxies. We can approximate the time since the last encounter for these galaxies, assuming that the galaxy perturber has an average diameter D$_{a}\sim$ 38 kpc and field velocity of $\sim$ 190 km s$^{-1}$. Thus a perturber of the same size would need $\sim$ 4 Gyr to traverse the distance equal to 20D$_{a}$. This time is similar to the one found for the AMIGA, SIGRID and 2MIG samples (we do find that some galaxies from our sample are also members of the SIGRID \citet{Melnyk2015} and 2MIG \citet{Karachentseva2010} samples). Secondly, we remove six galaxies which are within the projected radius of the Virgo cluster and within two times its velocity dispersion, adopting properties of the Virgo cluster given by \citet{Nicholls2011}. By removing galaxies which are in the vicinity of the cluster we discarded possible interlopers which could be infalling cluster galaxies. We additionally flagged three galaxies, one with a noisy \HI\ spectrum that could led to an error in the \HI\ mass estimation, one due to large uncertainty in its stellar mass and one with three H$\alpha$ clumps spreading outside of its stellar disc. With this procedure we created a sample of 16 isolated galaxies. The isolated galaxies and their properties are shown in Table \ref{isolated_properties}. We acknowledge that it is possible, in spite of these isolation criteria, to have  contamination in a sample of isolated galaxies due to low surface brightness galaxies. However within the 14.7$\arcmin$ FOV of these galaxies in SINGG, emission line galaxies below \textit{R}-band apparent magnitude of $\sim$20 were not detected \citep{Meurer2006}, thus contaminants should have negligible effects.

\begin{table*}
\captionsetup{width=1\columnwidth}
\caption{Summary of the ATCA and VLA \HI\ observations used in this paper. Columns: (1) Group name (HIPASS+ID); (2) Time on the source for given ATCA and VLA array configuration; (3) Used phase calibrator; (4) Central frequency of band in MHz. (5) Synthesized beam size minor and major axis, respectively; (6) RMS in the created data cube. Galaxy groups observed with the VLA in May 2013 are part of the project 13A-207, all of the ATCA observations were carried out in the project C2440, observed in 2011, 2015 (PI Kilborn); except J0205-55, which was observed in the projects: C1239 (750m Array) and C1294 (1.5km Array and EW Array) in 2004. Additionally, 1.5km baseline for J1250-20 was observed in 2017 (C2440, PI Dzudzar).}
\begin{tabular}{ccccccc}
\toprule
Group ID    & 1.5/750/EW [h] & Phase Calibrator & f$_{C}$ [MHz] & $\theta_{\textrm{min}} \times \theta_{\textrm{maj}}$ [$''$] & rms  [mJy beam$^{-1}$] \\
(1)		 & (2)		& (3)			   & (4)  & (5) &  (6) \B \\ 	\hline \hline
J0205-55 & 10.49/8.45/9.22   & PKS 0252-712 	& 1391  & 45.12 $\times$ 58.94& 1 \T \\
J0258-74 & 3.52/3.45/3.95    & PKS 0252-712 	& 1398  & 37.49 $\times$ 45.57& 1.5    \\
J0400-52 & 6.49/3.85/6.12	 & PKS 0420-625 	& 1371  & 44.96 $\times$ 56.76&  1.2  \\
J1051-17 & 8.84/2.46/4.34	 & PKS 1127-145 	& 1393  & 29.77 $\times$ 108.53 & 1.3   \\
J1250-20 & 7.09/2.07/3.74    & PKS 1245-197 	& 1384  & 30.79 $\times$ 102.8 & 2   \\
J2027-51 & 5.15/1.50/2.76    & PKS 1933-587 	& 1393  & 37.52 $\times$ 50.53& 1.65 \B  \\  \hline
 & VLA:DnC & & & \T \B \\ \hline
J1059-09   & 3.18   & J1130-1449 		& 1383	& 26.27 $\times$ 44.04 & 1.6 \T \\ 
J1026-19  & 2.98   & J1130-1449 		& 1379	& 26.10 $\times$ 46.34 & 1.4 \\
J1408-21  & 2.98   & J1337-1257 		& 1380	& 28.62 $\times$ 46.44 & 1.4   \\ 
\bottomrule
\end{tabular}                                   
\label{tab:Choir_ATCA} 
\end{table*}

\subsection{ATCA \& VLA observations and data reduction}
\label{reduction}

In this paper, we utilise observations obtained with the CFB 64M-32k (34 kHz resolution), CFB 1M-0.5k (0.5 kHz resolution) correlator configuration on the Compact Array Broadband Backend (CABB; \citealt{Wilson2011}) as well as a set of observations (HIPASSJ0205-55a) with the previous correlator (16 kHz resolution).

We also utilise observations obtained with the Karl G. Jansky Very Large Array (VLA, project 13A-207) in the hybrid DnC configuration, having a total bandwidth of 4 MHz, divided into 1024 channels with a width of 3.906 kHz.

The ATCA and VLA data were reduced using standard procedures with the MIRIAD \citep{Sault1995} and CASA \citep{McMulin2007} software packages, respectively. For the ATCA data we first removed radio frequency interference and then calibrated (bandpass, flux and phase) each data set separately. In the next step, we performed a continuum subtraction using a linear fit to line-free channels in each observational data set. We combined all baseline configurations for each data set in the Fourier transformation using task \texttt{INVERT}. This task was utilised excluding antenna six (CA06), using Brigg's robust parameter of 0.5 and a channel width of 5 km s$^{-1}$. The resulting cube for each data set was cleaned, restored and primary beam corrected. With the task \texttt{MOMENT} we created moment 0, 1 and 2 maps with a 3$\sigma$ clipping.

The VLA data were reduced in the same manner as the ATCA data using  CASA. Firstly, we flagged data, calibrated the observations and then subtracted the continuum. Secondly, we cleaned our data cubes using a robust parameter of 0.5 and then we corrected each data cube for the primary beam. Finally, we created moment 0, 1 and 2 maps with a 3$\sigma$ clipping.

In this work we utilise the obtained integrated galaxy properties: \HI\ fluxes, masses and diameters, and we show ATCA \HI\ intensity distribution of Choir HIPASSJ1250-20. The full sample of the moments maps will be published and analysed in a subsequent paper. 

\section{Results}
\label{Results}

\subsection{The HIPASSJ1250-20 group}
\label{J1250}
\begin{figure}
\centering
\includegraphics[width=84mm]{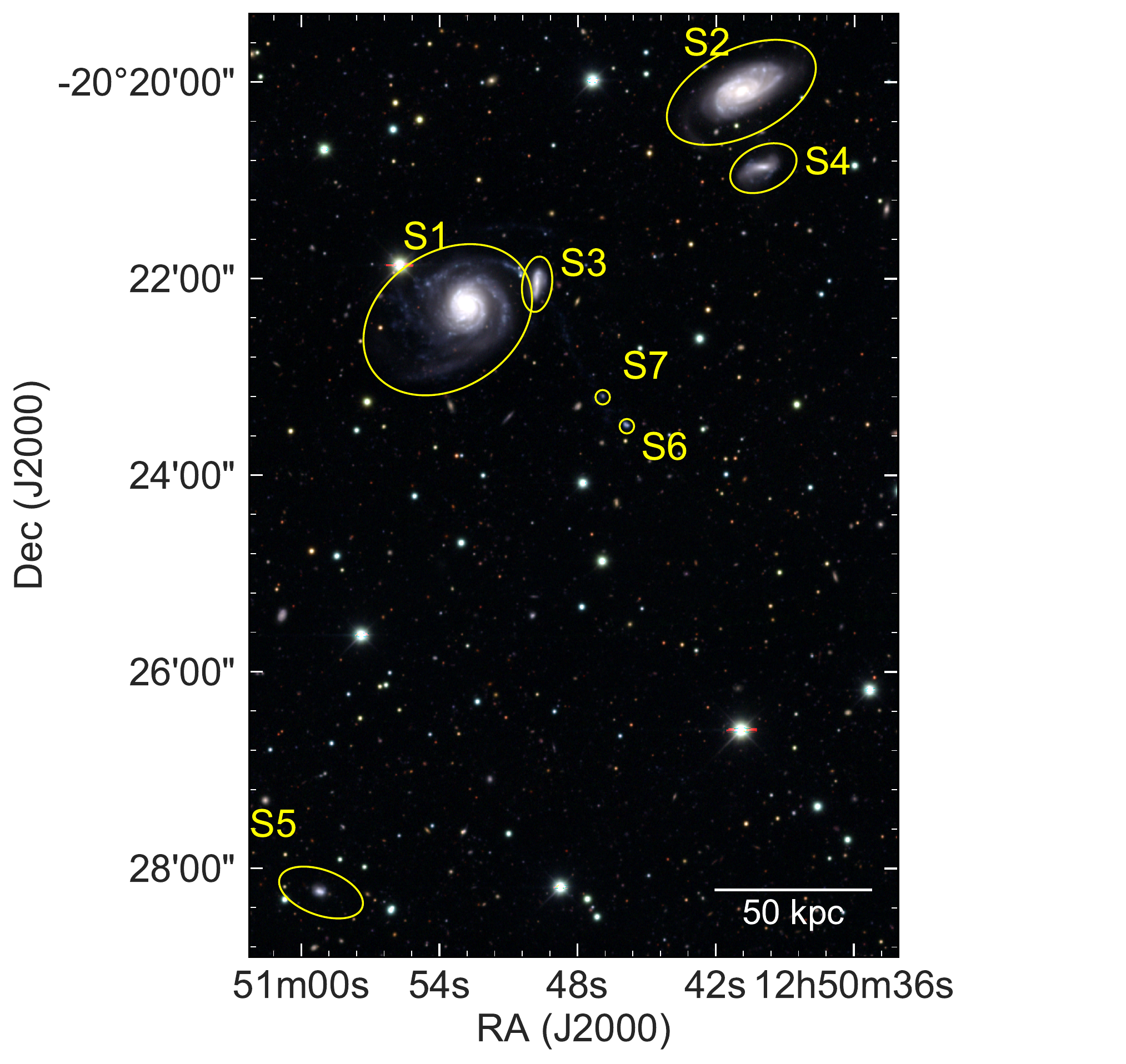}
\caption{The DECam \textit{g}, \textit{r} and \textit{i} band colour composite of the Choir group HIPASSJ1250-20. Galaxies within the yellow ellipses are the group members, marked from S1 to S7 as in \citet{Sweet2013}. The scale bar in the bottom right corner shows 50 kpc. North is up, east is to the left.}
\label{fig:group}
\end{figure}

We present the HIPASSJ1250-20 (hereafter J1250) Choir group and its complexity as an illustration for our discussion throughout this work.  The optical observations of J1250 were carried out with the Dark Energy Camera (DECam) on the CTIO Blanco 4-m telescope in the \textit{g}, \textit{r}, \textit{i} and \textit{z} bands. The exposure times of the observations were 3000, 1920, 3000 and 6486 seconds respectively, these are the deepest optical images of this group (based solely on the exposure time calculator, these images reach $\sim$26 mag arcsec$^{-2}$ with S/N$\sim$3). The optical images of similar depth for eight Choir groups will be included in the next paper. In Figure \ref{fig:group}, we show the DECam \textit{g}, \textit{r} and \textit{i} colour composite image of J1250 group and labels of the group members from S1 to S7. These members were presented in \citet{Sweet2013}: two large spiral galaxies (S1 and S2; respectively ESO575-G006 and ESO575-G004), three dwarf galaxies (S3, S4 and S5) and two small and compact H$\alpha$ emitters that they note as tidal dwarf candidates (S6 and S7). 

\begin{figure*}
\centering
\includegraphics[width=150mm]{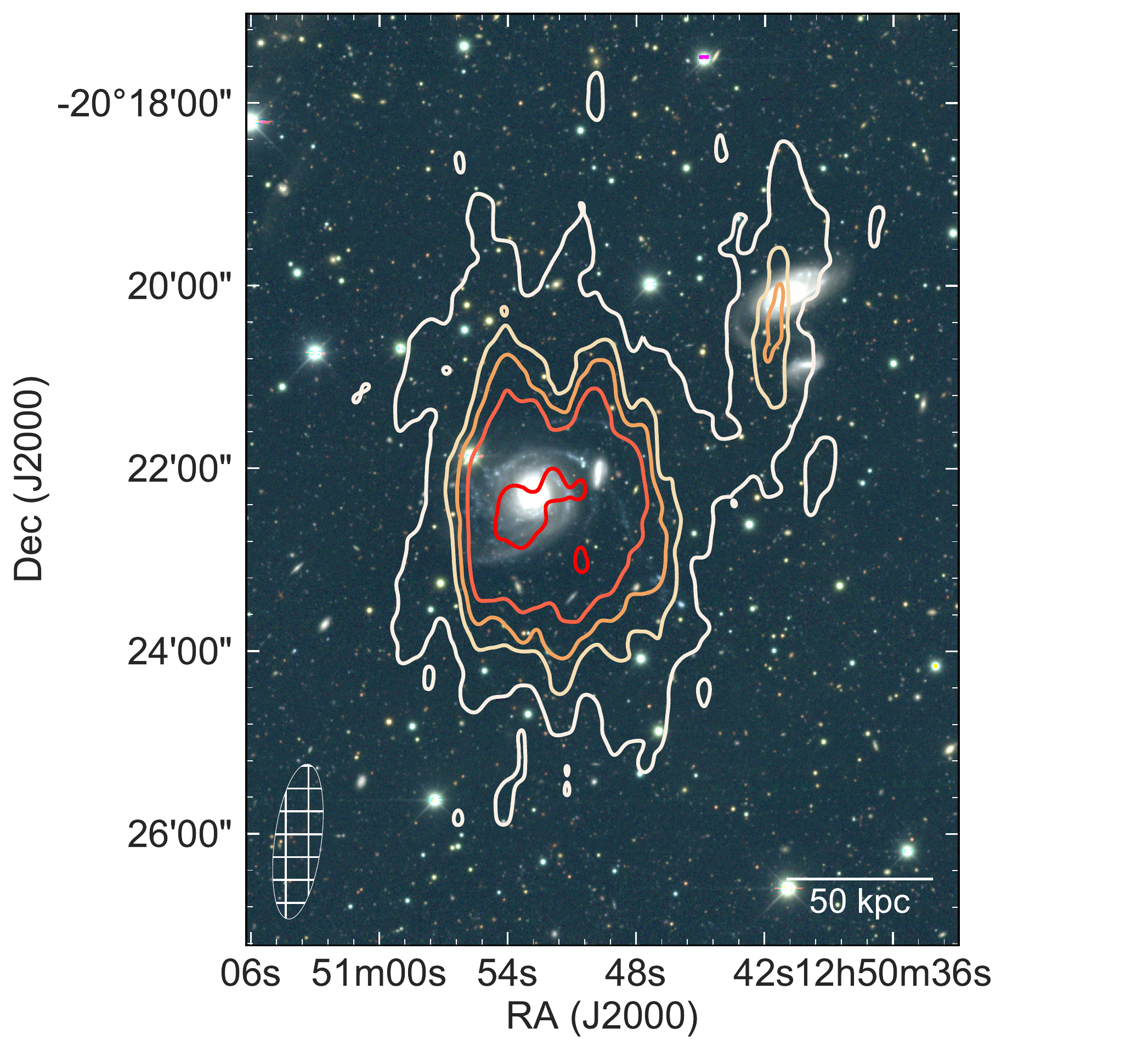}
\caption{ATCA \HI\ intensity distribution of HIPASSJ1250-20 galaxy group shown by the contours overlaid on the DECam \textit{g}, \textit{r}, \textit{i} colour composite image. The white contour is the lowest shown \HI\ column density and corresponds to a 3$\sigma$ detection over a velocity width of 15 km s$^{-1}$, equal to 3.5$\times$10$^{19}$ cm$^{-2}$ (0.1 Jy km s$^{-1}$). The other contours shown correspond to the \HI\ column density of 14, 20, 28 and 45 $\times$ 10$^{19}$ cm$^{-2}$ (0.4, 0.57, 0.8, 1.26 Jy km s$^{-1}$). The synthesized beam is shown in the bottom left corner. The scale bar in the bottom right corner shows 50 kpc. North is up, east is to the left.}
\label{fig:groupHI}
\end{figure*}

\begin{figure}
\centering
\includegraphics[width=1\columnwidth]{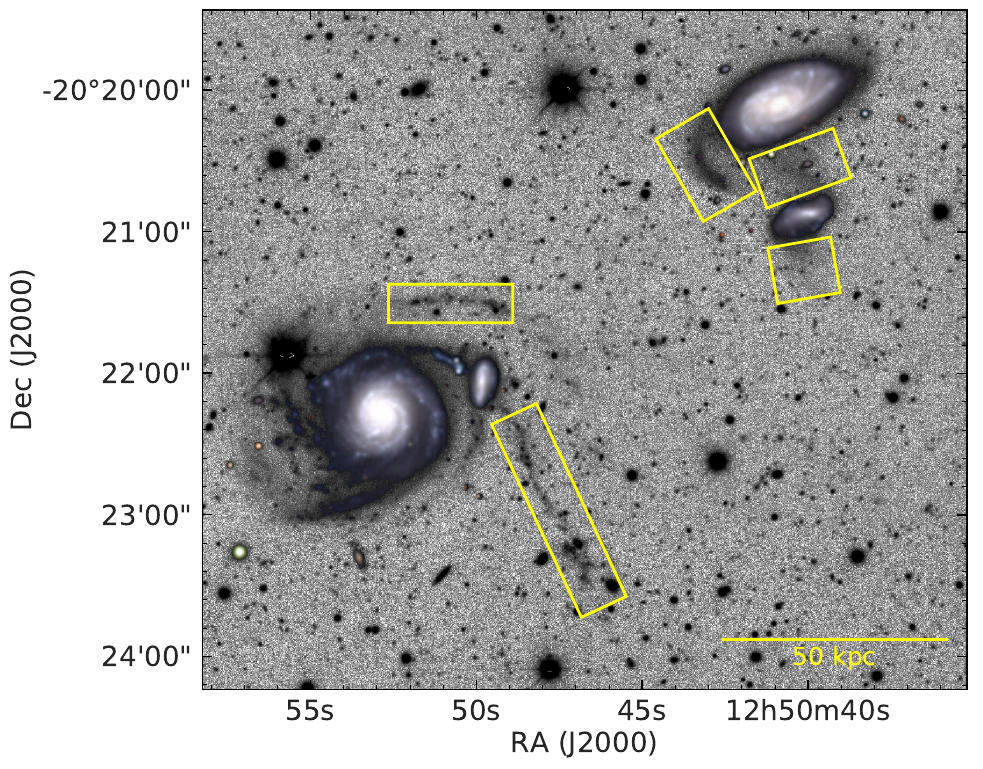}
\caption{The DECam \textit{g} band image, adaptively smoothed to show the faint stellar streams and the diffuse light around galaxies in HIPASSJ1250-20 group. We show a colour inset to enhance the disc of each galaxy. The yellow rectangles mark the faint stellar streams. The scale bar in the bottom right corner shows 50 kpc. North is up, east is to the left.}
\label{fig:stream}
\end{figure}

We observed J1250 group with the Australia Telescope Compact Array and mapped its \HI\ content. The \HI\ observations are described in Section \ref{reduction} and summarised in in Table \ref{tab:Choir_ATCA}. Figure \ref{fig:groupHI} shows the \HI\ column density contours overlaid on the optical colour composite image of J1250. The lowest shown \HI\ column density, the white contour, is a 3$\sigma$ detection and the \HI\ distribution reveals a possible interaction between S1 and S2 galaxy; we find similar interaction through \HI\ distribution in the HIPASSJ1408-21 group. We must note that because of the large elongation of the synthesized beam (as the source is at $-$20 degrees in Declination), this could be a result of the beam smearing. We find no evidence of interaction between S1 and S2 galaxy in the optical or H$\alpha$ imaging at the sensitivity with which they were imaged (while we do find interaction in optical and H$\alpha$ in the HIPASSJ1059-09 and HIPASSJ1026-19 group). We find that the of the \HI\ emission is off centre in S2 galaxy. The total \HI\ mass of J1250 is log M$_{\HI}$ [M$_{\odot}$] = 10.51$\pm$0.11 \citep{Sweet2013}, which is in agreement with the mass quoted in the HIPASS catalogue: log M$_{\HI}$ [M$_{\odot}$] = 10.39$\pm$0.11 \citep{Meyer2004}. The slight difference in obtained mass is due to the difference in adopted group distance, \citet{Meyer2004} uses 100 Mpc, while we are using 114 Mpc as in \citet{Sweet2013}. With our ATCA observations we recover the \HI\ flux detected with the single dish telescope and the sum of the masses of (S1 + S2) is log M$_{\HI}$ [M$_{\odot}$] = 10.53$\pm$0.2. Using the ATCA data we can separate the \HI\ flux in S1 (\HI\ emission between 7630 and 7860 km s$^{-1}$) and S2 (\HI\ emission between 7685 and 7880 km s$^{-1}$) galaxies and we find that the HIPASS source is mistakenly assigned to S2 galaxy in the HyperLeda database. We show that the majority of the \HI\ gas appears to be associated with S1 galaxy, moreover the S1 is an \HI-rich galaxy whilst S2 is an \HI-poor galaxy. The dwarf galaxies S3, S6 and S7 are within the \HI\ envelope associated with S1 galaxy. A small \HI\ component associated with S2 galaxy coincides also with S4 dwarf galaxy (top right part in Figure \ref{fig:groupHI}). Galaxy marked S5 is not detected within the ATCA primary beam. We do not have spectra of S3, S4, S5 and S6 galaxies however they are assumed to be the group members as they are detected in SINGG H$\alpha$ imaging. This imaging used filters with a bandwidth of $\sim$ 30 \AA\ which corresponds to $\sim$ 3000 km s$^{-1}$ \citep{Meurer2006, Sweet2013}.

Both, S1 and S2 galaxies have comparable stellar masses, while the \HI\ mass of S1 is $\sim$ 7.6 times larger than the \HI\ mass of S2. This is the one of the groups where two galaxies have an equal probability of being the central galaxy (see next Section). The gas-mass fraction of S1 is -0.73 dex and it is within the 2$\sigma$ scatter from the fit to the isolated galaxies while the gas-mass fraction of S2 is -1.42 dex, and it is more than 2$\sigma$ lower from the fit to the isolated galaxies (see Figure \ref{fig:gasfraction}). 

Figure \ref{fig:stream} shows the DECam \textit{g} band image which we adaptively smoothed to enhance the faint stellar streams around the galaxies in J1250 (Illustrating the discussion in Section \ref{gasstream}). The stellar streams around S1 galaxy (bottom left side in Figure \ref{fig:stream}) have a blue colour and they are visible in the H$\alpha$ emission (Figure B10 in \citet{Sweet2013}), also S6 and S7 galaxies are detected in GALEX All Sky Survey far ultraviolet and near ultraviolet images. The stellar streams around S1 galaxy are possibly caused by a disrupted satellite galaxy. The streams around S2 galaxy (top right side in Figure \ref{fig:stream}) have a red colour, they do not show emission in the H$\alpha$ imaging and they are possibly caused by a tidal interaction with S4 galaxy. 

In this paper we present results from the analysis of similar quality \HI\ data (listed in Table \ref{tab:Choir_ATCA}) for the other eight groups.

\subsection{Galaxy properties}
\label{properties}

Analysing the interferometric observations from ATCA and VLA, we detected \HI\ in 27 galaxies within nine Choir groups.
For each galaxy we obtain their \HI\ mass from their \HI\ spectra using the standard formula:
\begin{equation}
\hspace{2cm} \textrm{M}_{\textrm{HI}} \hspace{0.05cm} [\textrm{M}_{\odot}] = 2.365\times10^{5} \textrm{D}^{2} \textrm{F}_{\textrm{HI}},\\
\label{mass}
\end{equation}
where D is the distance to the galaxy (Mpc) and F$_{\textrm{\HI}}$ is the integrated flux density (Jy km s$^{-1}$). The distance to the each galaxy is assumed to be equal to the galaxy's group distance, as used in \citet{Sweet2013} and \citet{Meurer2006}, based on the multipole attractor model \citep{Mould2000}. To determine the errors on the integrated flux density we follow the prescriptions by \citet{Koribalski2004}, whilst we assume a 10 per cent error on the distance following \citet{Meurer2006}. We use standard error propagation on the integrated flux density and the distance to determine the uncertainty of the \HI\ mass. The Choir galaxy \HI\ properties are summarised in Table \ref{hi_properties}.

We use dust-corrected \textit{R}-band AB magnitudes \citep{Meurer2006} to determine stellar masses (M$_{\star}$) using the relation between M$_{\star}$/L$_{\textrm{R}}$ and absolute \textit{R}-band magnitude, as derived for the SINGG sample, equation (4) in \citet{Wong2016}. We note that using near-infrared bands for obtaining stellar masses would be more precise since they are less affected by dust extinction and recent star formation, however we do not have \textit{Ks} fluxes for the entire sample of galaxies. In order to check the reliability of \textit{R}-band stellar masses, we compute \textit{Ks} stellar masses where possible, using images from the the 2 Micron All Sky Survey (2MASS; \citealt{Skrutskie2006}) and deriving stellar masses as prescribed by \citet{Wen2013}. On average, there is good agreement between the stellar masses obtained with the \textit{R}-band and \textit{Ks}-band, with a mean difference of 0.3 dex.

We derive the star formation rates (SFR), as in \citet{Sweet2013}, using the H$\alpha$ luminosities which are corrected for Galactic extinction and [NII] contamination (see details of the flux measurement in \citealt{Meurer2006}). To check for any Active Galactic Nuclei (AGN) contamination we identify the dominant source of ionization in the Choir galaxies using the `BPT' emission line diagnostic diagram of \citet{Baldwin1981}. Using the emission line fluxes from \citet{Sweet2014}, we examine the position of the Choir galaxies in BPT diagram using the line ratios: [NII]/H$\alpha$ versus [OIII]/H$\beta$ \citep{Kauffmann2003, Kewley2001} and identify four galaxies whose emission is dominated by AGN. We also use available classifications from \citet{Sweet2013} which is noted from the NED (NASA/IPAC Extragalactic Database) and find three more galaxies dominated by AGN radiation. We excluded these galaxies from the analysis of the specific star formation. 

\begin{figure}
\centering
\includegraphics[width=\columnwidth]{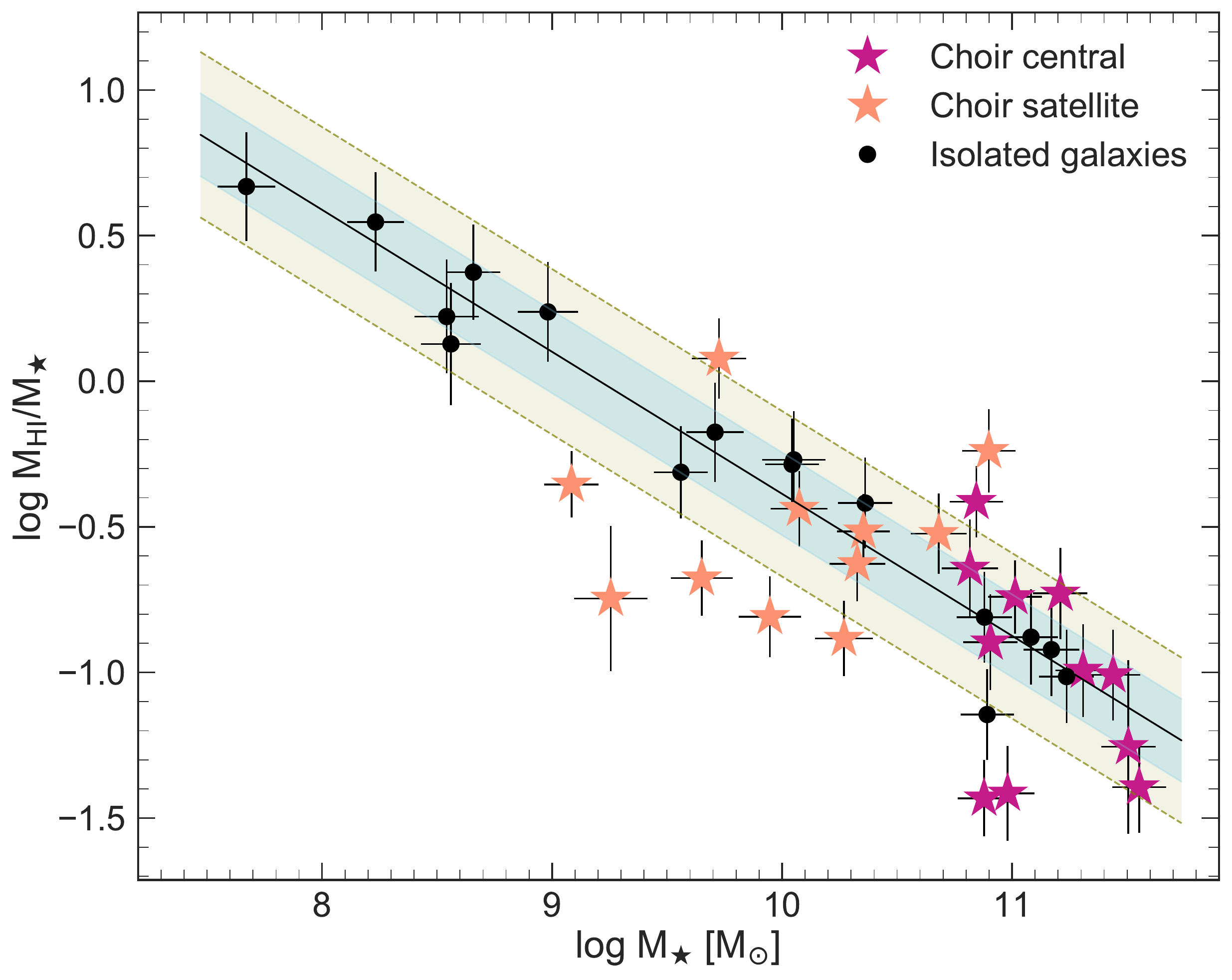}
\caption{\HI\ fraction versus stellar mass. The black dots are isolated galaxies. The magenta and orange stars correspond to the central and satellite galaxies in Choir groups, respectively. The solid line is a fit to the isolated galaxies, the blue shaded area corresponds to 1$\sigma$ scatter, while the yellow shaded area shows 2$\sigma$ scatter from the fit. }
\label{fig:gasfraction}
\end{figure}

\begin{figure}
\centering
\includegraphics[width=\columnwidth]{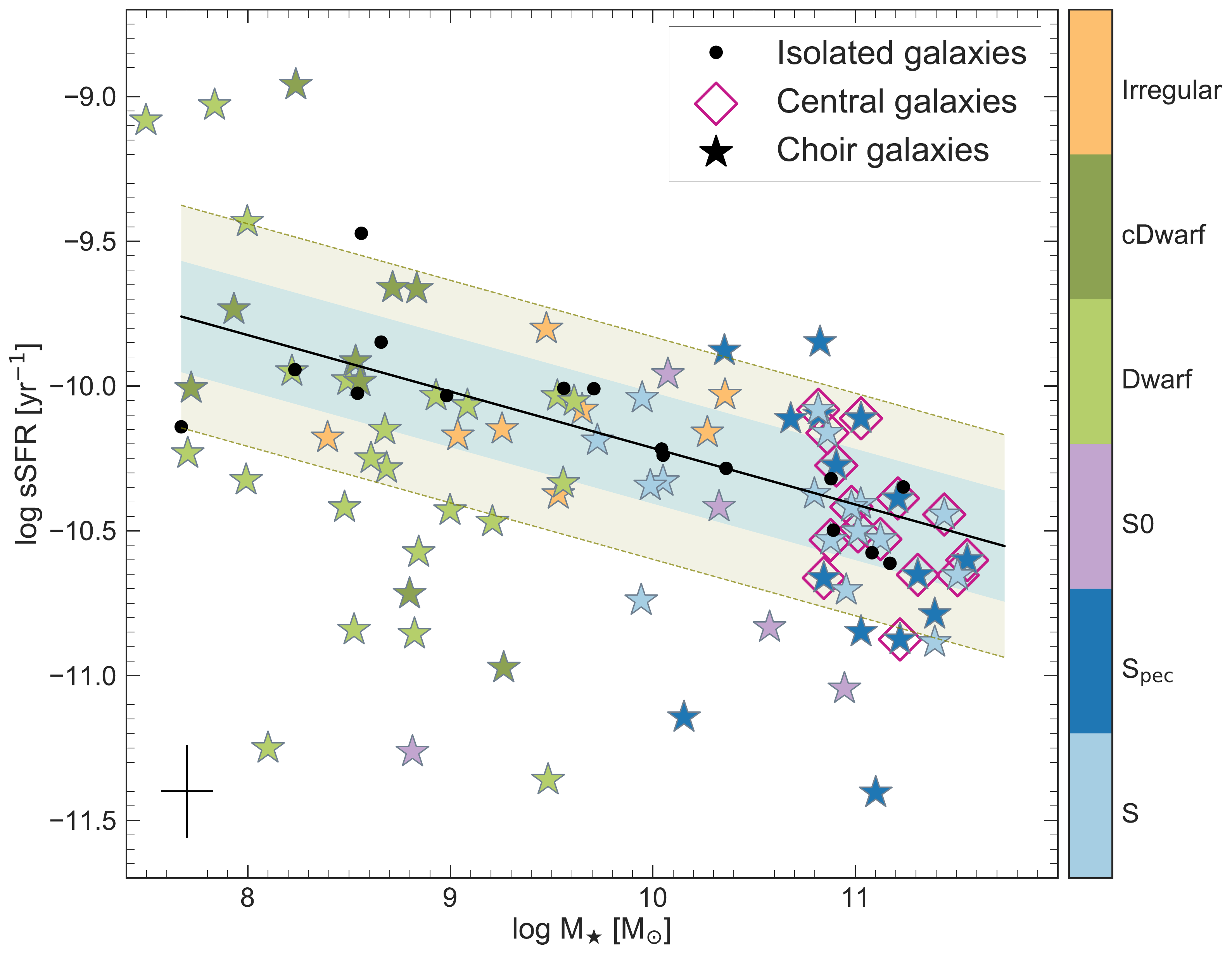}
\caption{Specific star formation rate versus stellar mass for all Choir galaxies. The black dots are isolated galaxies. The star symbols are all galaxies within Choir groups coloured by their morphology as indicated on colour bar (from top to bottom: Spiral, Spiral peculiar, Lenticular, Dwarf, Compact Dwarf and Irregular). The magenta diamonds around star symbols are central Choir galaxies. The solid line is a fit to the isolated galaxies; the blue shaded area corresponds to 1$\sigma$ scatter, while the yellow shaded area shows 2$\sigma$ scatter from the fit. The average error bar is shown in the bottom left corner. The morphologies of the isolated galaxies are shown in Table \ref{tab:Isolated}.}
\label{fig:sSFR}
\end{figure}

\begin{figure}
\centering
\includegraphics[width=\columnwidth]{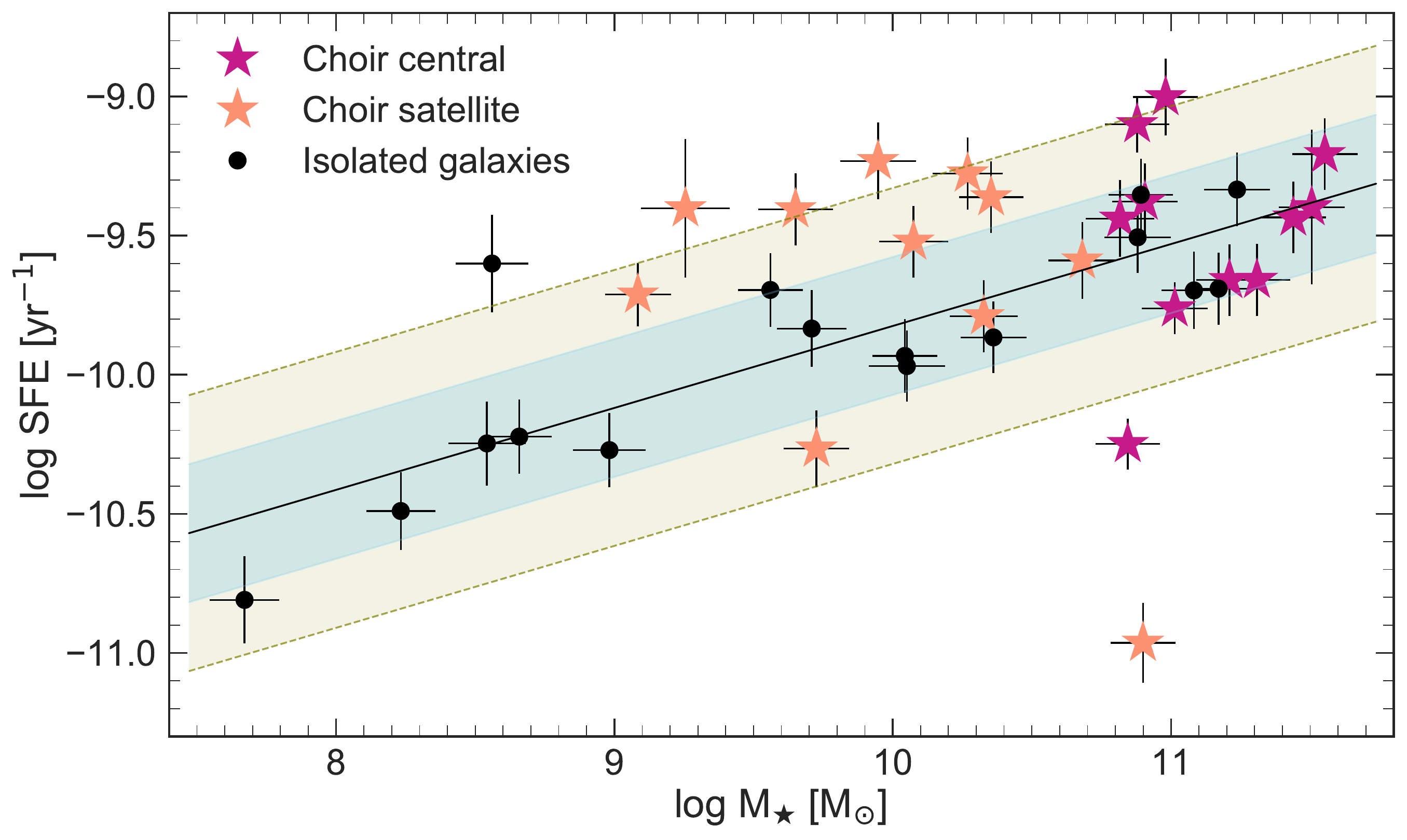}
\caption{Star formation efficiency, defined as star formation rate divided by \HI\ mass. The black dots are isolated galaxies. The magenta and orange stars correspond to the central and satellite galaxies in Choir groups, respectively. The solid line is a fit to the isolated galaxies; the blue shaded area corresponds to 1$\sigma$ scatter, while the yellow shaded area shows 2$\sigma$ scatter from the fit.}
\label{fig:SFE}
\end{figure}

We show the relations between our sample of isolated galaxies and Choir galaxies in terms of gas-mass fraction (M$_{\textrm{\HI}}$/M$_{\star}$), specific star formation rate (sSFR = SFR/M$_{\star}$) and star formation efficiency (SFE = SFR/M$_{\textrm{\HI}}$). These relations are analysed by dividing Choir galaxies into group central and satellite galaxies. We denote the brightest group galaxy (based on \textit{R}-band absolute magnitude) as the central galaxy. For groups with two galaxies of similar \textit{R}-band magnitudes within uncertainties: HIPASSJ1051-17, HIPASSJ2027-51 and HIPASSJ1250-20, two galaxies have an equal probability to be central galaxy. In these three cases we denote both galaxies as centrals. All other galaxies within the same group are treated as satellites. 

Figure \ref{fig:gasfraction} presents the \HI\ fraction versus stellar mass relation for the sample of isolated galaxies. We fit the isolated galaxies in this plane with \texttt{HyperFit}\footnote{\url{hyperfit.icrar.org}} package \citep{Robotham2015} and use the dispersion about this fit (vertical scatter) to parameterise how far off of this relationship the Choir galaxies deviate. In terms of gas-mass fraction, only one Choir central galaxy has a higher gas-mass fraction than the isolated galaxies. There are two centrals with smaller gas-mass fraction than the isolated galaxies (that is, which deviate more than 2$\sigma$ from the fitted relation) however, these two central galaxies are special cases. These two galaxies are part of groups (HIPASSJ1051-17 and HIPASSJ1250-20) where we have two central galaxies with similar stellar mass thus they have 50 per cent probability of being central galaxies in the first place. Satellites are more scattered around the relation; we find seven satellite galaxies (63 per cent) more than 2$\sigma$ from the relation. Five of these satellite galaxies have lower \HI\ content, and two have higher \HI\ content with respect to the similar mass isolated galaxies. These are indications that in groups with one central galaxy, the \HI\ gas will be depleted firstly in galaxies with the lower stellar masses (such satellites are in HIPASSJ1059-09, HIPASSJ0258-74 and HIPASSJ1026-19 group). In HIPASSJ1026-19 there is an ongoing tidal interaction between S1 and S2 thus tidal stripping maybe the main gas removal mechanism that is acting on S2. In groups with two central galaxies the gas consumption is more complicated; central galaxies can have similar or vastly different gas content.

Figure \ref{fig:sSFR} shows the specific star formation rate for the entire Choirs sample and isolated galaxies, where colours represent morphological classification of Choir galaxies. We show the \texttt{HyperFit} fit to the isolated galaxies and the associated 1$\sigma$ and 2$\sigma$ scatter. It has been shown that the suppression of star formation begins at group densities \citet{Lewis2002}. We do find that the majority of the dwarf satellite galaxies (M$_{\star}<$ 10$^{9.5}$ M$_{\odot}$) have lower specific star formation rate than the isolated dwarf galaxies. Comparing the star formation of central and satellite galaxies, we find that central galaxies are less scattered, only one deviate more than 2$\sigma$ from the fit. We find the galaxies that deviate more than 2$\sigma$ from the relation and have a high specific star formation rate are three dwarf satellites and two peculiar spirals. However, we find larger number of galaxies with a lower specific star formation rate (compared to galaxies of the same stellar mass found in isolation). These galaxies are mostly dwarf satellite galaxies, as well as lenticular and spirals satellite galaxies.

\begin{table}
\caption{The best-fit relations.}
\begin{tabular}{ll}
\toprule
{\bf HyperFit relation} & {\bf 1$\sigma$} \B \\ \hline \hline
log(M$_{\textrm{\HI}}$/M$_{\star}$) = (4.49$\pm$0.28) $-$ (0.49$\pm$0.03) $\times$ log M$_{\star}$ & 0.14 \T \\
log sSFR = ($-$8.26$\pm$0.39) $-$ (0.195$\pm$0.039) $\times$ log M$_{\star}$  & 0.19 \\
log SFE = ($-$12.77$\pm$0.50) $+$ (0.29$\pm$0.05) $\times$ log M$_{\star}$ & 0.25 \\
log M$_{\textrm{\HI}}$ = (4.59$\pm$0.28) $-$ (0.25$\pm$0.01) $\times$ M$_{R}$ & 0.143 \B \\ \hline
\end{tabular}
\begin{tablenotes}
\item The best-fit relations as obtained from HyperFit and 1$\sigma$ scatter about the fit. The relations are respectively: \HI\ fraction versus stellar mass (Figure \ref{fig:gasfraction}); Specific star formation rate versus stellar mass (Figure \ref{fig:sSFR}); Star formation efficiency versus stellar mass (Figure \ref{fig:SFE}) and the measured \HI\ mass versus the \textit{R}-band magnitude (Figure \ref{fig:scalingrelation}). 
\end{tablenotes}
\label{table:parameters}
\end{table}

\begin{table*}
\begin{threeparttable}
\captionsetup{width=1\columnwidth}
\caption{The mean properties of the Choir satellite, Choir central and isolated galaxies.}
\begin{tabular}{lccccc}
\toprule
{\bf Property} & {\bf Choir central} & {\bf Choir satellite} & \multicolumn{3}{c}{{\bf Isolated galaxies}}  \\
  &  {\bf galaxies}  &  {\bf galaxies}    & log M$_{\star}$ [M$_{\odot}$] \textless{} 8.8 & 8.8 \textless{} log M$_{\star}$ [M$_{\odot}$] \textless{} 10.5 & M$_{\star}$ [M$_{\odot}$]  \textgreater{} 10.5 \B \\ \hline \hline
 $\langle \textrm{log SFE} \rangle$ [yr$^{-1}$]        &        -9.5$\pm$0.3                 &       -9.68$\pm$0.49  &   -10.27$\pm$0.39     &   -9.93$\pm$0.18     &    -9.52$\pm$0.16             \T \\
 $\langle \textrm{log(M$_{\textrm{HI}}$/M$_{\star}$}) \rangle$ [dex]  &       -0.99$\pm$0.33                 &   -0.52$\pm$0.27    &     0.37$\pm$0.19     &  -0.28$\pm$0.21   &        -0.92$\pm$0.12    \\
 $\langle \textrm{\textrm{log sSFR}} \rangle$  [yr$^{-1}$]      &     -10.47$\pm$0.17                   &   -10.21$\pm$0.34      &   -9.89$\pm$0.23    &      -10.14$\pm$0.12       &    -10.47$\pm$0.12          \\
 $\langle \textrm{\textrm{t$_{\textrm{dep}}$}} \rangle$ [Gyr] &     4.3$\pm$4.5\tnote{$\dagger$} & 12.7$\pm$25.5\tnote{$\dagger$}   &    26.7$\pm$20.7   &  9.3$\pm$4.4 & 3.5$\pm$1.2  \B \\ \hline
\end{tabular}
\begin{tablenotes}
\item $\langle \textrm{log SFE} \rangle$ - The mean star formation efficiency; $\langle \textrm{log(M$_{\textrm{HI}}$/M$_{\star}$}) \rangle$ - The mean gas-mass fraction; $\langle \textrm{\textrm{log sSFR}} \rangle$ - The mean specific star formation rate. Uncertainties quoted in the table are standard deviations from the mean value. Isolated galaxies are separated into three bins based on their stellar mass. The bin with the log M$_{\star}$ [M$_{\odot}$] < 8.8 contains five low-mass isolated galaxies, these galaxies can be compared to low-mass Choir galaxies only in terms of specific star formation rate, however we do not have Choir galaxies of such stellar masses that are detected in \HI. The medium-mass isolated galaxies, six galaxies in the 8.8 \textless{} log M$_{\star}$ [M$_{\odot}$] \textless{} 10.5 bin are comparable (based on their stellar mass) to the Choir satellite galaxies. The bin with the log M$_{\star}$ [M$_{\odot}$] > 10.5 contains five isolated galaxies that are comparable (based on their stellar mass) to the Choir central galaxies.
\item[$\dagger$] Quoted standard deviation is larger than the mean value due to a small number of galaxies and their large scatter.
\end{tablenotes}
\label{table:meanproperties}
\end{threeparttable}
\end{table*}

\begin{figure*}
\centering
\includegraphics[width=1.6\columnwidth]{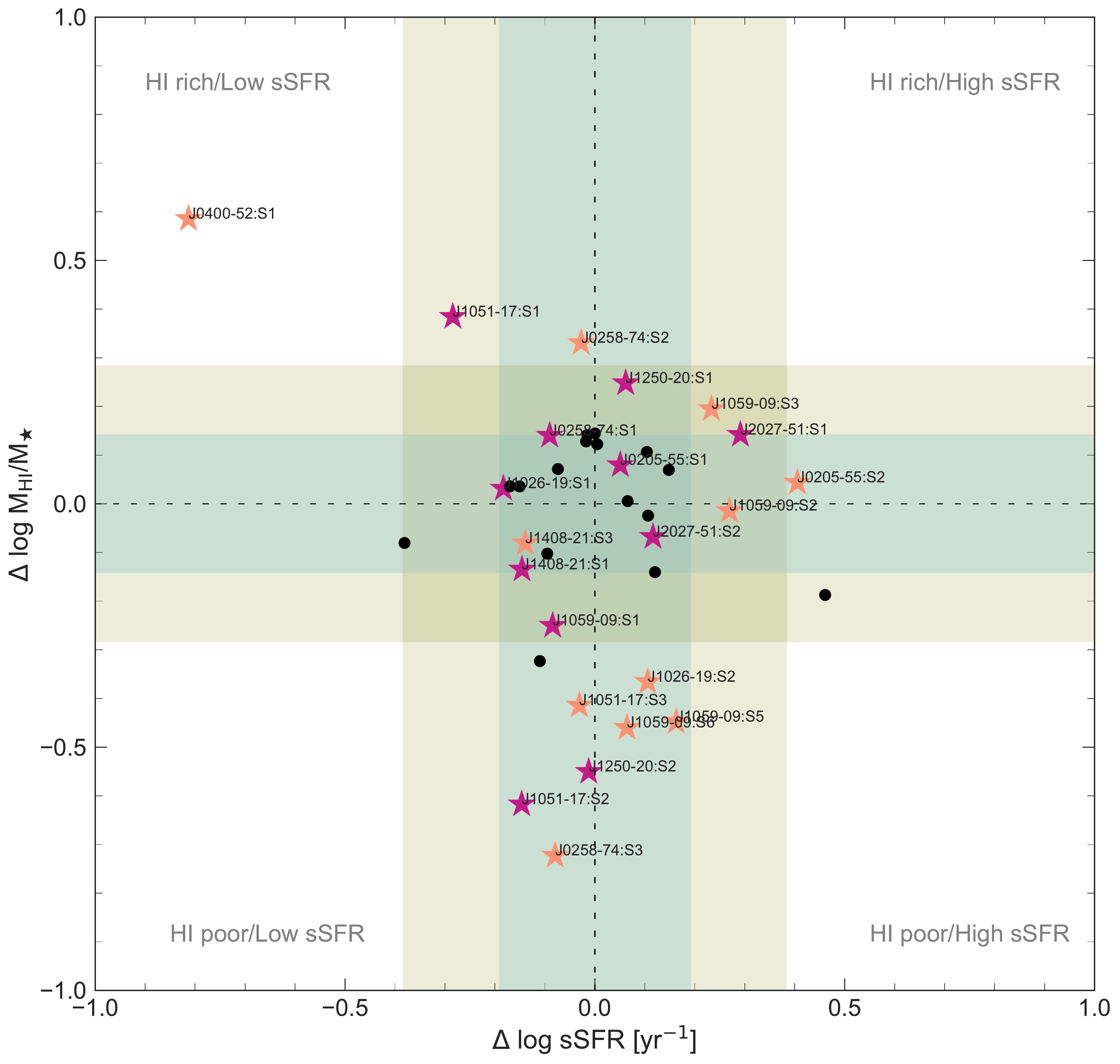}
\caption{\HI\ fraction offset versus specific star formation rate offset. The magenta and orange stars correspond to the central and satellite galaxies in Choir groups, respectively (galaxy labels are truncated versions of their ID names). The black dots are isolated galaxies. The dashed lines show zero lines, which correspond to the best-fit lines of gas-mass fraction and specific star formation rate, for the control sample of isolated galaxies (see Figure \ref{fig:gasfraction} and \ref{fig:sSFR}). The blue shaded area shows 1$\sigma$ scatter, while the yellow shaded area shows 2$\sigma$ scatter from the corresponding fit. Each galaxy is shown by its distance from the fit line to the \HI\ fraction and specific star formation rate. Figure has four quadrants as indicated in the corners. In the bottom part of the Figure we find gas-poor galaxies, we can see that they have average specific star formation rates, which complements our discussion in Section \ref{Results}, illustrating that high star formation efficiency of these galaxies is due to their lower gas-mass fraction. In contrast, galaxies that have low specific star formation efficiency (see Figure \ref{fig:SFE}) in this figure are in the upper left quadrant. The only outlier in the both properties is galaxy J0400-52:S1 (galaxy's morphology: SA(rs)cd pec), this is an \HI-rich galaxy with a low specific star formation rate (D\v{z}ud\v{z}ar et al. in prep.).} 
\label{fig:offset}
\end{figure*}

The star formation efficiency for Choirs and isolated galaxies is shown in Figure \ref{fig:SFE}. We fit the isolated galaxies using the \texttt{HyperFit} package. Two central galaxies (HIPASSJ1051-17:S1 and HIPASSJ1250-20:S2) deviate more than 2$\sigma$ from the fit. HIPASSJ1051-17:S1 exhibits lower star formation efficiency due to a higher gas-mass fraction, and HIPASSJ1250-20:S2 has higher specific star formation efficiency due to a lower gas-mass fraction. We illustrate \HI\ fraction offset versus specific star formation rate offset in Figure \ref{fig:offset}. On the x-axis the galaxy's offset from the fit to the specific star formation rate ($\Delta$ log sSFR) and on the y-axis the galaxy's offset from the fit to the \HI\ fraction ($\Delta$ log M$_{\textrm{\HI}}$/M$_{\star}$). The best-fit relations are placed in Table \ref{table:parameters}.

We summarise the mean properties of the Choir central, Choir satellite and isolated galaxies in the Table \ref{table:meanproperties}. The mean properties of isolated galaxies are separated into three bins of stellar mass which are matched to correspond to the stellar mass limits of Choir satellites (8.8 \textless{} log M$_{\star}$ [M$_{\odot}$] \textless{} 10.5) and Choir central galaxies (log M$_{\star}$ [M$_{\odot}$] > 10.5). The mean value of the gas-mass fraction for satellite galaxies and the ones in isolation differ, they are $-$0.52 and $-$0.28 dex respectively; however, due to the large scatter these values are in agreement within 1$\sigma$. Although there is no difference in the mean values, the scatter is considerably larger for the group galaxies, and actually $\sim$1/3 are somewhat \HI\ deficient (as described in Section \ref{Deficiency}).
Examining the \HI\ fraction, specific star formation rate and star formation efficiency of Choir galaxies yielded a number of galaxy outliers which should be investigated on a group by group basis, which we will do in a future paper.

\subsection{Size-mass relation}
\label{sizemass}
There is a tight correlation between the \HI\ diameter of a galaxy defined at a surface density of 1 M$_{\odot}$ pc$^{-2}$ and \HI\ mass \citep{Broeils1997, Verheijen2001}. The derived size-mass relation by \citet{Broeils1997} for 108 galaxies; log D$_{\textrm{\HI}}$ = 0.51 log(M$_{\textrm{\HI}}$) $-$ 3.32 was recently updated by \citet{Wang2016} using a sample of 562 galaxies who found roughly the same result: log D$_{\textrm{\HI}}$ = 0.506 log(M$_{\textrm{\HI}}$) $-$ 3.293. It has been shown that the D$_{\textrm{HI}}$-M$_{\textrm{HI}}$ relation holds for different galaxy morphologies and for \HI-rich or relatively \HI-poor galaxies \citep{Broeils1997, Wang2016}.

\begin{figure}
\centering
\includegraphics[width = 0.95\columnwidth]{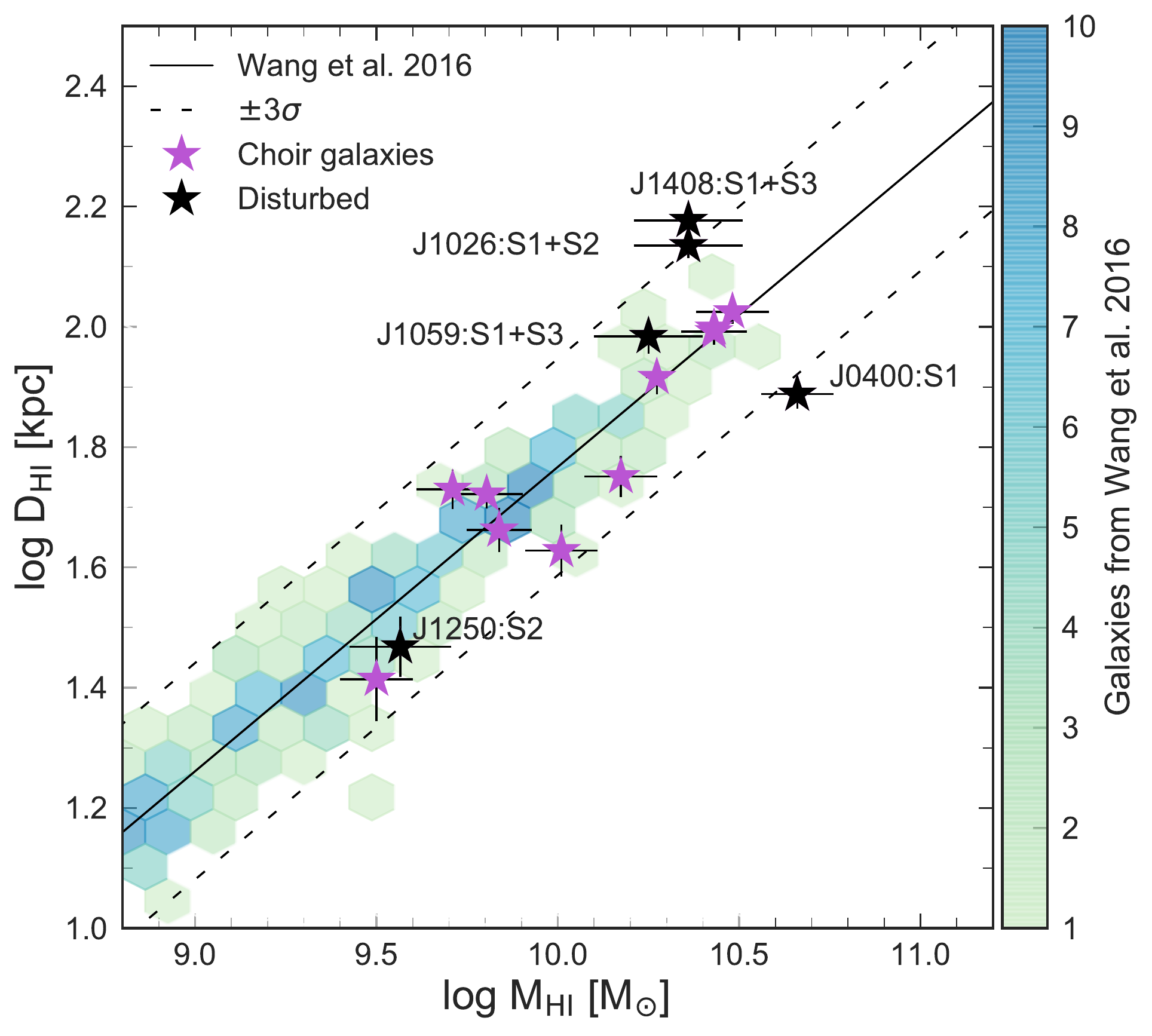}
\caption{Size-mass relation. The hexagonal bins in the background represent galaxies from \citet{Wang2017}, where colour bar shows how many galaxies are within the bin. The solid line \citep{Wang2016} shows \HI\ size-mass relation with dashed lines that mark 3$\sigma$ scatter. The Choir galaxies are marked with stars, where black stars are identified disturbed systems (galaxy labels are truncated versions of their ID names). The outliers of the 3$\sigma$ scatter are systems where at a surface density of 1 M$_{\odot}$ pc$^{-2}$ individual galaxies could not be resolved thus measurement includes two galaxies marked as J1408:S1+S3 and J1026:S1+S2. We identify one other disturbed galaxy: J0400:S1 which is just outside 3$\sigma$ limit. J1059:S1+S3 are two galaxies which are tidally interacting and yet are within size-mass scatter, as well as possible interacting galaxy J1250:S2, described in Section \ref{J1250}.}
\label{fig:DHIMHI}
\end{figure}

 We examine Choir galaxies and their position on the D$_{\textrm{HI}}$-M$_{\textrm{HI}}$ relation. We measure the \HI\ diameter from the \HI\ column density maps following prescriptions by \citet{Wang2016}. Firstly we de-project \HI\ column density maps by multiplying them by $\cos$ $i$ $\sim$ b/a, where $i$ is the inclination angle of the galaxy's disc and b/a is its axis ratio obtained from \textit{R}-band imaging \citep{Meurer2006, Sweet2013}. We map the contour density of 1.2 $\times$ 10$^{20}$ cm$^{-2}$ equal to 1 M$_{\odot}$ pc$^{-2}$, and measure the diameter of the contour by fitting it with an ellipse using the \texttt{KMPFIT} of the \texttt{Kapteyn python} package \citep{KapteynPackage}, see Appendix \ref{app2} for example maps and the obtained fit for HIPASSJ1250-20 group. We use standard error propagation of the uncertainties given by the fit to place errors on the \HI\ diameter. We then correct the obtained \HI\ diameters (D$_{\textrm{\HI,0}})$ for beam smearing effects (the best suitable for circular beam shape), as described in \citet{Wang2013, Wang2016}:
$\textrm{D}_{\textrm{\HI}} = \sqrt{\textrm{D}_{\textrm{\HI},0}^{2} - \textrm{B}_{\textrm{maj}}\times \textrm{B}_{\textrm{min}}}$, where the B$_{\textrm{maj}}$ and B$_{\textrm{min}}$ are respectively synthesized beam sizes of the major and minor axes, D$_{\textrm{\HI,0}}$ is uncorrected and D$_{\textrm{\HI}}$ is corrected \HI\ diameter. The correction for most of the Choir galaxies is small and it reduces the diameter up to 5 per cent. The only exception is the galaxy HIPASSJ1250-20:S2 for which the correction amounts to $\sim$ 32 per cent. This large correction is due to the fact that the major axis of the synthesized beam is larger than the measured \HI\ diameter at a surface density of 1 M$_{\odot}$ pc$^{-2}$. There were several galaxies for which the \HI\ diameter is not measured, these are small satellite galaxies and one small central galaxy HIPASSJ1051-17:S2 - which is part of a group with two centrals. These galaxies were excluded from Figure \ref{fig:DHIMHI} because their \HI\ emission was not detected at a surface density of 1 M$_{\odot}$ pc$^{-2}$.

Figure \ref{fig:DHIMHI} shows the relation between the total \HI\ mass and the diameter of the \HI\ disc of the measured Choir galaxies. The \HI\ size-mass relation prescribed by \citet{Wang2016} is shown as a solid line in Figure \ref{fig:DHIMHI}, and the dashed lines correspond to their 3$\sigma$ scatter which translates into 0.18 dex. We find three out of 15 measured Choir galaxies to be scattered outside 3$\sigma$ of the relation, and all three outliers are disturbed systems (black star symbols in Figure \ref{fig:DHIMHI}). The two systems with the largest diameters: HIPASSJ1408-21:S1+S3 and HIPASSJ1026-19:S1+S2 are merging galaxies and cannot be separated into individual galaxies at a surface density of 1 M$_{\odot}$ pc$^{-2}$. Thus the measured \HI\ diameter encompasses both of them. HIPASSJ0400-52:S1 is also disturbed, it shows an \HI\ offset from the stellar center and has a large asymmetry of the \HI\ distribution (D\v{z}ud\v{z}ar et al. in prep). Outliers from \cite{Wang2016} also show a broad range of abnormalities, such as highly \HI\ deficient galaxies, merger events and galaxies with an asymmetric \HI\ distribution within a disc. On the other hand, galaxies that have such abnormalities do not necessarily have to reside off the \HI\ size-mass relation. HIPASSJ1250-20:S2 and HIPASSJ1059:S1+S3 are such examples. The \HI\ distribution of HIPASSJ1250-10:S2 is offset from the stellar centre, and extends towards the larger companion (see Figure \ref{fig:groupHI}), and HIPASSJ1059:S1+S3 are tidally interacting galaxies.

The mean value of the D$_{\textrm{HI}}$ for the central Choir galaxies is $\sim$ 72 kpc, while for the satellites it is $\sim$ 51 kpc. The three largest central galaxies are HIPASSJ1250-20:S1, HIPASSJ1051-17:S1 and HIPASSJ0205-55:S1 with D$_{\textrm{HI}}$ sizes of 106, 100 and 98 kpc respectively; their size is comparable to the largest galaxies from the Bluedisk sample from \citet{Wang2013}. 

\subsection{Atomic gas depletion time}

Assuming continuous gas consumption with the current rate of the star formation determined as measured by H$\alpha$, we determined atomic gas depletion time (t$_{\mathrm{dep}}$) for Choir galaxies: 

\begin{equation}
\hspace{2.5cm} \text{t}_{\text{dep}} = \text{M}_{\text{\HI}} / \text{SFR}.\\
\label{timescale}
\end{equation}

\begin{figure}
\centering
\includegraphics[width = 0.88\columnwidth]{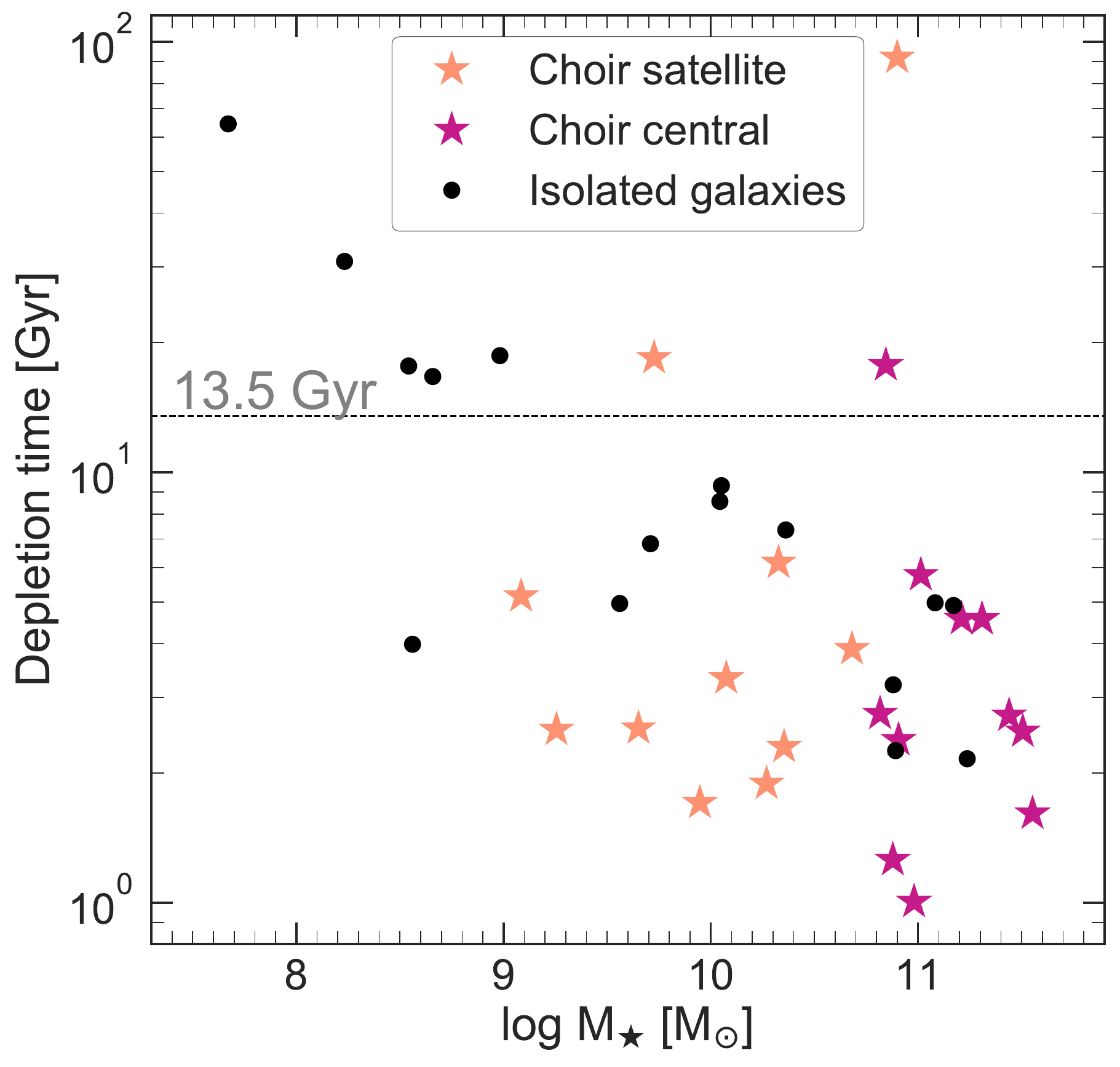}
\caption{The \HI\ depletion time versus stellar mass. The black dots are isolated galaxies. The magenta and orange stars correspond to the central and satellite galaxies in Choir groups, respectively. The horizontal dashed line marks Hubble time (13.5 Gyr). }
\label{fig:Depletion}
\end{figure}

Figure \ref{fig:Depletion} shows that all except one central and two satellite Choir galaxies have atomic depletion times smaller than the Hubble time (13.5 Gyr). Overall, the mean gas depletion time of all Choir galaxies ($\sim$ 8.5 Gyr) is smaller than that obtained for the sample of isolated galaxies ($\sim$ 12.9 Gyr), two thirds of the isolated galaxies have depletion times smaller than the Hubble time. We did not find any starburst time scales (smaller than 1 Gyr) that would indicate fast gas exhaustion and thus relatively rapid transformation of a gas-rich galaxy into a gas-poor one. The two centrals that have the smallest depletion time are HIPASSJ1250-20:S2 and HIPASSJ1051-17:S2, having depletion times of 1.0 Gyr and 1.3 Gyr respectively. The short depletion times of HIPASSJ1250-20:S2 and HIPASSJ1051-17:S2 are attributed to their low gas-mass fraction (as seen in Figure \ref{fig:gasfraction}: log(M$_{\HI}$/M$_{\star}$) $\sim$ $-$1.4 dex), whilst their specific star formation rates are comparable to those of isolated galaxies of the same stellar mass. 

\subsection{\HI\ deficiency}
\label{Deficiency}
We probe the \HI\ content of the Choir galaxies to determine whether it differs from the \HI\ content of the isolated galaxies. It has been shown that spiral galaxies have less \HI\ in high density environments, for instance, near a galaxy cluster center \citep{Chung2009} or in Hickson Compact Groups (HCG, \citealt{VerdesMontenegro2001}), than spiral galaxies that reside in the field. However, gas poor galaxies are also found in intermediate density environments \citep{Denes2016, Hess2013} as well as in loose groups \citep{Kilborn2005, Kilborn2009}. The \HI\ deficiency in high and intermediate density environments is related to the gas removal mechanisms either by gravitational interactions (tidal stripping e.g. \citealt{Yun1994}), or hydrodynamical interactions (ram pressure stripping e.g. \citealt{GunnGott1972,Rasmussen2006}). 

The \HI\ deficiency is defined as the difference between the derived \HI\ mass from a scaling relation and the observed \HI\ mass \citep{Hayens1984}:
\begin{equation}
\hspace{1cm} \text{DEF}_{\text{\HI}} = \text{log}(\text{M}_{\text{\HI}}^{\text{expected}}) - \text{log}(\text{M}_{\text{\HI}}^{\text{observed}}),\\
\label{defeq}
\end{equation}
where $\text{log}(\text{M}_{\text{\HI}}^{\text{expected}})$ is the logarithm of the derived \HI\ mass based on the optical properties of a galaxy and $\text{log}(\text{M}_{\text{\HI}}^{\text{observed}})$ is the logarithm of the measured \HI\ mass. A galaxy is considered to be \HI\ excess or \HI\ deficient with deficiency values below $-0.3$ dex or above $0.3$ dex \citep{VerdesMontenegro2001, Kilborn2009}. \citet{Lutz2017} uses a slightly higher value of $\pm 0.4$ dex whilst \citet{Denes2016} and \citet{Wolfinger2016} use $\pm 0.6$ dex to define the most \HI\ deficient and the most \HI\ excess, respectively. 

The global \HI\ deficiency of Choir groups, based on the total \HI\ group mass obtained from HIPASS, was studied by \cite{Sweet2013} and they found that these groups are not significantly \HI\ deficient. This is in contrast to Hickson Compact Groups where \cite{VerdesMontenegro2001} obtained average \HI\ deficiency of 0.4 dex (although HCGs are optically selected groups and morphologically more evolved). \citet{Sweet2013} concluded that the Choir groups represent an early stage of a group assembly, similar to the M81 group.

\begin{figure}
\centering
\includegraphics[width=0.95\columnwidth]{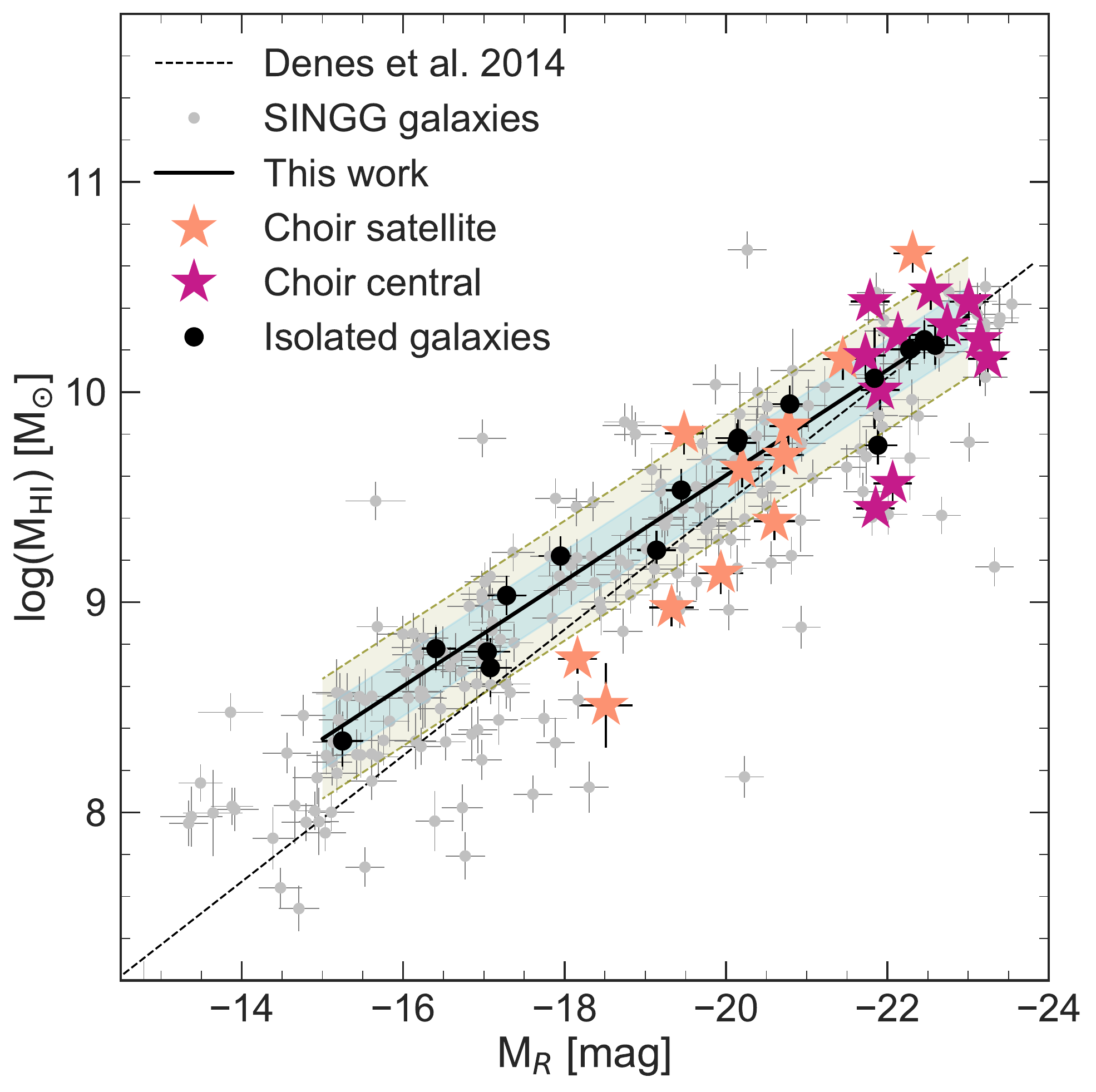}
\caption{The measured \HI\ mass versus the \textit{R}-band magnitudes for the SINGG galaxies. The grey points represent SINGG galaxy sample (single detected galaxies in the SINGG field of view). The black points are isolated galaxies. The magenta and orange stars correspond to the central and satellite galaxies in Choir groups, respectively. The solid line is the fit to the sample of isolated galaxies, the blue shaded area correspond to 1$\sigma$ scatter, while the yellow shaded area shows 2$\sigma$ scatter. The dashed line shows the scaling relation prescribed by \citet{Denes2014}. The 1$\sigma$ scatter from the scaling relation amounts to 0.143 dex.}
\label{fig:scalingrelation}
\end{figure}

In contrast, we derive the \HI\ deficiency for the individual Choir galaxies, comparing \HI\ masses obtained from the resolved ATCA and VLA imaging with  the presumed \HI\ masses based on the SINGG \textit{R}-band magnitudes. We obtain the \HI\ deficiency through the scaling relation from the \texttt{HyperFit} fit to the SINGG galaxies, see Figure \ref{fig:scalingrelation}. The obtained scaling relation is:
 
\vspace{-0.2cm}
\begin{equation}
\hspace{1.5cm} \text{log}\left ( \text{M}_{\text{HI}} \right ) = (4.59\pm0.28) - (0.25\pm0.01) \text{M}_{R}.\\
\label{denes}
\end{equation}

We find that our relation is slightlyshallower than the scaling relation of \citet{Denes2014} which was based on the optical photometry from HOPCAT (optical counterparts for \HI\ Parkes All Sky Survey Catalogue from \citealt{Doyle2005}). The difference in slope could be due to several reasons: a) An intrinsic difference between HOPCAT \textit{R}-band magnitude (from SuperCosmos plates \citet{Doyle2005}) and SINGG \textit{R}-band magnitudes (from digital recordings \citet{Meurer2006}); b) We are using more strict isolation criteria than \citet{Denes2014}, thus it is less contaminated than the HOPCAT sample.

We compare the optical and \HI\ properties of Choir galaxies to the entire SINGG galaxy sample and find that the Choir galaxies are scattered within its overall distribution (see Figure \ref{fig:scalingrelation}). We adopt the \HI\ deficiency criteria from the literature, where galaxies with DEF$_{\HI}$ above $-$0.3 dex are \HI\ excess, and below 0.3 dex are \HI\ deficient. We find that 52 per cent of our galaxies have normal \HI\ content (between $-0.3$ and $0.3$ dex). We find that one Choir central galaxy out of 11 is \HI\ excess galaxy, (DEF$_{\HI}$ $=$ $-$0.38 dex, as seen in Figure \ref{fig:scalingrelation}). Two centrals are \HI\ deficient, having DEF$_{\HI}$ of 0.56 and 0.62 dex. We find that two satellites are \HI\ excess, four are within the scatter and five are \HI\ deficient. HIPASSJ0258-74:S3 has more than four times less \HI\ than expected from the scaling relation, which makes it the most \HI\ deficient galaxy in our sample. Analysing separately central and satellite galaxies, we obtain average deficiency values for galaxies within Choir groups: DEF$_{\HI}$ $=$ 0.06 dex and DEF$_{\HI}$ $=$ 0.13 dex, respectively.

The deficiency of the entire SINGG sample is shifted towards being more gas-poor with the mean deficiency of $0.11$ dex. The obtained result for the SINGG sample is expected since galaxies in the sample are within different environments, while isolated galaxies are ``nurture-free" and thus their gas content is not influenced by the environment. Choir satellite galaxies have a similar or lower \HI\ mass than isolated galaxies of the same absolute magnitude, with two exceptions. Conversely, Choir central galaxies have similar \HI\ mass than isolated galaxies of the same absolute magnitude, with three exceptions. The mean deficiency of the isolated galaxies is similar to that of the Choir central galaxies, while satellites are on average more \HI\ deficient. This difference in \HI\ deficiency is indicating that satellites are going to be depleted first in gas-rich groups, and likely reason for \HI\ deficiency in Choir groups is tidal stripping.

\subsection{Physical processes}
\label{gasstream}
What physical processes could be acting in Choir groups? We find groups which are showing signs of past or ongoing gravitational interactions as well as groups without evidence of interaction. We comment on a possible gas accretion from the gas-rich mergers and gas refuelling from the cosmic web. We also determine the global stability parameter of the Choir galaxies to determine whether the angular momenta (an internal galaxy process) is responsible for keeping gas in the disc of gas-rich galaxies.

\subsubsection{Galaxy interactions and gas-accretion}

We visually examined deep\footnote{Preliminary surface brightness limits, based on the DECam exposure time calculator is between 25-27 mag arcsec$^{-2}$.} optical DECam (Dark Energy Camera on the CTIO Blanco 4-m telescope) \textit{g} band images and we found stellar streams as possible tracers of recent accretion of gas from minor gas-rich mergers (e.g. see Figure \ref{fig:stream}). In two groups faint stellar streams are visible: HIPASSJ0400-51 has faint features with \textit{g} band magnitude of $\sim$22 mag and surface brightness $\sim$25 mag arcsec$^{-2}$; HIPASSJ1250-20 has faint features with \textit{g} band magnitude of $\sim$23 mag and surface brightness $\sim$26 mag arcsec$^{-2}$, thus it is possible that they accreted gas through past gas-rich mergers. We presented HIPASSJ1250-20 group in Section \ref{J1250} as an example of a possible accretion onto S1 galaxy. 
Gas accretion from minor mergers is shown not to be significant source of refuelling the gas content in galaxies \citep{Sancisi2008, DiTeodoro2014} since it solely cannot be enough to sustain long star formation. Thus, the largest source of accretion in theory is the `cold mode' accretion \citep{Keres2005}, where gas is accreted from the cosmic web filaments. However, the current observations are not sensitive enough to probe this extremely low \HI\ column density $\sim$ 10$^{16}$ cm$^{-2}$ \citep{Popping2009} at which \HI\ gas within filaments is expected to be detected. 

In four Choir groups (HIPASSJ1026-19, HIPASSJ1059-09, HIPASSJ1250-20, HIPASSJ1408-21), tidal interactions are evident either in optical images or in \HI\ gas distribution or both, and yet these groups do not show extreme gas depletion. Groups with evident tidal interactions have galaxies with lower gas-mass fractions which could have experienced tidal stripping in the past. The best cases for ongoing tidal stripping as a main mechanism for gas removal is in group HIPASSJ1250-20 and HIPASSJ1026-19. Since the amount of the gas is not heavily depleted we can assume that these interactions are in an early phase.  We will present in depth analysis of these groups in a subsequent paper.

\subsubsection{Global stability parameter q}
\label{stability}
Rather than invoking accretion as gas-replenishment, since Choir groups are in an early stage of group formation, we can ask ourselves are the Choir galaxies instead retaining their gas content due to their intrinsic properties? 
\citet{Obreschkow2016} show that angular momentum regulates the \HI\ content of a galaxy, and show a relation between the mass fraction and the global stability parameter for a sample of isolated disc galaxies. Using a sample of \HI\ eXtreme (\HIX) galaxies, \cite{Lutz2017} have shown that extremely gas-rich galaxies accumulate and retain gas (building up large \HI\ discs) due to high baryonic specific angular momentum which makes them less efficient in forming stars.

We evaluate the global stability parameter for eight Choir galaxies: six centrals and two satellites using the approximation given by the \cite{Obreschkow2016} in order to determine whether gas-rich group galaxies are governed by this property. \citet{Obreschkow2016} define the global stability parameter q as: q $=$ j$\sigma$/(GM), where j is the baryonic specific angular momentum of the disc and its value can be approximated as r$_{\textrm{HI}}$v$_{\textrm{max}}$; $\sigma$ is a velocity dispersion and as in \citet{Lutz2018} we assume its value of 11 km s$^{-1}$; G is the gravitational constant and M is the baryonic mass. The baryonic mass is defined as sum of the stellar mass, \HI\ mass and the H$_{2}$ mass:  M = M$_{\star}$ + 1.35 (M$_{\textrm{HI}}$ + M$_{\textrm{H}_{2}}$). The factor 1.35 accounts for the helium and the H$_{2}$ fraction (M$_{\textrm{H}_{2}}/\textrm{M}$) is assumed to be 4 per cent \citep{Obreschkow2016}. 
In order to evaluate where galaxies lie on the mass fraction versus global stability parameter diagram we need the following galaxy properties: \textit{Ks}-band stellar mass, rotational velocities and the galaxy \HI\ radius. With the available data we can calculate stability parameter for eight Choir galaxies. We derive the \textit{Ks}-band magnitudes from the 2 Micron All Sky Survey (2MASS) images \citep{Skrutskie2006}, and calculate stellar masses following \citet{Wen2013}. We determine the rotational velocities using the 20\% widths of \HI\ emission line profiles following the prescription of \citet{Meyer2008}. The measurement of the \HI\ radius is described in the section \ref{sizemass}.

\begin{figure}
\centering
\includegraphics[width=\columnwidth]{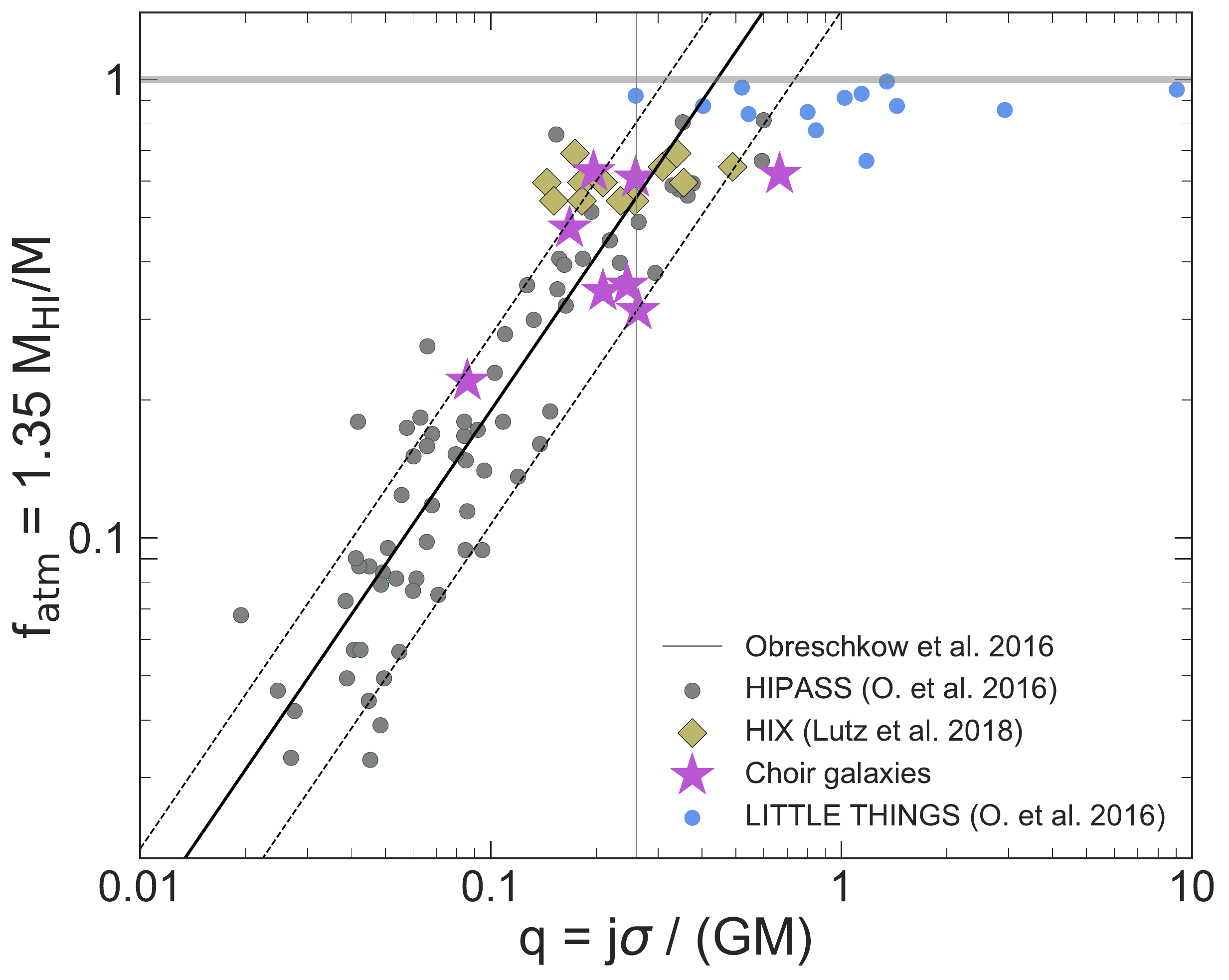}
\caption{Mass fraction (f$_{\textrm{atm}}=1.35 \textrm{M}_{\textrm{\HI}}/\textrm{M}$) as a function of a global stability parameter q $=$ j$\sigma$/(GM). The Choir galaxies are marked as purple stars. The black solid line corresponds to the relation given by \citet{Obreschkow2016} with a 40 per cent scatter marked with the dashed lines. The vertical line shows limit where for values of q > 1/(2$^{1/2}e$) and dwarf galaxies are close to f$_{\textrm{atm}}$=1 \citep{Obreschkow2016}. The grey diamonds are galaxies from the \HIX\ sample \citep{Lutz2018}. Galaxies from the HIPASS and LITTLE THINGS from \citet{Obreschkow2016} are shown with the grey and blue points, respectively.}
\label{fig:q}
\end{figure}

In Figure \ref{fig:q} we show that all gas-rich Choir galaxies follow the analytical model derived by \citet{Obreschkow2016} for isolated disc galaxies, and are in agreement with the central, star-forming, disc-dominated galaxies from the semi-analytic model of galaxy formation (DARK SAGE; \citealt{Stevens2018}). The Choir galaxies with high \HI\ mass fractions have similar q values to the \HIX\ galaxies.

In the DARK SAGE semi-analytic model galaxies are seen to lie below the f$_{\textrm{atm}}$-q relation due to merger-quenching (central galaxies) and ram-pressure stripping (satellite galaxies). These effects drastically reduce f$_{\textrm{atm}}$ in galaxies while q is only slightly affected \citep{Stevens2018}, which would cause galaxies to move downwards on the Figure \ref{fig:q}. Observationally, the study of the f$_{\textrm{atm}}$-q relation in compact groups and clusters have not been performed yet.

We conclude that, at least partially, the gas-rich Choir galaxies are keeping their gas in the disc due to higher specific angular momenta. This result shows that in small gas-rich groups, such as Choirs, the internal galaxy properties are dominating over a group impact on gas-rich galaxies since they follow the f$_{\textrm{atm}}$-q relation.

\section{Discussion}
\label{Discussion}

The main cause of disturbances in galaxy groups are tidal interactions, and it is still an open question whether ram-pressure stripping occurs in groups \citep{Westmeier2011, Rasmussen2012, Brown2016, Stevens2017}. Studying individual galaxies in large loose groups, \citet{Kilborn2009} found that the \HI\ deficient galaxies are likely to be within a projected distance of 1 Mpc from the group centre. As galaxy groups grow in number of members they are more likely to have \HI\ deficient galaxies \citep{Hess2013}. Based on the degree of \HI\ disturbance and deficiency in 72 systems, \citet{VerdesMontenegro2001} proposed an evolutionary scenario for Hickson compact groups (HCG) in phases: phase one is the least evolved and shows a relatively unperturbed \HI\ distribution; in phase two, tidal features become apparent due to galaxy interactions; and in phase three, \HI\ is stripped from some galaxies, ending up either in the intragroup medium or forming a large envelope around the group. Combining observations from the Very Large Array and Green Bank Telescope of 22 HCGs, \citealt{Borthakur2010} found that the amount of detected diffuse gas was larger in the final phase of the group evolution with respect to phase one or two. We are starting to combine the puzzle of galaxy group evolutionary stages, from young gas-rich groups that are coming together for the first time \citep{Stierwalt2017}, all the way to the old fossil groups with a large central elliptical galaxy but no obvious companions \citep{Ponman1994, Jones2003}.

Our \HI\ selected ``Choirs" groups usually consist of one or two large spiral galaxies and a number of smaller galaxies. The central galaxies have on average normal gas content, unlike that typically found in more compact groups or clusters. Here we have been considering possible scenarios that can explain their \HI\ gas content. \HI-rich central galaxies could be a result of \HI-rich mergers and/or gas accretion from the intrafilament medium \citep{Janowiecki2017, Kleiner2017, Wang2015}. However, gas accretion from minor mergers is thought not to be significant, \citet{Sancisi2008} and \citet{DiTeodoro2014} estimate (for spiral galaxies from the WHISP sample) a maximum gas accretion rate to be 0.1--0.2 and 0.28 M$_{\odot}$ yr$^{-1}$ respectively, which is not enough to sustain star formation. We find a number of dwarf galaxies around the large Choir spiral galaxies, however we do not detect \HI\ gas in all these dwarf galaxies. The \HI\ flux from the HIPASS survey yielded the total flux of each group. By a comparison of the total flux from the HIPASS and the resolved \HI\ fluxes of each detected galaxy in a group from the ATCA, we find that the majority of the \HI\ content is within the central galaxies. 

\citet{Sweet2013} stated that eight groups from the Choir sample can be considered as Local Group (LG) analogous in terms of their morphology and magnitude. They note that in terms of compactness Choirs are more compact than the LG and in terms of \HI\ content Choirs are more gas-rich than LG. Our resolved study of \HI\ content is in agreement with \citet{Sweet2013}. We make a hypothesis of how LG would look like if it was placed at a mean Choir distance of 87 Mpc. The sum of the \HI\ masses of Andromeda, M33, Milky Way and neglecting all other dwarf galaxies would be $\sim$13.4$\times$10$^{9}$ M$_{\odot}$ \citep{Cram1980, Carigan2006, Tavakoli2012, Nakanishi2016} which is above the 3$\sigma$ detection limit in HIPASS \citep{Sweet2013}. In order to detect both large spiral galaxies, Andromeda and Milky Way within the SINGG FOV, they would have to have a certain orientation in space. The separation between them is $\sim$784 kpc \citep{Stanek1998} which is $\sim$34 arcmin (face-on), at the distance of 87 Mpc, thus the separation is larger than SINGG FOV ($\sim$15 arcmin). We would be able to see them within single a SINGG FOV if we are viewing the Milky Way - Andromeda system under angle between $\sim$55 and $\sim$85 degrees, because in this case the projected separation between the galaxies would be between 2 and 13 arcmin. Only in the stated case, the LG would be considered as a group with two central galaxies, since Milky Way and Andromeda have similar \textit{R}-band magnitudes \citep{Sweet2013}. We would be able to see one smaller satellite M33 and a dwarf satellite Large Magellanic Cloud (LMC) in SINGG because they are massive enough and are star-forming galaxies \citep{Gratier2010, Garrison2018}, although we would probably not resolve \HI\ in LMC with our ATCA observations. In all other cases, we would observe Milky Way and Andromeda as single galaxies, (or they would be aligned one in front of the other) and their detection would be below the detection limit of 3$\sigma$ in HIPASS.

We compare our gas-rich galaxies to gas-rich galaxies from the \HIX\ sample \citep{Lutz2018} in terms of their global stability parameter q. We find that the Choir galaxies with the higher global stability parameter, comparable to that of \HIX\ galaxies, have stabilized the gas against the gravitational collapse which led to a build up of their large \HI\ sizes and masses. This led us to conclude that the evolution of these Choir galaxies is primarily governed by their internal properties and not by the group environment. 

With upcoming surveys, such as WALLABY \citep{Koribalski2015}, the number of gas-rich galaxy groups will increase and we will be able to conduct an in depth statistical study in order to understand the most common environment in the Universe. 

\section{Summary and Conclusions}
\label{Summary}
In this paper we present an analysis of integrated \HI\ properties for 27 galaxies within nine low mass, gas-rich, late-type dominated groups denoted ``Choirs". We compare the Choir group galaxies to a sample of isolated galaxies in terms of gas-mass fraction, specific star formation rate, star formation efficiency, depletion time and \HI\ deficiency. 

We find that central Choir galaxies have on average normal gas-mass fraction with respect to the isolated galaxies of the same stellar mass. The specific star formation rates of these central galaxies are similar to the galaxies in isolation. Using the \HI\ size-mass relation we find that the most massive galaxies have the largest \HI\ discs and fall neatly onto the \citet{Wang2016} relation. We find that all outliers are disturbed systems that are experiencing interactions. 
We find that satellite Choir galaxies have on average lower gas-mass fraction with respect to the isolated galaxies of the same stellar mass, with a number of exceptions. The specific star formation rate of Choir satellite galaxies is scattered with respect to the ones in isolation and we find that the majority of the dwarf satellite galaxies (M$_{\star} <$ 10$^{9.5}$ M$_{\odot}$) have lower specific star formation rate than the isolated dwarf galaxies.

We comment on the physical processes that are acting in Choir groups: 

    (a) We find possible traces of recent or ongoing minor mergers in HIPASSJ0400-51:S1 and HIPASSJ1250-20:S1 (see Figure \ref{fig:stream}) which could be responsible for an overall increase in their \HI\ content. We find traces of past interactions in HIPASSJ1051-17 as the S1 galaxy is a polar ring galaxy (Kilborn et al. in prep.).
    
    (b) In four galaxy groups: HIPASSJ1026-19, HIPASSJ1059-09, HIPASSJ1250-20 and HIPASSJ1408-21, tidal interactions are evident either in optical images, H$\alpha$ images or \HI\ gas distribution or in some groups all of them. Tidal interactions maybe the main gas removal mechanism in galaxies HIPASSJ1026-19:S2 and HIPASSJ1250-20:S2.
    
    (c) We do not find evidence of current tidal inderactions in HIPASSJ0205-55, HIPASSJ0258-74 and HIPASSJ2027-51.

We determine the specific angular momentum for eight Choir galaxies (see Figure \ref{fig:q}) and find that seven of them have comparable values to the galaxies from the \HIX\ sample \citep{Lutz2018}. It is possible that a higher specific angular momentum could explain how galaxies retain the large fraction of \HI\ gas in their discs. We find that Choir galaxies lie on the f$_{\textrm{atm}}$-q relation thus it is possible that normal secular evolution is prevailing over the environmental influences on these galaxies.

We provided further evidence that the Choirs are in the early stages of a group assembly as stated in \citet{Sweet2013}. This scenario is supported by the fact that we find galaxies with average gas content and specific star formation rate comparable to those galaxies found in isolation. In groups with two large galaxies of similar stellar mass, we find that either the \HI\ gas content is evenly distributed amongst them or the \HI\ gas content is vastly different: one galaxy being gas-rich, other galaxy being gas-poor. In groups with one large spiral galaxy and a number of smaller galaxies, we see that the \HI-poor galaxies are those of a smaller stellar mass. These are indications that the \HI\ gas will be depleted firstly in galaxies with the lower stellar mass. The overall \HI\ content of Choir groups is higher than that of HCGs; nevertheless, we use the proposed evolutionary scenario for HCGs by \citet{VerdesMontenegro2001}, and we see that some Choir groups presented in this paper would be placed into phase 1: groups that contain average gas content (and gas-rich galaxies) and do not exhibit signs of interaction; and early phase 2: groups that contain galaxies that interact and contain \HI\ deficient galaxies. We find no phase 3 Choir groups.

In a subsequent paper we will present an analysis of each group individually and we will discuss outliers, galaxy interactions and the distribution of atomic and molecular gas in Choir groups. The majority of galaxies in the local Universe are within environments such as Choirs thus it is important to study this group regime in order to have a complete understanding of a galaxy evolution.

\section{Acknowledgements}

We thank the anonymous referee whose very detailed and constructive comments and suggestions improved this paper.\\ 
RD is supported by a Swinburne University SUPRA postgraduate scholarship.\\
RD would like to thank Katinka Gereb for her help with the VLA data reduction using CASA; Katharina Lutz for providing values of the global stability parameter and mass fraction for the \HIX\ galaxies; and Chandrashekar Murugeshan for the helpful discussions. \\
This work benefited from the trip to KIAA conference, RD would like to thank the Astronomical Society of Australia for providing a student grant for this trip.\\
The Australia Telescope Compact Array is part of the Australia Telescope National Facility which is funded by the Australian Government for operation as a National Facility managed by CSIRO.\\
This paper includes archived data obtained through the Australia Telescope Online Archive (http://atoa.atnf.csiro.au).\\
Parts of this research were conducted by the Australian Research Council Centre of Excellence for All Sky Astrophysics in 3 Dimensions (ASTRO 3D), through project number CE170100013.\\
The National Radio Astronomy Observatory is a facility of the National Science Foundation operated under cooperative agreement by Associated Universities, Inc.\\
Based on observations at Cerro Tololo Inter-American Observatory, National Optical Astronomy Observatory (NOAO Prop. AAT/13A/02; PI: Sarah M. Sweet), which is operated by the Association of Universities for Research in Astronomy (AURA) under a cooperative agreement with the National Science Foundation. \\
This research has made use of the NASA/IPAC Ex- tragalactic Database (NED), which is operated by the Jet Propulsion Laboratory, California Institute of Technology, under contract with the National Aeronautics and Space Administration.\\
This publication makes use of data products from the Two Micron All Sky Survey, which is a joint project of the University of Massachusetts and the Infrared Processing and Analysis Center/California Institute of Technology, funded by the National Aeronautics and Space Administration and the National Science Foundation.\\
This research has made use of the NASA/ IPAC Infrared Science Archive, which is operated by the Jet Propulsion Laboratory, California Institute of Technology, under contract with the National Aeronautics and Space Administration.\\
This research has made use of the VizieR catalogue access tool, CDS, Strasbourg, France. The original description of the VizieR service was published in A\&AS 143, 23.\\
This research has made use of \texttt{python} \url{https://www.python.org} and python packages: \texttt{astropy} \citep{Astropy2013}, \texttt{matplotlib} \url{http://matplotlib.org/} \citep{Hunter2007}, \texttt{APLpy} \url{https://aplpy.github.io/}, \texttt{NumPy} \url{http://www.numpy.org/} \citep{VanDerWalt2011} and \texttt{SciPy} \url{https://www.scipy.org/} \citep{Jones2001}. 

\bibliographystyle{mnras}
\bibliography{Choir_paper} 

\begin{thebibliography}{}
\makeatletter
\relax
\def\mn@urlcharsother{\let\do\@makeother \do\$\do\&\do\#\do\^\do\_\do\%\do\~}
\def\mn@doi{\begingroup\mn@urlcharsother \@ifnextchar [ {\mn@doi@}
  {\mn@doi@[]}}
\def\mn@doi@[#1]#2{\def\@tempa{#1}\ifx\@tempa\@empty \href
  {http://dx.doi.org/#2} {doi:#2}\else \href {http://dx.doi.org/#2} {#1}\fi
  \endgroup}
\def\mn@eprint#1#2{\mn@eprint@#1:#2::\@nil}
\def\mn@eprint@arXiv#1{\href {http://arxiv.org/abs/#1} {{\tt arXiv:#1}}}
\def\mn@eprint@dblp#1{\href {http://dblp.uni-trier.de/rec/bibtex/#1.xml}
  {dblp:#1}}
\def\mn@eprint@#1:#2:#3:#4\@nil{\def\@tempa {#1}\def\@tempb {#2}\def\@tempc
  {#3}\ifx \@tempc \@empty \let \@tempc \@tempb \let \@tempb \@tempa \fi \ifx
  \@tempb \@empty \def\@tempb {arXiv}\fi \@ifundefined
  {mn@eprint@\@tempb}{\@tempb:\@tempc}{\expandafter \expandafter \csname
  mn@eprint@\@tempb\endcsname \expandafter{\@tempc}}}

\bibitem[\protect\citeauthoryear{{Argudo-Fern{\'a}ndez}
  et~al.,}{{Argudo-Fern{\'a}ndez} et~al.}{2013}]{Argundo2013}
{Argudo-Fern{\'a}ndez} M.,  et~al., 2013, \mn@doi [\aap]
  {10.1051/0004-6361/201321326}, \href
  {http://adsabs.harvard.edu/abs/2013A%26A...560A...9A} {560, A9}

\bibitem[\protect\citeauthoryear{{Astropy Collaboration} et~al.,}{{Astropy
  Collaboration} et~al.}{2013}]{Astropy2013}
{Astropy Collaboration} et~al., 2013, \mn@doi [\aap]
  {10.1051/0004-6361/201322068}, \href
  {http://adsabs.harvard.edu/abs/2013A%26A...558A..33A} {558, A33}

\bibitem[\protect\citeauthoryear{{Audcent-Ross} et~al.,}{{Audcent-Ross}
  et~al.}{2018}]{Audcent2018}
{Audcent-Ross} F.~M.,  et~al., 2018, \mn@doi [\mnras] {10.1093/mnras/sty1538},
  \href {http://adsabs.harvard.edu/abs/2018MNRAS.480..119A} {480, 119}

\bibitem[\protect\citeauthoryear{{Baldwin}, {Phillips}  \&
  {Terlevich}}{{Baldwin} et~al.}{1981}]{Baldwin1981}
{Baldwin} J.~A.,  {Phillips} M.~M.,   {Terlevich} R.,  1981, \mn@doi [\pasp]
  {10.1086/130766}, \href {http://adsabs.harvard.edu/abs/1981PASP...93....5B}
  {93, 5}

\bibitem[\protect\citeauthoryear{{Barnes} et~al.,}{{Barnes}
  et~al.}{2001}]{Barnes2001}
{Barnes} D.~G.,  et~al., 2001, \mn@doi [\mnras]
  {10.1046/j.1365-8711.2001.04102.x}, \href
  {http://adsabs.harvard.edu/abs/2001MNRAS.322..486B} {322, 486}

\bibitem[\protect\citeauthoryear{{Barsanti} et~al.,}{{Barsanti}
  et~al.}{2018}]{Barsanti2018}
{Barsanti} S.,  et~al., 2018, \mn@doi [\apj] {10.3847/1538-4357/aab61a}, \href
  {http://adsabs.harvard.edu/abs/2018ApJ...857...71B} {857, 71}

\bibitem[\protect\citeauthoryear{Bekki}{Bekki}{2008}]{Bekki2008}
Bekki K.,  2008, \mn@doi [\mnras: Letters] {10.1111/j.1745-3933.2008.00528.x},
  390, L24

\bibitem[\protect\citeauthoryear{{Berlind} et~al.,}{{Berlind}
  et~al.}{2006}]{Berlind2006}
{Berlind} A.~A.,  et~al., 2006, \mn@doi [\apjs] {10.1086/508170}, \href
  {http://adsabs.harvard.edu/abs/2006ApJS..167....1B} {167, 1}

\bibitem[\protect\citeauthoryear{{Bigiel} \& {Blitz}}{{Bigiel} \&
  {Blitz}}{2012}]{Bigiel2012}
{Bigiel} F.,  {Blitz} L.,  2012, \mn@doi [\apj] {10.1088/0004-637X/756/2/183},
  \href {http://adsabs.harvard.edu/abs/2012ApJ...756..183B} {756, 183}

\bibitem[\protect\citeauthoryear{{Blanton}, {Eisenstein}, {Hogg}, {Schlegel}
  \& {Brinkmann}}{{Blanton} et~al.}{2005}]{Blanton2005}
{Blanton} M.~R.,  {Eisenstein} D.,  {Hogg} D.~W.,  {Schlegel} D.~J.,
  {Brinkmann} J.,  2005, \mn@doi [\apj] {10.1086/422897}, \href
  {http://adsabs.harvard.edu/abs/2005ApJ...629..143B} {629, 143}

\bibitem[\protect\citeauthoryear{{Borthakur}, {Yun}  \&
  {Verdes-Montenegro}}{{Borthakur} et~al.}{2010}]{Borthakur2010}
{Borthakur} S.,  {Yun} M.~S.,   {Verdes-Montenegro} L.,  2010, \mn@doi [\apj]
  {10.1088/0004-637X/710/1/385}, \href
  {http://adsabs.harvard.edu/abs/2010ApJ...710..385B} {710, 385}

\bibitem[\protect\citeauthoryear{{Bosma}}{{Bosma}}{1981}]{Bosma1981}
{Bosma} A.,  1981, \mn@doi [\aj] {10.1086/113062}, \href
  {http://adsabs.harvard.edu/abs/1981AJ.....86.1791B} {86, 1791}

\bibitem[\protect\citeauthoryear{{Bosma}}{{Bosma}}{2017}]{Bosma2017}
{Bosma} A.,  2017, in {Knapen} J.~H.,  {Lee} J.~C.,   {Gil de Paz} A.,  eds,
  Astrophysics and Space Science Library Vol. 434, Outskirts of Galaxies.
  p.~209 (\mn@eprint {arXiv} {1612.05272}),
  \mn@doi{10.1007/978-3-319-56570-5_7}

\bibitem[\protect\citeauthoryear{{Broeils} \& {Rhee}}{{Broeils} \&
  {Rhee}}{1997}]{Broeils1997}
{Broeils} A.~H.,  {Rhee} M.-H.,  1997, \aap, \href
  {http://adsabs.harvard.edu/abs/1997A%26A...324..877B} {324, 877}

\bibitem[\protect\citeauthoryear{{Brough}, {Forbes}, {Kilborn}  \&
  {Couch}}{{Brough} et~al.}{2006}]{Brough2006}
{Brough} S.,  {Forbes} D.~A.,  {Kilborn} V.~A.,   {Couch} W.,  2006, \mn@doi
  [\mnras] {10.1111/j.1365-2966.2006.10542.x}, \href
  {http://adsabs.harvard.edu/abs/2006MNRAS.370.1223B} {370, 1223}

\bibitem[\protect\citeauthoryear{{Brown} et~al.,}{{Brown}
  et~al.}{2016}]{Brown2016}
{Brown} T.,  et~al., 2016, arXiv.org, pp 1275--1289

\bibitem[\protect\citeauthoryear{{Brown} et~al.,}{{Brown}
  et~al.}{2017}]{Brown2017}
{Brown} T.,  et~al., 2017, \mn@doi [\mnras] {10.1093/mnras/stw2991}, \href
  {http://adsabs.harvard.edu/abs/2017MNRAS.466.1275B} {466, 1275}

\bibitem[\protect\citeauthoryear{{Campbell}, {van den Bosch}, {Hearin},
  {Padmanabhan}, {Berlind}, {Mo}, {Tinker}  \& {Yang}}{{Campbell}
  et~al.}{2015}]{Campbell2015}
{Campbell} D.,  {van den Bosch} F.~C.,  {Hearin} A.,  {Padmanabhan} N.,
  {Berlind} A.,  {Mo} H.~J.,  {Tinker} J.,   {Yang} X.,  2015, \mn@doi [\mnras]
  {10.1093/mnras/stv1091}, \href
  {http://adsabs.harvard.edu/abs/2015MNRAS.452..444C} {452, 444}

\bibitem[\protect\citeauthoryear{{Carignan}, {Chemin}, {Huchtmeier}  \&
  {Lockman}}{{Carignan} et~al.}{2006}]{Carigan2006}
{Carignan} C.,  {Chemin} L.,  {Huchtmeier} W.~K.,   {Lockman} F.~J.,  2006,
  \mn@doi [\apjl] {10.1086/503869}, \href
  {http://adsabs.harvard.edu/abs/2006ApJ...641L.109C} {641, L109}

\bibitem[\protect\citeauthoryear{{Carollo} et~al.,}{{Carollo}
  et~al.}{2016}]{Carollo2016}
{Carollo} C.~M.,  et~al., 2016, \mn@doi [\apj] {10.3847/0004-637X/818/2/180},
  \href {http://adsabs.harvard.edu/abs/2016ApJ...818..180C} {818, 180}

\bibitem[\protect\citeauthoryear{{Chung}, {van Gorkom}, {Kenney}, {Crowl}  \&
  {Vollmer}}{{Chung} et~al.}{2009}]{Chung2009}
{Chung} A.,  {van Gorkom} J.~H.,  {Kenney} J.~D.~P.,  {Crowl} H.,   {Vollmer}
  B.,  2009, \mn@doi [\aj] {10.1088/0004-6256/138/6/1741}, \href
  {http://adsabs.harvard.edu/abs/2009AJ....138.1741C} {138, 1741}

\bibitem[\protect\citeauthoryear{{Coziol} \& {Plauchu-Frayn}}{{Coziol} \&
  {Plauchu-Frayn}}{2007}]{Coziol2007}
{Coziol} R.,  {Plauchu-Frayn} I.,  2007, \mn@doi [\aj] {10.1086/513514}, \href
  {http://adsabs.harvard.edu/abs/2007AJ....133.2630C} {133, 2630}

\bibitem[\protect\citeauthoryear{{Cram}, {Roberts}  \& {Whitehurst}}{{Cram}
  et~al.}{1980}]{Cram1980}
{Cram} T.~R.,  {Roberts} M.~S.,   {Whitehurst} R.~N.,  1980, \aaps, \href
  {http://adsabs.harvard.edu/abs/1980A%26AS...40..215C} {40, 215}

\bibitem[\protect\citeauthoryear{{Davies} et~al.,}{{Davies}
  et~al.}{2015}]{Davies2015}
{Davies} L.~J.~M.,  et~al., 2015, \mn@doi [\mnras] {10.1093/mnras/stv1241},
  \href {http://adsabs.harvard.edu/abs/2015MNRAS.452..616D} {452, 616}

\bibitem[\protect\citeauthoryear{D{\'e}nes, Kilborn  \& Koribalski}{D{\'e}nes
  et~al.}{2014}]{Denes2014}
D{\'e}nes H.,  Kilborn V.~A.,   Koribalski B.~S.,  2014, arXiv.org, pp 667--681

\bibitem[\protect\citeauthoryear{{D{\'e}nes}, {Kilborn}, {Koribalski}  \&
  {Wong}}{{D{\'e}nes} et~al.}{2016}]{Denes2016}
{D{\'e}nes} H.,  {Kilborn} V.~A.,  {Koribalski} B.~S.,   {Wong} O.~I.,  2016,
  \mn@doi [\mnras] {10.1093/mnras/stv2391}, \href
  {http://adsabs.harvard.edu/abs/2016MNRAS.455.1294D} {455, 1294}

\bibitem[\protect\citeauthoryear{{Di Teodoro} \& {Fraternali}}{{Di Teodoro} \&
  {Fraternali}}{2014}]{DiTeodoro2014}
{Di Teodoro} E.~M.,  {Fraternali} F.,  2014, \mn@doi [\aap]
  {10.1051/0004-6361/201423596}, \href
  {http://adsabs.harvard.edu/abs/2014A%26A...567A..68D} {567, A68}

\bibitem[\protect\citeauthoryear{{Doyle} et~al.,}{{Doyle}
  et~al.}{2005}]{Doyle2005}
{Doyle} M.~T.,  et~al., 2005, \mn@doi [\mnras]
  {10.1111/j.1365-2966.2005.09159.x}, \href
  {http://adsabs.harvard.edu/abs/2005MNRAS.361...34D} {361, 34}

\bibitem[\protect\citeauthoryear{{Dressler}}{{Dressler}}{1980}]{Dressler1980}
{Dressler} A.,  1980, \mn@doi [\apj] {10.1086/157753}, \href
  {http://adsabs.harvard.edu/abs/1980ApJ...236..351D} {236, 351}

\bibitem[\protect\citeauthoryear{{Eke} et~al.,}{{Eke} et~al.}{2004}]{Eke2004}
{Eke} V.~R.,  et~al., 2004, \mn@doi [\mnras]
  {10.1111/j.1365-2966.2004.07408.x}, \href
  {http://adsabs.harvard.edu/abs/2004MNRAS.348..866E} {348, 866}

\bibitem[\protect\citeauthoryear{{Elagali}, {Wong}, {Oh}, {Staveley-Smith},
  {Koribalski}, {Bekki}  \& {Zwaan}}{{Elagali} et~al.}{2018}]{Elagali2018}
{Elagali} A.,  {Wong} O.~I.,  {Oh} S.-H.,  {Staveley-Smith} L.,  {Koribalski}
  B.~S.,  {Bekki} K.,   {Zwaan} M.,  2018, \mn@doi [\mnras]
  {10.1093/mnras/sty741}, \href
  {http://adsabs.harvard.edu/abs/2018MNRAS.476.5681E} {476, 5681}

\bibitem[\protect\citeauthoryear{{Garrison-Kimmel} et~al.,}{{Garrison-Kimmel}
  et~al.}{2018}]{Garrison2018}
{Garrison-Kimmel} S.,  et~al., 2018, preprint, \href
  {http://adsabs.harvard.edu/abs/2018arXiv180604143G} {} (\mn@eprint {arXiv}
  {1806.04143})

\bibitem[\protect\citeauthoryear{{Giovanelli} \& {Haynes}}{{Giovanelli} \&
  {Haynes}}{1985}]{Giovanelli1985}
{Giovanelli} R.,  {Haynes} M.~P.,  1985, \mn@doi [\apj] {10.1086/163170}, \href
  {http://adsabs.harvard.edu/abs/1985ApJ...292..404G} {292, 404}

\bibitem[\protect\citeauthoryear{{G{\'o}mez} et~al.,}{{G{\'o}mez}
  et~al.}{2003}]{Gomez2003}
{G{\'o}mez} P.~L.,  et~al., 2003, \mn@doi [\apj] {10.1086/345593}, \href
  {http://adsabs.harvard.edu/abs/2003ApJ...584..210G} {584, 210}

\bibitem[\protect\citeauthoryear{{Goto}, {Yamauchi}, {Fujita}, {Okamura},
  {Sekiguchi}, {Smail}, {Bernardi}  \& {Gomez}}{{Goto} et~al.}{2003}]{Goto2003}
{Goto} T.,  {Yamauchi} C.,  {Fujita} Y.,  {Okamura} S.,  {Sekiguchi} M.,
  {Smail} I.,  {Bernardi} M.,   {Gomez} P.~L.,  2003, \mn@doi [\mnras]
  {10.1046/j.1365-2966.2003.07114.x}, \href
  {http://adsabs.harvard.edu/abs/2003MNRAS.346..601G} {346, 601}

\bibitem[\protect\citeauthoryear{{Gratier} et~al.,}{{Gratier}
  et~al.}{2010}]{Gratier2010}
{Gratier} P.,  et~al., 2010, \mn@doi [\aap] {10.1051/0004-6361/201014441},
  \href {http://adsabs.harvard.edu/abs/2010A%26A...522A...3G} {522, A3}

\bibitem[\protect\citeauthoryear{{Gunn} \& {Gott}}{{Gunn} \&
  {Gott}}{1972}]{GunnGott1972}
{Gunn} J.~E.,  {Gott} III J.~R.,  1972, \mn@doi [\apj] {10.1086/151605}, \href
  {http://adsabs.harvard.edu/abs/1972ApJ...176....1G} {176, 1}

\bibitem[\protect\citeauthoryear{{Haynes} \& {Giovanelli}}{{Haynes} \&
  {Giovanelli}}{1984}]{Hayens1984}
{Haynes} M.~P.,  {Giovanelli} R.,  1984, \mn@doi [\aj] {10.1086/113573}, \href
  {http://adsabs.harvard.edu/abs/1984AJ.....89..758H} {89, 758}

\bibitem[\protect\citeauthoryear{{Hess} \& {Wilcots}}{{Hess} \&
  {Wilcots}}{2013}]{Hess2013}
{Hess} K.~M.,  {Wilcots} E.~M.,  2013, \mn@doi [\aj]
  {10.1088/0004-6256/146/5/124}, \href
  {http://adsabs.harvard.edu/abs/2013AJ....146..124H} {146, 124}

\bibitem[\protect\citeauthoryear{{Hess}, {Cluver}, {Yahya}, {Leisman}, {Serra},
  {Lucero}, {Passmoor}  \& {Carignan}}{{Hess} et~al.}{2017}]{Hess2017}
{Hess} K.~M.,  {Cluver} M.~E.,  {Yahya} S.,  {Leisman} L.,  {Serra} P.,
  {Lucero} D.~M.,  {Passmoor} S.~S.,   {Carignan} C.,  2017, \mn@doi [\mnras]
  {10.1093/mnras/stw2338}, \href
  {http://adsabs.harvard.edu/abs/2017MNRAS.464..957H} {464, 957}

\bibitem[\protect\citeauthoryear{{Hickson}}{{Hickson}}{1982}]{Hickson1982}
{Hickson} P.,  1982, \mn@doi [\apj] {10.1086/159838}, \href
  {http://adsabs.harvard.edu/abs/1982ApJ...255..382H} {255, 382}

\bibitem[\protect\citeauthoryear{{Hoffman}, {Salpeter}, {Farhat}, {Roos},
  {Williams}  \& {Helou}}{{Hoffman} et~al.}{1996}]{Hoffman1996}
{Hoffman} G.~L.,  {Salpeter} E.~E.,  {Farhat} B.,  {Roos} T.,  {Williams} H.,
  {Helou} G.,  1996, \mn@doi [\apjs] {10.1086/192314}, \href
  {http://adsabs.harvard.edu/abs/1996ApJS..105..269H} {105, 269}

\bibitem[\protect\citeauthoryear{Hunter}{Hunter}{2007}]{Hunter2007}
Hunter J.~D.,  2007, \mn@doi [Computing In Science \& Engineering]
  {10.1109/MCSE.2007.55}, 9, 90

\bibitem[\protect\citeauthoryear{Janowiecki, Catinella, Cortese, Saintonge,
  Brown  \& Wang}{Janowiecki et~al.}{2017}]{Janowiecki2017}
Janowiecki S.,  Catinella B.,  Cortese L.,  Saintonge A.,  Brown T.,   Wang J.,
   2017, arXiv.org, p. stx046

\bibitem[\protect\citeauthoryear{Jones, Oliphant, Peterson  et~al.}{Jones
  et~al.}{2001}]{Jones2001}
Jones E.,  Oliphant T.,  Peterson P.,   et~al., 2001, {SciPy}: Open source
  scientific tools for {Python}, \url {http://www.scipy.org/}

\bibitem[\protect\citeauthoryear{{Jones}, {Ponman}, {Horton}, {Babul},
  {Ebeling}  \& {Burke}}{{Jones} et~al.}{2003}]{Jones2003}
{Jones} L.~R.,  {Ponman} T.~J.,  {Horton} A.,  {Babul} A.,  {Ebeling} H.,
  {Burke} D.~J.,  2003, \mn@doi [\mnras] {10.1046/j.1365-8711.2003.06702.x},
  \href {http://adsabs.harvard.edu/abs/2003MNRAS.343..627J} {343, 627}

\bibitem[\protect\citeauthoryear{{Karachentseva}}{{Karachentseva}}{1973}]{Karachentseva1973}
{Karachentseva} V.~E.,  1973, Soobshcheniya Spetsial'noj Astrofizicheskoj
  Observatorii, \href {http://adsabs.harvard.edu/abs/1973SoSAO...8....3K} {8}

\bibitem[\protect\citeauthoryear{{Karachentseva}, {Mitronova}, {Melnyk}  \&
  {Karachentsev}}{{Karachentseva} et~al.}{2010}]{Karachentseva2010}
{Karachentseva} V.~E.,  {Mitronova} S.~N.,  {Melnyk} O.~V.,   {Karachentsev}
  I.~D.,  2010, \mn@doi [Astrophysical Bulletin] {10.1134/S1990341310010013},
  \href {http://adsabs.harvard.edu/abs/2010AstBu..65....1K} {65, 1}

\bibitem[\protect\citeauthoryear{{Kauffmann} et~al.,}{{Kauffmann}
  et~al.}{2003}]{Kauffmann2003}
{Kauffmann} G.,  et~al., 2003, \mn@doi [\mnras]
  {10.1111/j.1365-2966.2003.07154.x}, \href
  {http://adsabs.harvard.edu/abs/2003MNRAS.346.1055K} {346, 1055}

\bibitem[\protect\citeauthoryear{{Kauffmann}, {White}, {Heckman}, {M{\'e}nard},
  {Brinchmann}, {Charlot}, {Tremonti}  \& {Brinkmann}}{{Kauffmann}
  et~al.}{2004}]{Kauffmann2004}
{Kauffmann} G.,  {White} S.~D.~M.,  {Heckman} T.~M.,  {M{\'e}nard} B.,
  {Brinchmann} J.,  {Charlot} S.,  {Tremonti} C.,   {Brinkmann} J.,  2004,
  \mn@doi [\mnras] {10.1111/j.1365-2966.2004.08117.x}, \href
  {http://adsabs.harvard.edu/abs/2004MNRAS.353..713K} {353, 713}

\bibitem[\protect\citeauthoryear{{Kere{\v s}}, {Katz}, {Weinberg}  \&
  {Dav{\'e}}}{{Kere{\v s}} et~al.}{2005}]{Keres2005}
{Kere{\v s}} D.,  {Katz} N.,  {Weinberg} D.~H.,   {Dav{\'e}} R.,  2005, \mn@doi
  [\mnras] {10.1111/j.1365-2966.2005.09451.x}, \href
  {http://adsabs.harvard.edu/abs/2005MNRAS.363....2K} {363, 2}

\bibitem[\protect\citeauthoryear{{Kewley}, {Dopita}, {Sutherland}, {Heisler}
  \& {Trevena}}{{Kewley} et~al.}{2001}]{Kewley2001}
{Kewley} L.~J.,  {Dopita} M.~A.,  {Sutherland} R.~S.,  {Heisler} C.~A.,
  {Trevena} J.,  2001, \mn@doi [\apj] {10.1086/321545}, \href
  {http://adsabs.harvard.edu/abs/2001ApJ...556..121K} {556, 121}

\bibitem[\protect\citeauthoryear{{Kilborn}, {Koribalski}, {Forbes}, {Barnes}
  \& {Musgrave}}{{Kilborn} et~al.}{2005}]{Kilborn2005}
{Kilborn} V.~A.,  {Koribalski} B.~S.,  {Forbes} D.~A.,  {Barnes} D.~G.,
  {Musgrave} R.~C.,  2005, \mn@doi [\mnras] {10.1111/j.1365-2966.2004.08450.x},
  \href {http://adsabs.harvard.edu/abs/2005MNRAS.356...77K} {356, 77}

\bibitem[\protect\citeauthoryear{{Kilborn}, {Forbes}, {Barnes}, {Koribalski},
  {Brough}  \& {Kern}}{{Kilborn} et~al.}{2009}]{Kilborn2009}
{Kilborn} V.~A.,  {Forbes} D.~A.,  {Barnes} D.~G.,  {Koribalski} B.~S.,
  {Brough} S.,   {Kern} K.,  2009, \mn@doi [\mnras]
  {10.1111/j.1365-2966.2009.15587.x}, \href
  {http://adsabs.harvard.edu/abs/2009MNRAS.400.1962K} {400, 1962}

\bibitem[\protect\citeauthoryear{{Kleiner}, {Pimbblet}, {Jones}, {Koribalski}
  \& {Serra}}{{Kleiner} et~al.}{2017}]{Kleiner2017}
{Kleiner} D.,  {Pimbblet} K.~A.,  {Jones} D.~H.,  {Koribalski} B.~S.,   {Serra}
  P.,  2017, \mn@doi [\mnras] {10.1093/mnras/stw3328}, \href
  {http://adsabs.harvard.edu/abs/2017MNRAS.466.4692K} {466, 4692}

\bibitem[\protect\citeauthoryear{{Knobel}, {Lilly}, {Woo}  \& {Kova{\v
  c}}}{{Knobel} et~al.}{2015}]{Knobel2015}
{Knobel} C.,  {Lilly} S.~J.,  {Woo} J.,   {Kova{\v c}} K.,  2015, \mn@doi
  [\apj] {10.1088/0004-637X/800/1/24}, \href
  {http://adsabs.harvard.edu/abs/2015ApJ...800...24K} {800, 24}

\bibitem[\protect\citeauthoryear{{Koribalski}}{{Koribalski}}{2015}]{Koribalski2015}
{Koribalski} B.~S.,  2015, in {Ziegler} B.~L.,  {Combes} F.,  {Dannerbauer} H.,
    {Verdugo} M.,  eds,  IAU Symposium Vol. 309, Galaxies in 3D across the
  Universe. pp 39--46 (\mn@eprint {arXiv} {1604.08773}),
  \mn@doi{10.1017/S1743921314009272}

\bibitem[\protect\citeauthoryear{{Koribalski} et~al.,}{{Koribalski}
  et~al.}{2004}]{Koribalski2004}
{Koribalski} B.~S.,  et~al., 2004, \mn@doi [\aj] {10.1086/421744}, \href
  {http://adsabs.harvard.edu/abs/2004AJ....128...16K} {128, 16}

\bibitem[\protect\citeauthoryear{{Lewis} et~al.,}{{Lewis}
  et~al.}{2002}]{Lewis2002}
{Lewis} I.,  et~al., 2002, \mn@doi [\mnras] {10.1046/j.1365-8711.2002.05558.x},
  \href {http://adsabs.harvard.edu/abs/2002MNRAS.334..673L} {334, 673}

\bibitem[\protect\citeauthoryear{{Lutz} et~al.,}{{Lutz}
  et~al.}{2017}]{Lutz2017}
{Lutz} K.~A.,  et~al., 2017, \mn@doi [\mnras] {10.1093/mnras/stx053}, \href
  {http://adsabs.harvard.edu/abs/2017MNRAS.467.1083L} {467, 1083}

\bibitem[\protect\citeauthoryear{{Lutz} et~al.,}{{Lutz}
  et~al.}{2018}]{Lutz2018}
{Lutz} K.~A.,  et~al., 2018, preprint, \href
  {http://adsabs.harvard.edu/abs/2018arXiv180204043L} {} (\mn@eprint {arXiv}
  {1802.04043})

\bibitem[\protect\citeauthoryear{{McMullin}, {Waters}, {Schiebel}, {Young}  \&
  {Golap}}{{McMullin} et~al.}{2007}]{McMulin2007}
{McMullin} J.~P.,  {Waters} B.,  {Schiebel} D.,  {Young} W.,   {Golap} K.,
  2007, in {Shaw} R.~A.,  {Hill} F.,   {Bell} D.~J.,  eds,  Astronomical
  Society of the Pacific Conference Series Vol. 376, Astronomical Data Analysis
  Software and Systems XVI. p.~127

\bibitem[\protect\citeauthoryear{{Melnyk}, {Mitronova}  \&
  {Karachentseva}}{{Melnyk} et~al.}{2014}]{Melnyk2014}
{Melnyk} O.,  {Mitronova} S.,   {Karachentseva} V.,  2014, \mn@doi [\mnras]
  {10.1093/mnras/stt2225}, \href
  {http://adsabs.harvard.edu/abs/2014MNRAS.438..548M} {438, 548}

\bibitem[\protect\citeauthoryear{{Melnyk}, {Karachentseva}  \&
  {Karachentsev}}{{Melnyk} et~al.}{2015}]{Melnyk2015}
{Melnyk} O.,  {Karachentseva} V.,   {Karachentsev} I.,  2015, \mn@doi [\mnras]
  {10.1093/mnras/stv950}, \href
  {http://adsabs.harvard.edu/abs/2015MNRAS.451.1482M} {451, 1482}

\bibitem[\protect\citeauthoryear{{Meurer} et~al.,}{{Meurer}
  et~al.}{2006}]{Meurer2006}
{Meurer} G.~R.,  et~al., 2006, \mn@doi [\apjs] {10.1086/504685}, \href
  {http://adsabs.harvard.edu/abs/2006ApJS..165..307M} {165, 307}

\bibitem[\protect\citeauthoryear{{Meyer} et~al.,}{{Meyer}
  et~al.}{2004}]{Meyer2004}
{Meyer} M.~J.,  et~al., 2004, \mn@doi [\mnras]
  {10.1111/j.1365-2966.2004.07710.x}, \href
  {http://adsabs.harvard.edu/abs/2004MNRAS.350.1195M} {350, 1195}

\bibitem[\protect\citeauthoryear{{Meyer}, {Zwaan}, {Webster}, {Schneider}  \&
  {Staveley-Smith}}{{Meyer} et~al.}{2008}]{Meyer2008}
{Meyer} M.~J.,  {Zwaan} M.~A.,  {Webster} R.~L.,  {Schneider} S.,
  {Staveley-Smith} L.,  2008, \mn@doi [\mnras]
  {10.1111/j.1365-2966.2008.13424.x}, \href
  {http://adsabs.harvard.edu/abs/2008MNRAS.391.1712M} {391, 1712}

\bibitem[\protect\citeauthoryear{{Mihos}, {Keating}, {Holley-Bockelmann},
  {Pisano}  \& {Kassim}}{{Mihos} et~al.}{2012}]{Mihos2012}
{Mihos} J.~C.,  {Keating} K.~M.,  {Holley-Bockelmann} K.,  {Pisano} D.~J.,
  {Kassim} N.~E.,  2012, \mn@doi [\apj] {10.1088/0004-637X/761/2/186}, \href
  {http://adsabs.harvard.edu/abs/2012ApJ...761..186M} {761, 186}

\bibitem[\protect\citeauthoryear{{Mould} et~al.,}{{Mould}
  et~al.}{2000}]{Mould2000}
{Mould} J.~R.,  et~al., 2000, \mn@doi [\apj] {10.1086/308304}, \href
  {http://adsabs.harvard.edu/abs/2000ApJ...529..786M} {529, 786}

\bibitem[\protect\citeauthoryear{{Nakanishi} \& {Sofue}}{{Nakanishi} \&
  {Sofue}}{2016}]{Nakanishi2016}
{Nakanishi} H.,  {Sofue} Y.,  2016, \mn@doi [\pasj] {10.1093/pasj/psv108},
  \href {http://adsabs.harvard.edu/abs/2016PASJ...68....5N} {68, 5}

\bibitem[\protect\citeauthoryear{{Nicholls}, {Dopita}, {Jerjen}  \&
  {Meurer}}{{Nicholls} et~al.}{2011}]{Nicholls2011}
{Nicholls} D.~C.,  {Dopita} M.~A.,  {Jerjen} H.,   {Meurer} G.~R.,  2011,
  \mn@doi [\aj] {10.1088/0004-6256/142/3/83}, \href
  {http://adsabs.harvard.edu/abs/2011AJ....142...83N} {142, 83}

\bibitem[\protect\citeauthoryear{{Obreschkow}, {Glazebrook}, {Kilborn}  \&
  {Lutz}}{{Obreschkow} et~al.}{2016}]{Obreschkow2016}
{Obreschkow} D.,  {Glazebrook} K.,  {Kilborn} V.,   {Lutz} K.,  2016, \mn@doi
  [\apjl] {10.3847/2041-8205/824/2/L26}, \href
  {http://adsabs.harvard.edu/abs/2016ApJ...824L..26O} {824, L26}

\bibitem[\protect\citeauthoryear{{Odekon} et~al.,}{{Odekon}
  et~al.}{2016}]{Odekon2016}
{Odekon} M.~C.,  et~al., 2016, \mn@doi [\apj] {10.3847/0004-637X/824/2/110},
  \href {http://adsabs.harvard.edu/abs/2016ApJ...824..110O} {824, 110}

\bibitem[\protect\citeauthoryear{{Oosterloo}, {Zhang}, {Lucero}  \&
  {Carignan}}{{Oosterloo} et~al.}{2018}]{Osterloo2018}
{Oosterloo} T.~A.,  {Zhang} M.-L.,  {Lucero} D.~M.,   {Carignan} C.,  2018,
  preprint, \href {http://adsabs.harvard.edu/abs/2018arXiv180308263O} {}
  (\mn@eprint {arXiv} {1803.08263})

\bibitem[\protect\citeauthoryear{{Pisano}, {Wilcots}  \& {Elmegreen}}{{Pisano}
  et~al.}{2000}]{Pisano2000}
{Pisano} D.~J.,  {Wilcots} E.~M.,   {Elmegreen} B.~G.,  2000, \mn@doi [\aj]
  {10.1086/301464}, \href {http://adsabs.harvard.edu/abs/2000AJ....120..763P}
  {120, 763}

\bibitem[\protect\citeauthoryear{{Pisano}, {Barnes}, {Staveley-Smith},
  {Gibson}, {Kilborn}  \& {Freeman}}{{Pisano} et~al.}{2011}]{Pisano2011}
{Pisano} D.~J.,  {Barnes} D.~G.,  {Staveley-Smith} L.,  {Gibson} B.~K.,
  {Kilborn} V.~A.,   {Freeman} K.~C.,  2011, \mn@doi [\apjs]
  {10.1088/0067-0049/197/2/28}, \href
  {http://adsabs.harvard.edu/abs/2011ApJS..197...28P} {197, 28}

\bibitem[\protect\citeauthoryear{{Ponman}, {Allan}, {Jones}, {Merrifield},
  {McHardy}, {Lehto}  \& {Luppino}}{{Ponman} et~al.}{1994}]{Ponman1994}
{Ponman} T.~J.,  {Allan} D.~J.,  {Jones} L.~R.,  {Merrifield} M.,  {McHardy}
  I.~M.,  {Lehto} H.~J.,   {Luppino} G.~A.,  1994, \mn@doi [\nat]
  {10.1038/369462a0}, \href {http://adsabs.harvard.edu/abs/1994Natur.369..462P}
  {369, 462}

\bibitem[\protect\citeauthoryear{{Popping}, {Dav{\'e}}, {Braun}  \&
  {Oppenheimer}}{{Popping} et~al.}{2009}]{Popping2009}
{Popping} A.,  {Dav{\'e}} R.,  {Braun} R.,   {Oppenheimer} B.~D.,  2009,
  \mn@doi [\aap] {10.1051/0004-6361/200911811}, \href
  {http://adsabs.harvard.edu/abs/2009A%26A...504...15P} {504, 15}

\bibitem[\protect\citeauthoryear{{Postman} \& {Geller}}{{Postman} \&
  {Geller}}{1984}]{Postman1984}
{Postman} M.,  {Geller} M.~J.,  1984, \mn@doi [\apj] {10.1086/162078}, \href
  {http://adsabs.harvard.edu/abs/1984ApJ...281...95P} {281, 95}

\bibitem[\protect\citeauthoryear{{Press} \& {Schechter}}{{Press} \&
  {Schechter}}{1974}]{Press1974}
{Press} W.~H.,  {Schechter} P.,  1974, \mn@doi [\apj] {10.1086/152650}, \href
  {http://adsabs.harvard.edu/abs/1974ApJ...187..425P} {187, 425}

\bibitem[\protect\citeauthoryear{{Rasmussen}, {Ponman}  \&
  {Mulchaey}}{{Rasmussen} et~al.}{2006}]{Rasmussen2006}
{Rasmussen} J.,  {Ponman} T.~J.,   {Mulchaey} J.~S.,  2006, \mn@doi [\mnras]
  {10.1111/j.1365-2966.2006.10492.x}, \href
  {http://adsabs.harvard.edu/abs/2006MNRAS.370..453R} {370, 453}

\bibitem[\protect\citeauthoryear{Rasmussen et~al.,}{Rasmussen
  et~al.}{2012}]{Rasmussen2012}
Rasmussen J.,  et~al., 2012, The Astrophysical Journal, 747, 31

\bibitem[\protect\citeauthoryear{{Reda}, {Forbes}, {Beasley}, {O'Sullivan}  \&
  {Goudfrooij}}{{Reda} et~al.}{2004}]{Reda2004}
{Reda} F.~M.,  {Forbes} D.~A.,  {Beasley} M.~A.,  {O'Sullivan} E.~J.,
  {Goudfrooij} P.,  2004, \mn@doi [\mnras] {10.1111/j.1365-2966.2004.08250.x},
  \href {http://adsabs.harvard.edu/abs/2004MNRAS.354..851R} {354, 851}

\bibitem[\protect\citeauthoryear{{Robotham} \& {Obreschkow}}{{Robotham} \&
  {Obreschkow}}{2015}]{Robotham2015}
{Robotham} A.~S.~G.,  {Obreschkow} D.,  2015, \mn@doi [\pasa]
  {10.1017/pasa.2015.33}, \href
  {http://adsabs.harvard.edu/abs/2015PASA...32...33R} {32, e033}

\bibitem[\protect\citeauthoryear{Rots, Bosma, M.~van~der Hulst, Athanassoula
  \& C.~Crane}{Rots et~al.}{1990}]{Rots1990}
Rots A.,  Bosma A.,  M.~van~der Hulst J.,  Athanassoula E.,   C.~Crane P.,
  1990, 100, 387

\bibitem[\protect\citeauthoryear{{Sancisi}, {Fraternali}, {Oosterloo}  \& {van
  der Hulst}}{{Sancisi} et~al.}{2008}]{Sancisi2008}
{Sancisi} R.,  {Fraternali} F.,  {Oosterloo} T.,   {van der Hulst} T.,  2008,
  \mn@doi [\aapr] {10.1007/s00159-008-0010-0}, \href
  {http://adsabs.harvard.edu/abs/2008A%26ARv..15..189S} {15, 189}

\bibitem[\protect\citeauthoryear{{Saulder}, {van Kampen}, {Chilingarian},
  {Mieske}  \& {Zeilinger}}{{Saulder} et~al.}{2016}]{Saulder2016}
{Saulder} C.,  {van Kampen} E.,  {Chilingarian} I.~V.,  {Mieske} S.,
  {Zeilinger} W.~W.,  2016, \mn@doi [\aap] {10.1051/0004-6361/201526711}, \href
  {http://adsabs.harvard.edu/abs/2016A%26A...596A..14S} {596, A14}

\bibitem[\protect\citeauthoryear{{Sault}, {Teuben}  \& {Wright}}{{Sault}
  et~al.}{1995}]{Sault1995}
{Sault} R.~J.,  {Teuben} P.~J.,   {Wright} M.~C.~H.,  1995, in {Shaw} R.~A.,
  {Payne} H.~E.,   {Hayes} J.~J.~E.,  eds,  Astronomical Society of the Pacific
  Conference Series Vol. 77, Astronomical Data Analysis Software and Systems
  IV. p.~433 (\mn@eprint {} {astro-ph/0612759})

\bibitem[\protect\citeauthoryear{{Serra} et~al.,}{{Serra}
  et~al.}{2012}]{Serra2012}
{Serra} P.,  et~al., 2012, \mn@doi [\mnras] {10.1111/j.1365-2966.2012.20219.x},
  \href {http://adsabs.harvard.edu/abs/2012MNRAS.422.1835S} {422, 1835}

\bibitem[\protect\citeauthoryear{{Skrutskie} et~al.,}{{Skrutskie}
  et~al.}{2006}]{Skrutskie2006}
{Skrutskie} M.~F.,  et~al., 2006, \mn@doi [\aj] {10.1086/498708}, \href
  {http://adsabs.harvard.edu/abs/2006AJ....131.1163S} {131, 1163}

\bibitem[\protect\citeauthoryear{{Spindler} et~al.,}{{Spindler}
  et~al.}{2018}]{Spindler2018}
{Spindler} A.,  et~al., 2018, \mn@doi [\mnras] {10.1093/mnras/sty247}, \href
  {http://adsabs.harvard.edu/abs/2018MNRAS.476..580S} {476, 580}

\bibitem[\protect\citeauthoryear{{Stanek} \& {Garnavich}}{{Stanek} \&
  {Garnavich}}{1998}]{Stanek1998}
{Stanek} K.~Z.,  {Garnavich} P.~M.,  1998, \mn@doi [\apjl] {10.1086/311539},
  \href {http://adsabs.harvard.edu/abs/1998ApJ...503L.131S} {503, L131}

\bibitem[\protect\citeauthoryear{{Stevens} \& {Brown}}{{Stevens} \&
  {Brown}}{2017}]{Stevens2017}
{Stevens} A.~R.~H.,  {Brown} T.,  2017, \mn@doi [\mnras]
  {10.1093/mnras/stx1596}, \href
  {http://adsabs.harvard.edu/abs/2017MNRAS.471..447S} {471, 447}

\bibitem[\protect\citeauthoryear{{Stevens}, {Lagos}, {Obreschkow}  \&
  {Sinha}}{{Stevens} et~al.}{2018}]{Stevens2018}
{Stevens} A.~R.~H.,  {Lagos} C.~d.~P.,  {Obreschkow} D.,   {Sinha} M.,  2018,
  preprint, \href {http://adsabs.harvard.edu/abs/2018arXiv180607402S} {}
  (\mn@eprint {arXiv} {1806.07402})

\bibitem[\protect\citeauthoryear{{Stierwalt}, {Liss}, {Johnson}, {Patton},
  {Privon}, {Besla}, {Kallivayalil}  \& {Putman}}{{Stierwalt}
  et~al.}{2017}]{Stierwalt2017}
{Stierwalt} S.,  {Liss} S.~E.,  {Johnson} K.~E.,  {Patton} D.~R.,  {Privon}
  G.~C.,  {Besla} G.,  {Kallivayalil} N.,   {Putman} M.,  2017, \mn@doi [Nature
  Astronomy] {10.1038/s41550-016-0025}, \href
  {http://adsabs.harvard.edu/abs/2017NatAs...1E..25S} {1, 0025}

\bibitem[\protect\citeauthoryear{{Sweet} et~al.,}{{Sweet}
  et~al.}{2013}]{Sweet2013}
{Sweet} S.~M.,  et~al., 2013, \mn@doi [\mnras] {10.1093/mnras/stt747}, \href
  {http://adsabs.harvard.edu/abs/2013MNRAS.433..543S} {433, 543}

\bibitem[\protect\citeauthoryear{{Sweet}, {Drinkwater}, {Meurer}, {Bekki},
  {Dopita}, {Kilborn}  \& {Nicholls}}{{Sweet} et~al.}{2014}]{Sweet2014}
{Sweet} S.~M.,  {Drinkwater} M.~J.,  {Meurer} G.,  {Bekki} K.,  {Dopita} M.~A.,
   {Kilborn} V.,   {Nicholls} D.~C.,  2014, \mn@doi [\apj]
  {10.1088/0004-637X/782/1/35}, \href
  {http://adsabs.harvard.edu/abs/2014ApJ...782...35S} {782, 35}

\bibitem[\protect\citeauthoryear{{Tavakoli}}{{Tavakoli}}{2012}]{Tavakoli2012}
{Tavakoli} M.,  2012, preprint, \href
  {http://adsabs.harvard.edu/abs/2012arXiv1207.6150T} {} (\mn@eprint {arXiv}
  {1207.6150})

\bibitem[\protect\citeauthoryear{{Terlouw} \& {Vogelaar}}{{Terlouw} \&
  {Vogelaar}}{2015}]{KapteynPackage}
{Terlouw} J.~P.,  {Vogelaar} M.~G.~R.,  2015, {Kapteyn Package, version 2.3}.
{Kapteyn Astronomical Institute}, Groningen

\bibitem[\protect\citeauthoryear{{Tonry}, {Blakeslee}, {Ajhar}  \&
  {Dressler}}{{Tonry} et~al.}{2000}]{Tonry2000}
{Tonry} J.~L.,  {Blakeslee} J.~P.,  {Ajhar} E.~A.,   {Dressler} A.,  2000,
  \mn@doi [\apj] {10.1086/308409}, \href
  {http://adsabs.harvard.edu/abs/2000ApJ...530..625T} {530, 625}

\bibitem[\protect\citeauthoryear{{Tully}}{{Tully}}{1987}]{Tully1987}
{Tully} R.~B.,  1987, \mn@doi [\apj] {10.1086/165629}, \href
  {http://adsabs.harvard.edu/abs/1987ApJ...321..280T} {321, 280}

\bibitem[\protect\citeauthoryear{{Van Der Walt}, {Colbert}  \&
  {Varoquaux}}{{Van Der Walt} et~al.}{2011}]{VanDerWalt2011}
{Van Der Walt} S.,  {Colbert} S.~C.,   {Varoquaux} G.,  2011, preprint, \href
  {http://adsabs.harvard.edu/abs/2011arXiv1102.1523V} {} (\mn@eprint {arXiv}
  {1102.1523})

\bibitem[\protect\citeauthoryear{{Verdes-Montenegro}, {Yun}, {Williams},
  {Huchtmeier}, {Del Olmo}  \& {Perea}}{{Verdes-Montenegro}
  et~al.}{2001}]{VerdesMontenegro2001}
{Verdes-Montenegro} L.,  {Yun} M.~S.,  {Williams} B.~A.,  {Huchtmeier} W.~K.,
  {Del Olmo} A.,   {Perea} J.,  2001, \mn@doi [\aap]
  {10.1051/0004-6361:20011127}, \href
  {http://adsabs.harvard.edu/abs/2001A%26A...377..812V} {377, 812}

\bibitem[\protect\citeauthoryear{{Verdes-Montenegro}, {Sulentic}, {Lisenfeld},
  {Leon}, {Espada}, {Garcia}, {Sabater}  \& {Verley}}{{Verdes-Montenegro}
  et~al.}{2005}]{VM2005}
{Verdes-Montenegro} L.,  {Sulentic} J.,  {Lisenfeld} U.,  {Leon} S.,  {Espada}
  D.,  {Garcia} E.,  {Sabater} J.,   {Verley} S.,  2005, \mn@doi [\aap]
  {10.1051/0004-6361:20042280}, \href
  {http://adsabs.harvard.edu/abs/2005A%26A...436..443V} {436, 443}

\bibitem[\protect\citeauthoryear{{Verheijen} \& {Sancisi}}{{Verheijen} \&
  {Sancisi}}{2001}]{Verheijen2001}
{Verheijen} M.~A.~W.,  {Sancisi} R.,  2001, \mn@doi [\aap]
  {10.1051/0004-6361:20010090}, \href
  {http://adsabs.harvard.edu/abs/2001A%26A...370..765V} {370, 765}

\bibitem[\protect\citeauthoryear{Wang et~al.,}{Wang et~al.}{2013}]{Wang2013}
Wang J.,  et~al., 2013, arXiv.org, pp 270--294

\bibitem[\protect\citeauthoryear{{Wang} et~al.,}{{Wang}
  et~al.}{2015}]{Wang2015}
{Wang} J.,  et~al., 2015, \mn@doi [\mnras] {10.1093/mnras/stv1767}, \href
  {http://adsabs.harvard.edu/abs/2015MNRAS.453.2399W} {453, 2399}

\bibitem[\protect\citeauthoryear{Wang, Koribalski, Serra, van~der Hulst,
  Roychowdhury, Kamphuis  \& Chengalur}{Wang et~al.}{2016}]{Wang2016}
Wang J.,  Koribalski B.~S.,  Serra P.,  van~der Hulst T.,  Roychowdhury S.,
  Kamphuis P.,   Chengalur J.~N.,  2016, arXiv.org, pp 2143--2151

\bibitem[\protect\citeauthoryear{{Wang}, {Koribalski}, {Serra}, {van der
  Hulst}, {Roychowdhury}, {Kamphuis}  \& {Chengalur}}{{Wang}
  et~al.}{2017}]{Wang2017}
{Wang} J.,  {Koribalski} B.~S.,  {Serra} P.,  {van der Hulst} T.,
  {Roychowdhury} S.,  {Kamphuis} P.,   {Chengalur} J.~N.,  2017, VizieR Online
  Data Catalog, \href {http://adsabs.harvard.edu/abs/2017yCat..74602143W} {746}

\bibitem[\protect\citeauthoryear{{Wen}, {Wu}, {Zhu}, {Lam}, {Wu}, {Wicker}  \&
  {Zhao}}{{Wen} et~al.}{2013}]{Wen2013}
{Wen} X.-Q.,  {Wu} H.,  {Zhu} Y.-N.,  {Lam} M.~I.,  {Wu} C.-J.,  {Wicker} J.,
  {Zhao} Y.-H.,  2013, \mn@doi [\mnras] {10.1093/mnras/stt939}, \href
  {http://adsabs.harvard.edu/abs/2013MNRAS.433.2946W} {433, 2946}

\bibitem[\protect\citeauthoryear{{Westmeier}, {Braun}  \&
  {Koribalski}}{{Westmeier} et~al.}{2011}]{Westmeier2011}
{Westmeier} T.,  {Braun} R.,   {Koribalski} B.~S.,  2011, \mn@doi [\mnras]
  {10.1111/j.1365-2966.2010.17596.x}, \href
  {http://adsabs.harvard.edu/abs/2011MNRAS.410.2217W} {410, 2217}

\bibitem[\protect\citeauthoryear{{White} \& {Rees}}{{White} \&
  {Rees}}{1978}]{White1978}
{White} S.~D.~M.,  {Rees} M.~J.,  1978, \mn@doi [\mnras]
  {10.1093/mnras/183.3.341}, \href
  {http://adsabs.harvard.edu/abs/1978MNRAS.183..341W} {183, 341}

\bibitem[\protect\citeauthoryear{{Whitmore}, {Gilmore}  \& {Jones}}{{Whitmore}
  et~al.}{1993}]{Whitmore1993}
{Whitmore} B.~C.,  {Gilmore} D.~M.,   {Jones} C.,  1993, \mn@doi [\apj]
  {10.1086/172531}, \href {http://adsabs.harvard.edu/abs/1993ApJ...407..489W}
  {407, 489}

\bibitem[\protect\citeauthoryear{{Wilson} et~al.,}{{Wilson}
  et~al.}{2011}]{Wilson2011}
{Wilson} W.~E.,  et~al., 2011, \mn@doi [\mnras]
  {10.1111/j.1365-2966.2011.19054.x}, \href
  {http://adsabs.harvard.edu/abs/2011MNRAS.416..832W} {416, 832}

\bibitem[\protect\citeauthoryear{{Wolfinger}, {Kilborn}, {Ryan-Weber}  \&
  {Koribalski}}{{Wolfinger} et~al.}{2016}]{Wolfinger2016}
{Wolfinger} K.,  {Kilborn} V.~A.,  {Ryan-Weber} E.~V.,   {Koribalski} B.~S.,
  2016, \mn@doi [\pasa] {10.1017/pasa.2016.31}, \href
  {http://adsabs.harvard.edu/abs/2016PASA...33...38W} {33, e038}

\bibitem[\protect\citeauthoryear{{Wolter}, {Esposito}, {Mapelli}, {Pizzolato}
  \& {Ripamonti}}{{Wolter} et~al.}{2015}]{Wolter2015}
{Wolter} A.,  {Esposito} P.,  {Mapelli} M.,  {Pizzolato} F.,   {Ripamonti} E.,
  2015, \mn@doi [\mnras] {10.1093/mnras/stv054}, \href
  {http://adsabs.harvard.edu/abs/2015MNRAS.448..781W} {448, 781}

\bibitem[\protect\citeauthoryear{{Wong}, {Meurer}, {Zheng}, {Heckman},
  {Thilker}  \& {Zwaan}}{{Wong} et~al.}{2016}]{Wong2016}
{Wong} O.~I.,  {Meurer} G.~R.,  {Zheng} Z.,  {Heckman} T.~M.,  {Thilker} D.~A.,
    {Zwaan} M.~A.,  2016, \mn@doi [\mnras] {10.1093/mnras/stw993}, \href
  {http://adsabs.harvard.edu/abs/2016MNRAS.460.1106W} {460, 1106}

\bibitem[\protect\citeauthoryear{{Yang}, {Mo}, {van den Bosch}, {Pasquali},
  {Li}  \& {Barden}}{{Yang} et~al.}{2007}]{Yang2007}
{Yang} X.,  {Mo} H.~J.,  {van den Bosch} F.~C.,  {Pasquali} A.,  {Li} C.,
  {Barden} M.,  2007, \mn@doi [\apj] {10.1086/522027}, \href
  {http://adsabs.harvard.edu/abs/2007ApJ...671..153Y} {671, 153}

\bibitem[\protect\citeauthoryear{Yun, Ho  \& Lo}{Yun et~al.}{1994}]{Yun1994}
Yun M.~S.,  Ho P. T.~P.,   Lo K.~Y.,  1994, Nature, 372, 530

\makeatother
\end{thebibliography}

\appendix
\section{Additional tables: properties of the Choir and isolated galaxies. }

\onecolumn

\begin{landscape}
\begin{table*}
\centering
\begin{threeparttable}
\caption{\HI\ properties of the Choir galaxies obtained with the ATCA observations.}
\label{hi_properties}
\begin{tabular}{llllccccccccc}
\toprule
Galaxy ID& R.A. & Dec. & D  & M$_{R}$ & log M$_{\star}$ & log M$_{\HI\ }$& DEF$_{\HI\ }$ & SFR  & t$_{\textrm{dep}}$  & D$_{\textrm{\HI}}$  & q & Morphology \\ 
HIPASS+ &	[hh mm ss.s]	 &  [dd mm ss.s] &[Mpc] &	[mag]  & [M$_\odot$] & [M$_\odot$] & [dex] & [M$_{\odot}$ yr$^{-1}$] & [Gyr] & [kpc] & \\
(1) & (2) & (3) & (4) & (5) & (6) & (7) & (8) & (9) & (10) & (11) & (12) & (13) \B \\ \hline \hline

J0205-55:S1 & 02 05 05.48 & -55 06 42.54 & 93 &-23.01$\pm$0.32 &  11.44$\pm$0.16     & 10.43$\pm$0.09   & -0.07 & 9.9$\pm$1.9 & 2.7 & 98.5$\pm$4.9 & 0.09 &SAB(r)bc \T \\ 
J0205-55:S2 & 02 04 50.78 & -55 13 01.55 & 93 &-20.77$\pm$0.26 &  10.35$\pm$0.13      & 9.84$\pm$0.09   &  -0.04 & 2.99$\pm$0.61 & 2.3 & 47.6$\pm$3.6 & 0.2 & SB(s)cd pec \\
J0258-74:S1 & 02 58 06.48 & -74 27 22.79 & 70 &-22.13$\pm$0.24 &  11.01$\pm$0.12     & 10.27$\pm$0.03   & -0.13 & 3.2$\pm$0.6  & 5.8 & 82.4$\pm$5.3 & 0.26 & SAB(rs)bc HII\\
J0258-74:S2 & 02 58 52.43 & -74 25 53.25 & 70 &-19.48$\pm$0.27 &  9.73$\pm$0.12      & 9.8$\pm$0.1   &  -0.33 & 0.35$\pm$0.07 & 18.4 & 52.8$\pm$2.9 & $\cdots$ & S \\
J0258-74:S3 & 02 58 42.76 & -74 26 03.55 & 70 &-18.51$\pm$0.33 &  9.26$\pm$0.16      & 8.5$\pm$0.2   &  0.72 & 0.13$\pm$0.03 & 2.5 & $\cdots$ &  $\cdots$ & S \\
J0400-52:S1 & 04 00 40.82 & -52 44 02.71 & 151 &-22.31$\pm$0.25 &  11.10$\pm$0.12     & 10.66$\pm$0.09   &  -0.48 & 0.5$\pm$0.4 & 91.9 & 77.3$\pm$3.8 & 0.26 & SA(rs)cd pec \\
J1250-20:S1 & 12 50 52.84 & -20 22 15.65 & 114 &-22.54$\pm$0.24 &  11.21$\pm$0.12     & 10.48$\pm$0.09   & -0.24 & 6.6$\pm$1.3 & 4.6 & 106.4$\pm$4.7 & 0.17 & SA(s)bc pec HII\\
J1250-20:S2 & 12 50 40.91 & -20 20 06.22 & 114 &-22.06$\pm$0.24 &  10.98$\pm$0.12     & 9.6$\pm$0.1   &  0.56 & 3.7$\pm$0.8 & 1.0 & 29.2$\pm$9.1 &$\cdots$ & S\\
J2027-51:S1 & 20 28 06.39 & -51 41 29.83 &87 &-21.73$\pm$0.25 &  10.82$\pm$0.13     & 10.2$\pm$0.1   & -0.14 &  5.4$\pm$1.1  & 2.7 & 56.4$\pm$4.2 & 0.24 & (R')SA(s)bc\\
J2027-51:S2 & 20 27 31.97 & -51 39 20.81 & 87 &-21.91$\pm$0.24 &  10.91$\pm$0.12     & 10.0$\pm$0.1   &  0.07 & 4.3$\pm$0.9 & 2.4 & 42.4$\pm$3.8 & 0.21 & SAB(s)bc pec\\
J2027-51:Sm\tnote{*} & 20 28 08.4 & -51 28 31 & 87 &$\cdots$ &  $\cdots$              & 9.49$\pm$0.13   &   $\cdots$ & $\cdots$ & $\cdots$ & 25.9$\pm$3.8 & $\cdots$ & [Irr]\\
J1051-17:S1 & 10 51 37.45 & -17 07 29.23 & 83 &-21.78$\pm$0.25 &  10.84$\pm$0.12      & 10.43$\pm$0.02   &  -0.38 & 1.5$\pm$0.3 & 17.7 & 99.6$\pm$6.3 & 0.67 & (R? + PR?) \\
J1051-17:S2 & 10 51 15.11 & -17 00 29.44 & 83 &-21.85$\pm$0.24    &  10.88$\pm$0.12  & 9.45$\pm$0.05   &  0.62 & 2.2$\pm$0.4 & 1.3 & $\cdots$ & $\cdots$ & SABb?\\
J1051-17:S3 & 10 51 35.94 & -16 59 16.80 & 83 &-18.16$\pm$0.26    &  9.08$\pm$0.13   & 8.73$\pm$0.07   &  0.41 & 0.104$\pm$0.021 & 5.2 & $\cdots$ & $\cdots$ & dS\\
J1051-17:NED1\tnote{*} & 10 52 15.9 & -17 07 48.6 &83 &$\cdots$    &    $\cdots$             & 9.48$\pm$0.03   &   $\cdots$ & $\cdots$  & $\cdots$ & $\cdots$ & $\cdots$ & SA(s)a \\
J1051-17:NED6\tnote{*} & 10 50 49.6 & -17 14 07 &  83  &$\cdots$    &    $\cdots$             & 9.01$\pm$0.03   &  $\cdots$ & $\cdots$ & $\cdots$ & $\cdots$ & $\cdots$ & [Irr?] \\
J1059-09:S1 & 10 59 16.25 & -09 47 38.16 &122 &-23.24$\pm$0.24    & 11.55$\pm$0.12   & 10.16$\pm$0.10   &  0.26 & 8.9$\pm$1.8 & 1.6 & 96.3$\pm$6.3 $\dagger$ & $\cdots$ & SAB(rs)b pec\\
J1059-09:S2 & 10 59 06.77 & -09 45 04.38 & 122 &-20.20$\pm$0.24   & 10.07$\pm$0.13   & 9.6$\pm$0.1   &  0.02 & 1.3$\pm$0.3 & 3.3 & 50.4$\pm$3.8 & $\cdots$ & Irr\\
J1059-09:S3 & 10 59 15.61 & -09 48 59.41 & 122 &-21.45$\pm$0.25    & 10.68$\pm$0.12   & 10.2$\pm$0.1   &  -0.19 & 3.7$\pm$0.7 & 3.9 &  & $\cdots$ & Sab pec sp\\
J1059-09:S5 & 10 59 30.98 & -09 44 25.26 & 122 &-19.94$\pm$0.28    &  9.95$\pm$0.14   & 9.1$\pm$0.1   &  0.45 & 0.8$\pm$0.2 & 1.7 & $\cdots$ & $\cdots$ & S\\
J1059-09:S6 & 10 59 08.46 & -09 43 14.49 & 122 &-19.33$\pm$0.28    &  9.65$\pm$0.14   & 8.9$\pm$0.1   &  0.46 & 0.37$\pm$0.07 & 2.5 & $\cdots$ & $\cdots$ & Irr\\
J1026-19:S1 & 10 26 40.81 & -19 03 04.01 & 135 &-22.74$\pm$0.25    & 11.31$\pm$0.12   & 10.3$\pm$0.1   & -0.02 & 4.5$\pm$0.9 & 4.6 & \multirow{2}{*}{136.4$\pm$6.5} $\dagger$ & $\cdots$ &SAB(s)bc: pec Sbrst \\
J1026-19:S2 & 10 26 50.07 & -19 04 31.77 & 135 &-20.60$\pm$0.26    &  10.27$\pm$0.13  & 9.4$\pm$0.1   &  0.37 & 1.3$\pm$0.3 & 1.9 & & $\cdots$ &Irr \\
J1026-19:S5\tnote{a} & 10 26 42.07 & -19 07 35.07 & 135 &-16.94$\pm$0.41 & 8.5$\pm$0.2 & 9.3$\pm$0.1   &  -0.44 & 0.03$\pm$0.01 & $\cdots$ & $\cdots$ & $\cdots$ & dIrr\\
J1408-21:S1 & 14 08 42.04 & -21 35 49.82 & 128 &-23.15$\pm$0.24 &   11.51$\pm$0.12    & 10.25$\pm$0.22   & 0.14 & 7.1$\pm$1.4 & 2.5 & \multirow{2}{*}{134.7$\pm$3.7} $\dagger$ & $\cdots$ &SB(rl)c \\
J1408-21:S3 & 14 08 41.04 & -21 37 40.97 & 128 &-20.72$\pm$0.25 &  10.33$\pm$0.12     & 9.7$\pm$0.1   &  0.08 & 0.8$\pm$0.2 & 6.2 & & $\cdots$ & S0 \\
J1408-21:Sm\tnote{*} & 14 08 46.6 & -21 27 09 & 128 &$\cdots$ &  $\cdots$             & 9.5$\pm$0.2   &  $\cdots$ & $\cdots$ & $\cdots$ & $\cdots$ & $\cdots$ & [Edge-on]\\
  
\bottomrule
\end{tabular}
\begin{tablenotes}
\item Columns: (1) Galaxy ID as assigned in SINGG \citep{Meurer2006, Sweet2013}; (2) Right Ascension (J2000); (3) Declination (J2000); (4) Group distance \citep{Sweet2013}; (5) Absolute dust-corrected \textit{R}-band magnitude (see \citealt{Meurer2006} for details); (6) Stellar mass, derived from the \textit{R}-band magnitude, following \citet{Wong2016}, Eq. 4; (7) \HI\ mass measured from the ATCA/VLA data cubes (see Eq. \ref{mass}); (8) The \HI\ deficiency parameter (see Eq. \ref{defeq} in the text); (9) The star formation rate, derived from the H$\alpha$ fluxes obtained from the SINGG survey (as in \citet{Meurer2006} and \citet{Sweet2013}); (10) Atomic gas depletion time (see Eq. \ref{timescale} in the text); (11) The measured galaxy's \HI\ diameter at a surface density of 1 M$_{\odot}$ pc$^{-2}$ (described in Section \ref{sizemass}); (12) The global stability parameter derived as described in the Section \ref{stability}; (13) Adopted morphology from \citet{Sweet2013}, for new members classification is based on the NED (NASA/IPAC Extragalactic Database) classification where available, or [our classification].
\item[*] Discovered new group member based on the \HI\ emission.
\item[a] Galaxy detected only with natural weighted cubes, thus not used in the analysis.
\item[$\dagger$]Measured \HI\ diameters of HIPASSJ1059-09:S1+S3, HIPASSJ1026-19:S1+S2 and HIPASSJ1408-21:S1+S3 respectively, as described in Section \ref{sizemass}. 

\label{tab:Choirs} 
\end{tablenotes}
\end{threeparttable}
\end{table*}
\end{landscape}

\newpage

\begin{landscape}
\begin{table*}
\captionsetup{width=1\columnwidth}
\centering
\begin{threeparttable}
\caption{The sample of isolated galaxies and their properties.}
\label{isolated_properties}
\begin{tabular}{lllcccccccc}
\toprule
Galaxy ID & R.A. & Dec. & D &log M$_{\HI\ }$ & M$_{R}$ & log M$_{\star}$  & DEF$_{\HI\ }$ & SFR & t$_{\textrm{dep}}$ & Morphology\\ 
HIPASS+& [hh mm ss.s]	 &  [dd mm ss] & [Mpc] &	[M$_\odot$] & [mag] & [M$_\odot$] & [dex]	& [M$_{\odot}$ yr$^{-1}$] & [Gyr] & \\
(1) & (2) & (3) & (4) & (5) & (6) & (7) & (8) & (9) & (10) & (11) \B \\ \hline \hline
J0328-08\tnote{a}& 03 28 05.6 & -08 23 28    & 17.4 & 9.78$\pm$0.09 & -20.15$\pm$0.28  &10.05$\pm$0.14 &  -0.14 & 0.65$\pm$0.13 & 9.3     & SA(s)cd    \T \\
J0412+02&04 12 47.3	&+02 22 44               & 70.6 & 10.2$\pm$0.1 & -22.27$\pm$0.24    &11.08$\pm$0.12 & -0.03 & 3.211$\pm$0.643 & 5.0     & SB(s)b     \\
J0430-01\tnote{b}&04 30 55.8&	-01 59 13    & 36.1 & 9.5$\pm$0.1 & -19.45$\pm$0.26    &9.71$\pm$0.12 &  -0.07 & 0.501$\pm$0.101 & 6.8     & SAB(s)b pec    \\
J0512-57&05 12 09.8	&-57 24 34               & 19.3 & 9.25$\pm$0.09 & -19.14$\pm$0.24  &9.56$\pm$0.12 &   0.14 & 0.357$\pm$0.07 & 5.0     & SB(s)m    \\
J0515-41&05 15 01.9	&-41 21 33               & 14.5 & 9.03$\pm$0.09 & -17.28$\pm$0.24  &8.66$\pm$0.12 &   -0.11 & 0.065$\pm$0.013 & 16.7     & IB(s)m     \\
J1019-17&10 19 39.0	&-17 43 48               & 11 & 8.34$\pm$0.12 & -15.25$\pm$0.26  &7.67$\pm$0.13 &   0.07 & 0.003$\pm$0.01 & 64.5    & Sm    \\
J1051-19&10 51 24.6	&-19 53 07               & 31 & 9.76$\pm$0.09 & -20.14$\pm$0.24  &10.04$\pm$0.12 &  -0.12 & 0.67$\pm$0.13 & 8.6     & SB(s)cd     \\
J1120-21&11 20 12.3	&-21 28 32               & 18.3 & 9.22$\pm$0.09 & -17.95$\pm$0.27  &8.98$\pm$0.13 &   -0.13 & 0.089$\pm$0.018 & 18.7     & (R')SAc pec    \\
J1123-08&11 23 28.1	&-08 38 52               & 56.5 & 10.22$\pm$0.09 & -22.59$\pm$0.24  &11.24$\pm$0.12 & -0.03 & 7.732$\pm$1.547 & 2.2      & SB(r)bc     \\
J1558-10&15 58 22.3&-10 31 48                & 14.8 & 8.78$\pm$0.11 & -16.41$\pm$0.25  &8.23$\pm$0.12 &   -0.08 & 0.019$\pm$0.039 & 30.9     & BCD     \\
J1621-02\tnote{ab}&16 21 45.8&	-02 16 34    & 26.3 & 9.75$\pm$0.09 & -21.88$\pm$0.24  &10.89$\pm$0.12 &  0.33 & 2.476$\pm$0.495  & 2.3      & SA(s)cd    \\
J1954-58 & 19 54 23.4	& -58 42 47  & 31.1 & 10.07$\pm$0.09 &-21.85$\pm$0.24   &10.88$\pm$0.12 & 0.18 & 3.63$\pm$0.73 & 3.2 &  SAB(rs)c   \\
J2039-63\tnote{c}&20 39 07.3	&-63 48 50   & 23.6 & 8.69$\pm$0.14 & -17.08$\pm$0.27  &8.56$\pm$0.13 &   -0.14 & 0.122$\pm$0.024 & 4.0      & dS0p     \\
J2127-60&21 27 23.2&	-60 01 22            & 24.9 & 9.94$\pm$0.09 & -20.79$\pm$0.25  &10.36$\pm$0.12 &  -0.03 & 1.195$\pm$0.239 & 7.3     & SAB(rs)c    \\
J2152-55&21 52 12.3	&-55 35 12               & 43.5 & 10.25$\pm$0.09 & -22.46$\pm$0.25  &11.17$\pm$0.12 &  0.1 & 3.621$\pm$0.724 & 4.9     & SAB(rs)bc     \\
J2157-60&21 57 52.4	&-60 18 27               & 24.2 & 8.76$\pm$0.12 & -17.04$\pm$0.29  &8.54$\pm$0.14 & -0.06 & 0.033$\pm$0.007 & 17.7     & IB(s)m     \\
\bottomrule&
\end{tabular}
\begin{tablenotes}
\item Columns: (1) Galaxy ID as assigned in SINGG \citep{Meurer2006, Sweet2013}; (2) Right Ascension (J2000); (3) Declination (J2000); (4) Galaxy distance, adopted from SINGG (Meurer et al. in prep.); (5) \HI\ mass from the HIPASS data, adapted from \citet{Meurer2006}; (6) Absolute dust-corrected \textit{R}-band magnitude; (7) Stellar mass, derived from the \textit{R}-band magnitude, following \citet{Wong2016}; (8) The \HI\ deficiency parameter (Eq. \ref{defeq}, described in Section \ref{Deficiency}); (9) The star formation rate, derived from the SINGG H$\alpha$ fluxes (as in \citet{Meurer2006} and \citet{Sweet2013}); (10) Atomic gas depletion time (see Eq. \ref{timescale} in the text); (11) Morphology based on the NED (NASA/IPAC Extragalactic Database) classification.

\item[a] Part of the 2MIG (2MASS Isolated Galaxies) sample \citep{Karachentseva2010}.
\item[b] Part of the CIG (Catalogue of Isolated Galaxies) sample \citep{Karachentseva1973}.
\item[c] Part of the SIGRID (Small Isolated Gas-Rich Irregular Dwarfs) sample \citep{Nicholls2011}.

\label{tab:Isolated} 
\end{tablenotes}
\end{threeparttable}
\end{table*}
\end{landscape}

\newpage
\twocolumn
\section{The deprojection and measurement of the \HI\ diameters in HIPASSJ1250-20 group}
\label{app2}

\begin{figure}
    \centering
        \begin{tabular}{c}
        \smallskip
            \begin{subfigure}[t]{0.43\textwidth}
                \centering
                \includegraphics[width=0.9\textwidth]{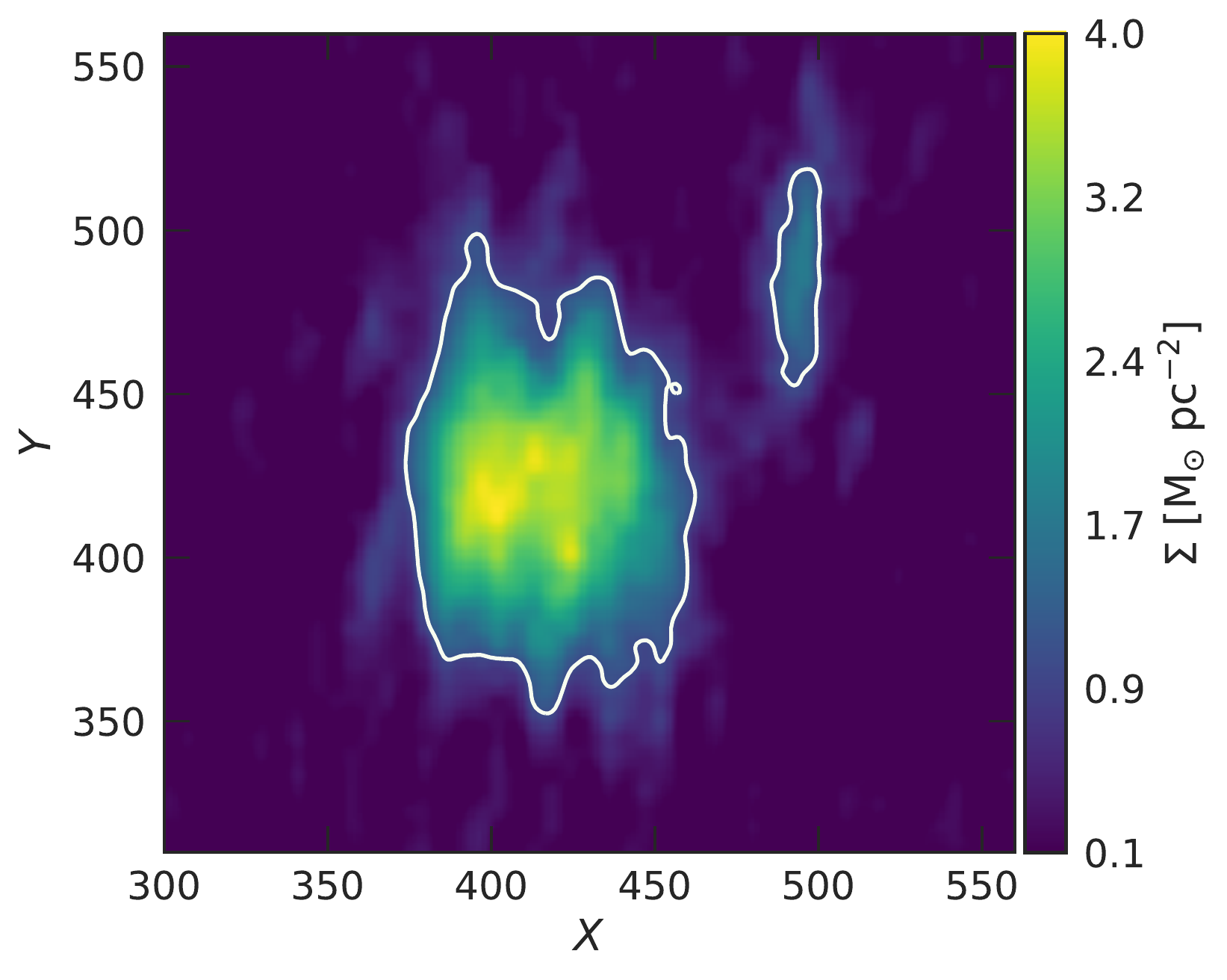}
                \caption{The observed HIPASSJ1250-20 intensity map.}
            \end{subfigure}\\
            \begin{subfigure}[t]{0.43\textwidth}
                \centering
                \includegraphics[width=0.9\textwidth]{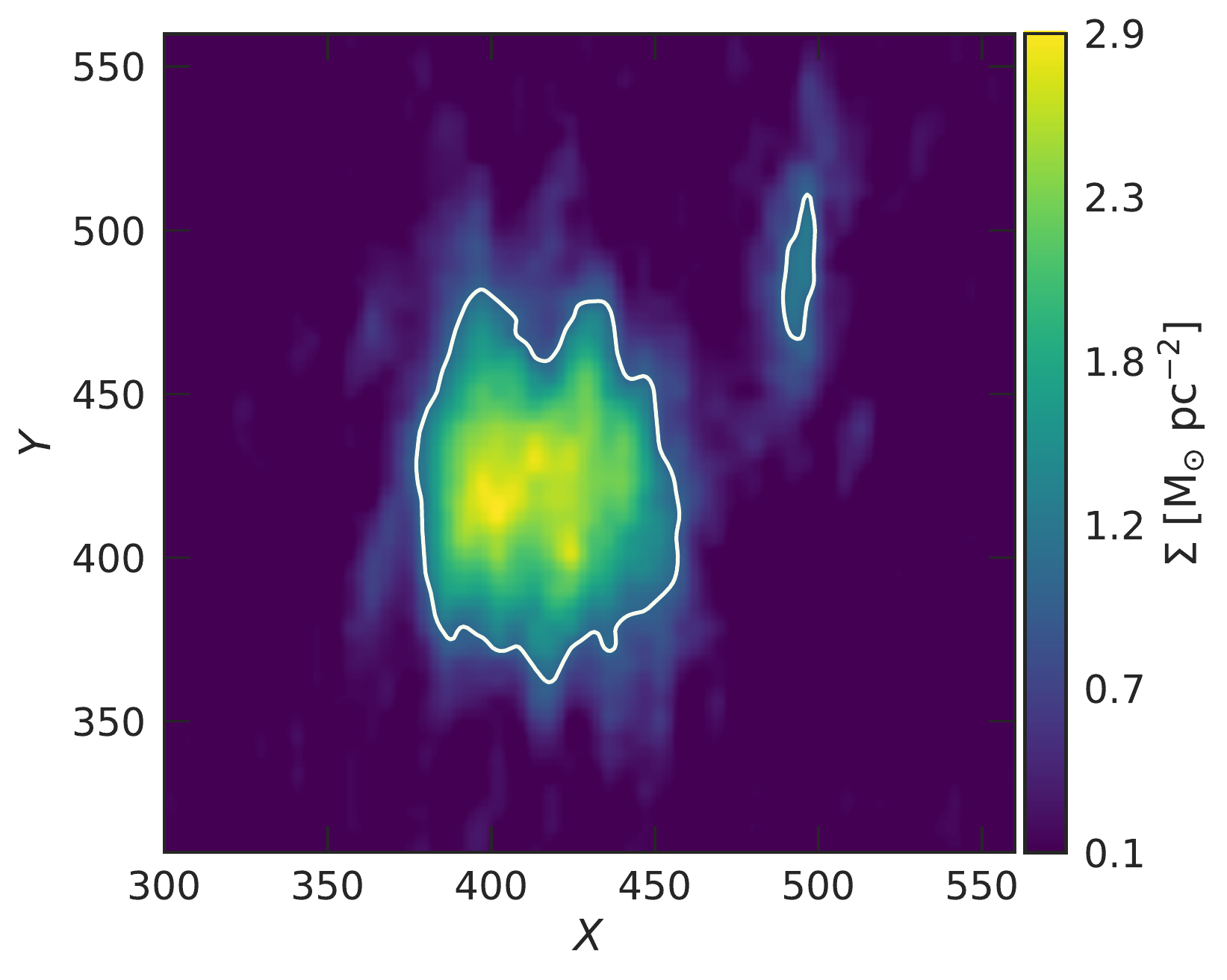}
                \caption{The deprojected HIPASSJ1250-20 intensity map.}
            \end{subfigure}
        \end{tabular}\\
    \caption{The \HI\ intensity maps of the HIPASSJ1250-20: (a) Observed; (b) Deprojected (as described in Section \ref{sizemass}). The white contour shows the \HI\ surface density of 1 M$_{\odot}$ pc$^{-2}$. The colour bar shows the \HI\ surface density from the lowest value above 3$\sigma$ detection (0.1 M$_{\odot}$ pc$^{-2}$) to the maximum value.}
    \label{intensitymaps}
\end{figure}

\begin{figure}
    \centering
        \begin{tabular}{c}
        \smallskip
            \begin{subfigure}[t]{0.4\textwidth}
                \centering
                \includegraphics[width=0.9\textwidth]{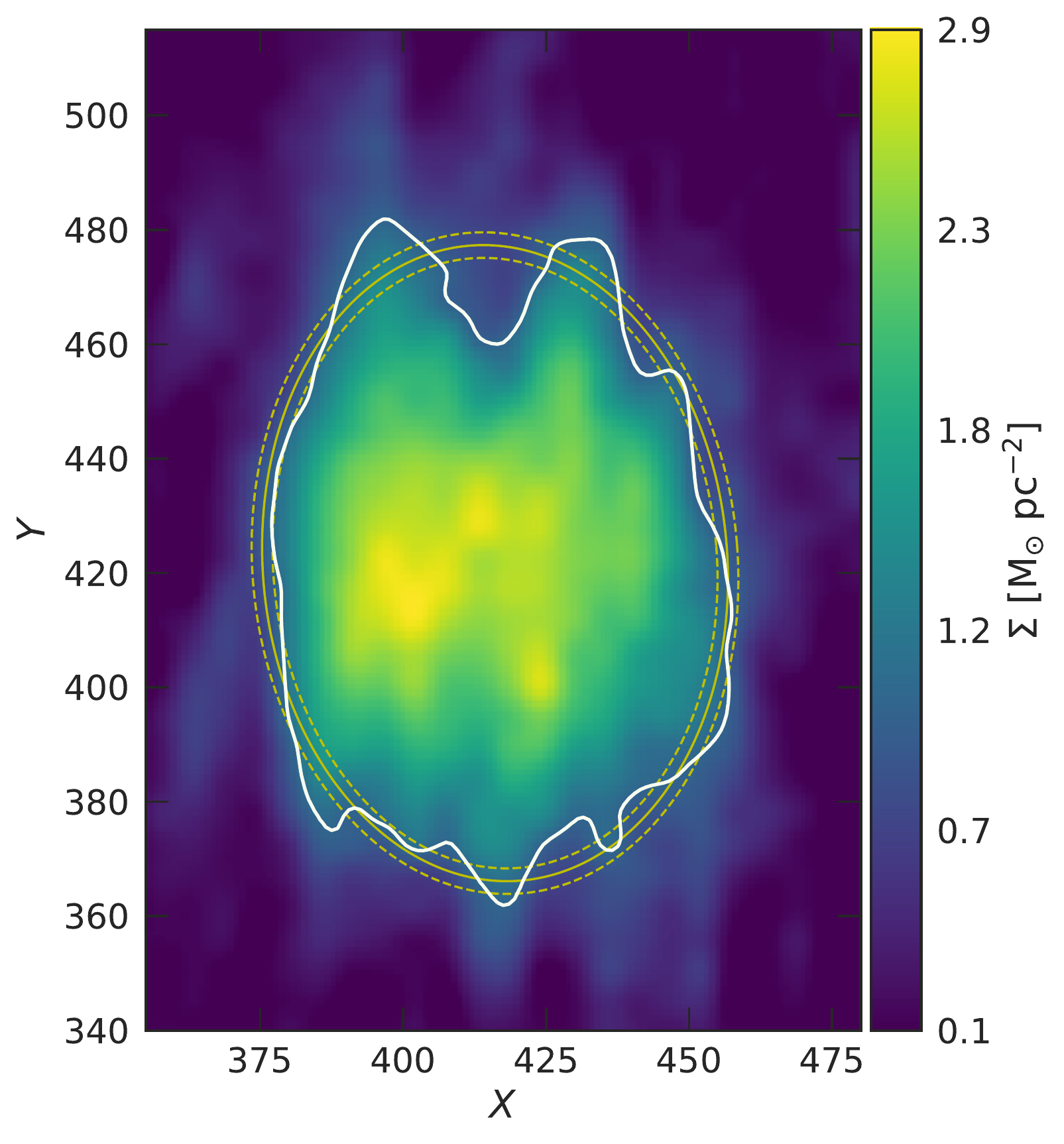}
                \caption{HIPASSJ1250-20:S1}
            \end{subfigure}\\
            \begin{subfigure}[t]{0.4\textwidth}
                \centering
                \includegraphics[width=0.9\textwidth]{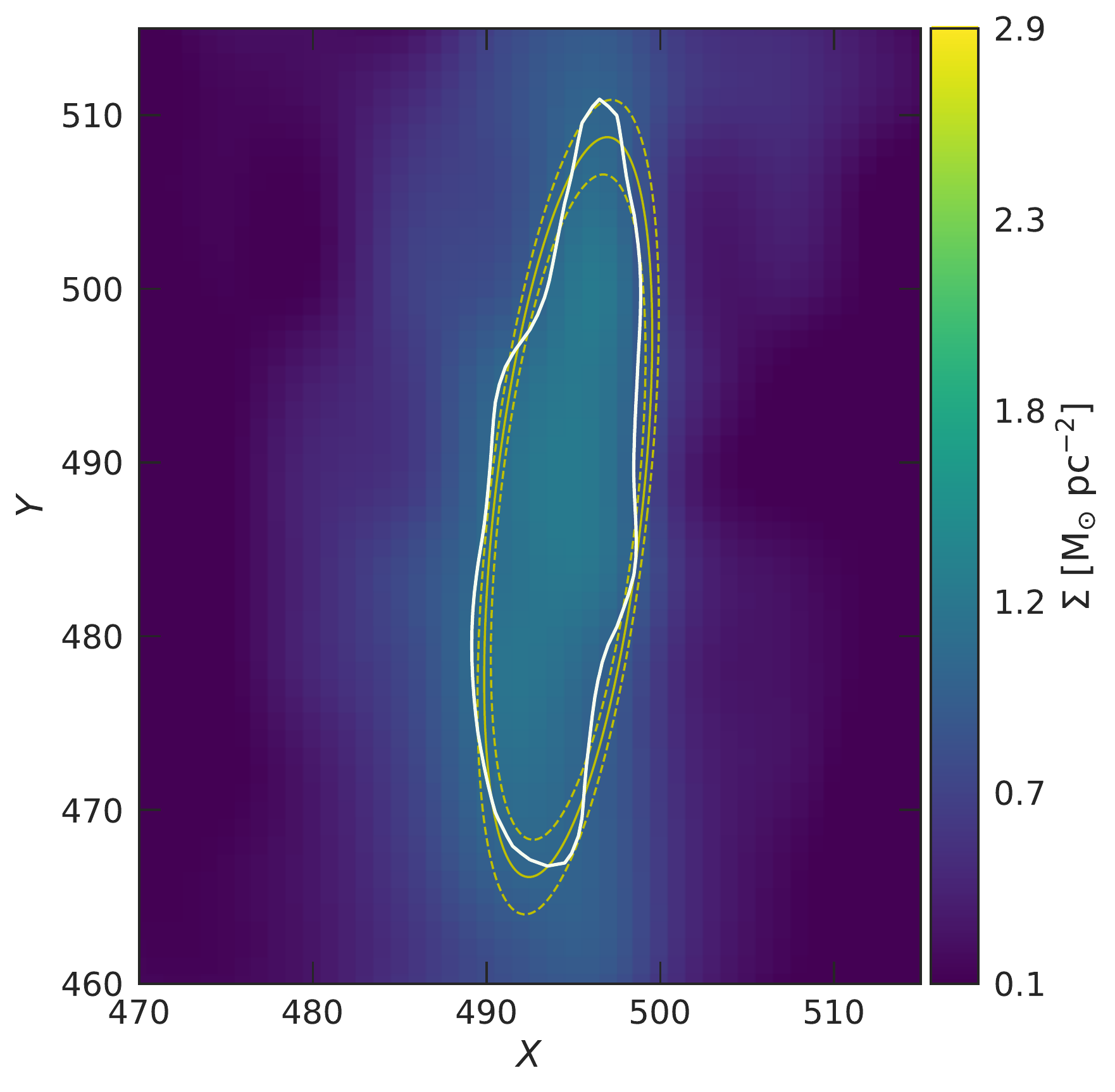}
                \caption{HIPASSJ1250-20:S2}
            \end{subfigure}
        \end{tabular}\\
    \caption{The measurement of the \HI\ diameter at a surface density of 1 M$_{\odot}$ pc$^{-2}$ using deprojected intensity maps. The white contour shows the \HI\ surface density of 1 M$_{\odot}$ pc$^{-2}$ . The yellow solid ellipse is the fit result from the \texttt{KMPFIT} package with the uncertainty shown as dashed ellipses. The colour bar shows the \HI\ surface density from the lowest value above 3$\sigma$ detection (0.1 M$_{\odot}$ pc$^{-2}$) to the maximum value (2.9 M$_{\odot}$ pc$^{-2}$).}
    \label{diametersS1S2}
\end{figure}

The method of measuring \HI\ diameters is described in Section \ref{sizemass}, here we show the observed and deprojected intensity maps (Figure \ref{intensitymaps} (a) and (b) respectively) of HIPASSJ1250-20 group and the results from the measurement of the \HI\ diameters (Figure \ref{diametersS1S2}: (a) HIPASSJ1250-20:S1; (b) HIPASSJ1250-20). In Figure \ref{intensitymaps} and Figure \ref{diametersS1S2}, images have pixel scale (X and Y), where one pixel corresponds to $\sim$1 kpc. Images used here are regridded using DECam optical image as a map, this was performed in MIRIAD with the task \texttt{REGRID}. The measured D$_{\HI}$ without the regridding has larger uncertainty, also the smaller galaxies (such as S2) due to larger beam size, have large beam corrections. The measured D$_{\HI}$ using normal and regridded images are in agreement: D$_{\HI}$[S1$_{\textrm{normal}}$]=105.7$\pm$8.7; D$_{\HI}$[S1$_{\textrm{regrid}}$]=106.4$\pm$4.7;
D$_{\HI}$[S2$_{\textrm{normal}}$]=24.1$\pm$15.2;
D$_{\HI}$[S2$_{\textrm{regrid}}$]=29.2$\pm$9.1.

\end{document}